\documentclass[12pt]{article}
\pdfoutput=1
\usepackage[square,comma,numbers,sort&compress]{natbib}
\usepackage{graphicx,epstopdf,amssymb,amsfonts,amsmath,amsthm,array,
mathrsfs,amscd}
\usepackage[vcentermath]{youngtab}
\usepackage{float}
\newcommand{\pbs}[1]{\let\temp=\\#1\let\\=\temp}
\numberwithin{equation}{section}
\def\be{\begin{equation}}\def\ee{\end{equation}}
\def\cvp{\raise 2pt\hbox{,}} 
 \def\tr{\mathop{\rm tr}\nolimits}
\def\im{\mathop{\rm Im}\nolimits}
\def\re{\mathop{\rm Re}\nolimits}  

 \def\d{{\text d}}

\def\La{\Lambda}\def\km{k_{\mu}}\def\lag{\mathscr L}\def\pol{\mathscr P}
\def\rs{\rho_{\text s}}\def\rd{\rho_{\text d}}\def\rn{\rho_{0}}
\def\Ap{\mathsf{A}_{\km,\delta}^{+}}\def\Am{\mathsf{A}_{\km,\delta}^{-}}
\def\Acoef{\mathsf{A}_{\km,\delta}^{m,k,p}}
\def\eN{\underset{N}{=}}
\def\Ga{G^{\text a}}\def\Ge{G^{\text e}}

\def\cmp#1#2#3{{\it Comm.\ Math.\ Phys.\ }{\bf #1} (#2) #3}

\def\epjc#1#2#3{{\it Eur.\ Phys.\ J.\ }{\bf C #1} (#2) #3}

\def\imath#1#2#3{{\it Invent math }{\bf #1} (#2) #3}

\begin{document}
%
%
{\pagestyle{empty}
\begin{flushright} CERN-TH-2016-023 \end{flushright}
\parskip 0in
\

\vfill

\begin{center}
%
%

{\LARGE The Analytic Renormalization Group}

\vspace{0.4in}

Frank F{\scshape errari}
\\

\medskip

{
\it Service de Physique Th\'eorique et Math\'ematique\\
Universit\'e libre de Bruxelles (ULB) and International Solvay Institutes\\
Campus de la Plaine, CP 231, B-1050 Bruxelles, Belgique

\smallskip

\it Theoretical Physics Department\\
CERN, CH-1211 Gen\`eve, Suisse}

%

\smallskip
{\tt frank.ferrari@cern.ch}
\end{center}
\vfill\noindent

Finite temperature Euclidean two-point functions in quantum mechanics or quantum field theory are characterized by a discrete set of Fourier coefficients $G_{k}$, $k\in\mathbb Z$, associated with the Matsubara frequencies $\nu_{k}=2\pi k/\beta$. We show that analyticity implies that the coefficients $G_{k}$ must satisfy an infinite number of model-independent linear equations that we write down explicitly. In particular, we construct ``Analytic Renormalization Group'' linear maps $\mathsf A_{\mu}$ which, for any choice of cut-off $\mu$, allow to express the low energy Fourier coefficients for $|\nu_{k}|<\mu$ (with the possible exception of the zero mode $G_{0}$), together with the real-time correlators and spectral functions, in terms of the high energy Fourier coefficients for $|\nu_{k}|\geq\mu$. Operating a simple numerical algorithm, we show that the exact universal linear constraints on $G_{k}$ can be used to systematically improve any random approximate data set obtained, for example, from Monte-Carlo simulations. Our results are illustrated on several explicit examples.
 
\vfill

\medskip
%
\begin{flushleft}
\today
\end{flushleft}
%
\newpage\pagestyle{plain}
\baselineskip 16pt
\setcounter{footnote}{0}

}


%
\section{\label{IntroSec} General presentation}

Consider the space $\mathscr M$ of arbitrary two-point functions between bosonic operators $A$ and $B$ in Quantum Mechanics or Quantum Field Theory, at finite temperature $T=1/\beta$.\footnote{We focus on the case of bosonic operators in the present paper. The case of fermionic operators can be discussed along the same lines, with minor and straightforward modifications.} As is well-known and will be reviewed in details in Section \ref{DefSec}, the space $\mathscr M$ can be presented in several equivalent ways. One can consider various real-time two-point functions (advanced, retarded, time-ordered, etc.), which turn out to be all related to each other, since their Fourier transforms can  be expressed in terms of a unique spectral function $\rho(\omega)$. Alternatively, one can work with the Euclidean-time two-point function $G(\tau)$. By the KMS condition, $G$ is periodic and can be expanded in Fourier series, 
\be\label{GFourier} G(\tau) = \frac{1}{\beta}\sum_{k\in\mathbb Z} G_{k}e^{-i\nu_{k}\tau}\, ,\ee
where the Matsubara frequencies are defined by
\be\label{Matdef} \nu_{k}= 2\pi k/\beta\, .\ee
We shall often refer to the set of Fourier coefficients $(G_{k})_{k\in\mathbb Z}$ as the ``data'' which encodes the two-point function. In a generic strongly coupled quantum mechanical model, this data can only be computed numerically, using Monte-Carlo numerical simulations. Analytic non-perturbative methods exist only in rare occasions.\footnote{See e.g.\ \cite{FerBH} for a recent example from which the investigations presented in this paper originated.}

By Carlson's theorem \cite{Rubel}, the real-time and Euclidean-time points of view are equivalent: the continuous spectral function $\rho(\omega)$ can be expressed in terms of the discrete set of Fourier coefficients $G_{k}$ and vice-versa, under some very general assumptions that are valid in all known interesting physical theories.\footnote{In Quantum Field Theory, two-point functions of local operators do not in general satisfy the hypothesis of Carlson's theorem, due to the usual UV divergences at coinciding points. These divergences are governed by the Operator Product Expansion which, in asymptotically free theories, can be reliably computed in perturbation theory. This problem is handled in a standard way: one either considers smeared versions of the local operators or, more generally, one subtracts explicitly the diverging piece in the correlator using the OPE.} The map between the real-time and the Euclidean-time formalism is quite interesting and will be discussed very explicitly below.

The two-point functions must satisfy general well-known constraints that follow straightforwardly from the definitions and the spectral decomposition, see Section \ref{DefSec}. For example, on top of being $\beta$-periodic, $G(\tau)$ is analytic except at the points $\tau = k\beta$, $k\in\mathbb Z$, where it is discontinuous if $A$ and $B$ do not commute. This implies in particular that $G_{k}=O(1/k)$ at large $|k|$. The Fourier coefficients also satisfy reality and positivity constraints depending on the reality properties of $A$ and $B$. We shall call $\mathscr F$ the real vector space of $\beta$-periodic functions satisfying all these standard model-independent constraints.

One of the main goal of the present work is to show that $\mathscr M$ is a linear subspace of $\mathscr F$ of infinite codimension. This may come as a surprise. It means that a typical set of Fourier coefficients $(G_{k})_{k\in\mathbb Z}$ satisfying all the usual constraints is actually inconsistent! Our central result is to show that \emph{the Fourier coefficients must always obey an infinite set of universal, model-independent, linear equations}. For reasons that will become clear below, we call these equations ``Analytic Renormalization Group'' (ARG) equations.  We shall write down these equations very explicitly in Section \ref{ARGSec} and use them extensively in Sections \ref{ExSec} and \ref{useARGSec}.

A spectacular concrete application of the existence of the ARG equations is as follows. Suppose that we have at our disposal an approximate data set $(\Ga_{k})_{k\in\mathbb Z}$, obtained, in a strongly coupled model of interest, by using Monte-Carlo simulations, possibly combined with perturbation theory at high energies. This data set corresponds to a point $\Ga\in\mathscr F$, randomly chosen in a small neighborhood of the exact, but unknown, result $\Ge\in\mathscr M$. The point $\Ga$ will never belong to $\mathscr M$, because a random approximate data set always violate the ARG equations (actually, this violation is always massive, even if the precision of the data is excellent; see below). It is then possible to use the ARG equations to systematically improve the approximate data, by suitably projecting $\Ga$ onto $\mathscr M$. An explicit  algorithm implementing this idea will be presented and tested in Section \ref{useARGSec}. In spite of its simplicity, our algorithm is able to improve the accuracy of typical random approximate data by a factor of 2 to 4! The startling feature is that the procedure is totally model-independent and can be applied straightforwardly to Monte-Carlo data in any quantum mechanical system.

The fundamental ingredient at the basis of the ARG equations is \emph{analyticity}, which is itself a consequence of causality. The fact that analyticity yields non-trivial constraints on real-time correlation functions is of course well-known. For example, the famous Plemelj-Kramers-Kronig identities relate the real and imaginary parts of the Fourier transform of the retarded two-point function,
\be\label{KK} \re\tilde\chi_{\text r}(\omega) =\frac{1}{\pi}\text{P}\!\!\int_{-\infty}^{+\infty}\!\frac{\im \tilde\chi_{\text r}(\omega')}{\omega'-\omega}\,\d\omega'\, ,\quad \im\tilde\chi_{\text r}(\omega) =-\frac{1}{\pi}\text{P}\!\!\int_{-\infty}^{+\infty}\!\frac{\re \tilde\chi_{\text r}(\omega')}{\omega'-\omega}\,\d\omega'\, .\ee
The starting point of our work is the very same analyticity property at the basis of \eqref{KK}, but we use it in a more sophisticated way to derive the infinite set of linear constraints that the Fourier coefficients $G_{k}$ must satisfy.

An interesting aspect of the construction is to make an unexpected link, valid in any quantum mechanical system, between the concept of Renormalization Group (RG) and analyticity. The fundamental idea of the RG is to describe the physics below a certain RG scale $\mu$ in terms of a Wilsonian action $S_{\mu}$ that takes into account the physics at scales greater than $\mu$. When $\mu$ is lowered, the action $S_{\mu}$ flows to a natural description of the physics at low energies. This flow is constrained by the obvious fact that physics is independent of the arbitrary RG scale. This yields the RG flow equations. The Analytic version of the RG that we find (the ARG) can be described as follows. To an arbitrary RG scale $\mu$, we associate the integer $\km>0$ (that we also call the RG scale by abuse of language) such that
\be\label{kmdef} \nu_{\km-1}\leq\mu < \nu_{\km}\, .\ee
We also introduce a strictly positive integer $\delta$, that we call the ``index'' of the ARG. The ARG states that there exists linear maps $\mathsf{A}^{+}_{\km,\delta}$ and $\mathsf{A}^{-}_{\km,\delta}$ allowing to express the low energy Fourier coefficients, for $1\leq k <\km$ and $-\km<k\leq -1$ respectively, in terms of the high energy Fourier coefficients $G_{\km + \delta q}$ and $G_{-\km-\delta q}$, $q\in\mathbb N$, respectively:
\begin{align}\label{ARGmapi} \bigl( G_{1},G_{2},\ldots,G_{\km-1}\bigr)& = \Ap\bigl(G_{\km},G_{\km + \delta}, G_{\km+2\delta},G_{\km+3\delta},\ldots\bigr)\, ,\\
\label{ARGmapii} \bigl( G_{-1},G_{-2},\ldots,G_{-\km+1}\bigr) &= \Am\bigl(G_{-\km},G_{-\km - \delta}, G_{-\km-2\delta},G_{-\km-3\delta},\ldots\bigr)\, .
\end{align}
Note that the zero mode $G_{0}$ plays a special role since it cannot be obtained, in general, from the ARG. This subtlety is related to the phenomenon of Bose-Einstein condensation (see e.g.\ \cite{FerBH}).

The above ARG maps $\mathsf{A}_{\km,\delta}^{\pm}$ are not the most general one can build. The most general maps allow to reconstruct the analytic continuations of the Fourier transforms of the retarded and advanced two-point functions, $\tilde\chi_{\text r}(z)$ and $\tilde\chi_{\text a}(z)$, in the upper and lower half complex plane respectively. This, in particular, automatically provides explicit maps between the real-time and the Euclidean-time formalism. The general relations are of the form
\begin{align}\label{ARGmapi2} \tilde\chi_{\text r}(\omega+i\Omega)& = \Ap (\omega,\Omega)\bigl(G_{\km},G_{\km + \delta}, G_{\km+2\delta},G_{\km+3\delta},\ldots\bigr)\, ,\\
\label{ARGmapii2} \tilde\chi_{\text a}(\omega-i\Omega) &= \Am(\omega,\Omega)\bigl(G_{-\km},G_{-\km - \delta}, G_{-\km-2\delta},G_{-\km-3\delta},\ldots\bigr)\, ,
\end{align}
for any $0<\Omega\leq\mu$. Explicit formulas for $\mathsf{A}_{\km,\delta}^{\pm}(\omega,\Omega)$ are given in Section 3. From \eqref{ARGmapi2} and \eqref{ARGmapii2}, one can straightforwardly obtain similar ARG maps expressing  the real-time correlators $\chi_{\text r}(t)$ and $\chi_{\text a}(t)$, the spectral function, or any other real-time two-point function in terms of the Euclidean Fourier coefficients above any RG scale $\km$. The ARG maps \eqref{ARGmapi} and \eqref{ARGmapii} can also be easily obtained from \eqref{ARGmapi2} and \eqref{ARGmapii2}. Full details will be given below.

The low energy Fourier coefficients $G_{k}$, for $|k|<\km$, obviously do not depend on the arbitrary choice of RG scale $\km$ and index $\delta$. On the other hand, the right-hand sides of the equations \eqref{ARGmapi} and \eqref{ARGmapii} depend explicitly, and non-trivially, on  $\km$ and $\delta$. As usual, this yields renormalization group equations. The same remark applies to the equations \eqref{ARGmapi2} and \eqref{ARGmapii2}. 
More generally, by abuse of language, we call ``ARG equation'' any universal linear relation between the Fourier coefficients like \eqref{ARGmapi} or \eqref{ARGmapii}.

Our results rely in an absolutely crucial way on a generalization of mathematical techniques first introduced in a remarkable paper by Cuniberti et al.\ \cite{Cuniberti}.\footnote{See also \cite{Burnier} for a concrete discussion of the construction in \cite{Cuniberti}.} The aim of \cite{Cuniberti} was to provide explicit formulas for the reconstruction of the real-time correlators from the Euclidean-time correlators, a notoriously difficult and important problem. One particular aspect of our results is to provide a useful generalization and simplification of the reconstruction procedure of \cite{Cuniberti}. 

The plan of the paper is as follows. In Section \ref{DefSec}, we review basic facts on two-point functions in quantum mechanics: definitions of real-time and Euclidean-time correlators, spectral decompositions and spectral function, the resolvent, analytic properties and Carlson theorem. Section \ref{ARGSec} is devoted to the derivation of the ARG maps and equations. We start in \ref{BISec} explaining a simple idea at the basis of the ARG. We then present in \ref{MathSec} some useful mathematical results on Laguerre polynomials, Pollaczek polynomials and the relation between them. These results are used in \ref{gARGSec} to build the general ARG maps $\mathsf A^{\pm}_{\km,\delta}(\omega,\Omega)$. We specialize to the case of the maps $\smash{\mathsf A^{\pm}_{\km,\delta}}$ in \ref{AmapSec} and to the reconstruction of the real-time correlators from the Euclidean data in \ref{realSec}. This eventually yields a very general multi-parameter continuous family of ARG equations. In Section \ref{ExSec}, we discuss the numerical implementation of the ARG. We use in particular the example of the damped harmonic oscillator to illustrate our results. In Section \ref{useARGSec}, we explain how the ARG equations can be used to systematically improve any given random approximate data set. This is certainly the newest, most surprising and most central concrete application of our work. We provide a simple numerical algorithm that we test successfully on several examples. Finally, we briefly summarize our results and suggest future directions of research in Section \ref{SecConc}.

\section{\label{DefSec} Preliminaries}
\subsection{Basic definitions}

Let $H$ be the Hamiltomian. To simplify some formulas, we do as if the spectrum of $H$ were discrete. The generalization to the case of a continuous spectrum is completely straightforward and the required modifications will be taken into account in our discussion.\footnote{Note also that most systems with a continuous spectrum can be obtained by taking the appropriate thermodynamic limit of a compact system with a discrete spectrum.} Let $\{|p\rangle\}$ be an orthonormal basis of energy eigenstates, $H|p\rangle = E_{p}|p\rangle$. 

The partition function at temperature $T=1/\beta$ is
\be\label{Zdef} Z=\tr e^{-\beta H}=\sum_{p}e^{-\beta E_{p}}\, .\ee
The expectation value of any operator $\mathscr O$ at temperature $T$ is defined by
\be\label{vevdef}\langle\mathscr O\rangle_{\beta} = \frac{1}{Z}\tr \bigl(e^{-\beta H}\mathscr O\bigr)=\frac{1}{Z}\sum_{p}e^{-\beta E_{p}}\langle p|\mathscr O|p\rangle\, .\ee
The real-time and Euclidean-time evolutions are defined as usual by
\be\label{timeevol} \mathscr O(t) = e^{itH}\mathscr O e^{-itH}\, ,\quad 
\mathscr O_{\text E}(\tau) = e^{\tau H}\mathscr O e^{-\tau H}\ee
respectively.

We consider two particular bosonic operators $A$ and $B$ and denote their matrix elements as 
\be\label{ABmelem} A_{pq} = \langle p | A|q\rangle\, ,\quad B_{pq} = \langle p | B|q\rangle\, .\ee
The spectral function is defined by
\be\label{specdef} \rho(\omega) = \frac{1}{Z}\sum_{\substack{p,q\\E_{p}\not = E_{q}}}\bigl( e^{-\beta E_{p}}-e^{-\beta E_{q}}\bigr) A_{pq} B_{qp}\delta (\omega + E_{p}-E_{q}) + \beta n_{0}\omega\delta(\omega)\, ,\ee
where the zero-frequency contribution\footnote{This contribution is associated with the phenomenon of Bose-Einstein condensation. See \cite{FerBH} for a recent discussion.} reads
\be\label{n0def}n_{0}=\frac{1}{Z}\sum_{\substack{p,q\\E_{p} = E_{q}}}e^{-\beta E_{p}}A_{pq}B_{qp}\, .\ee
By using the $\delta$-function constraint, we may also rewrite \eqref{specdef} as
\be\label{rewrite}\rho(\omega)= \frac{1-e^{-\beta\omega}}{Z}\sum_{p,q}e^{-\beta E_{p}}A_{pq}B_{qp}\delta(\omega+E_{p}-E_{q})\, ,\ee
where the sum over energy eigenstates is now unconstrained and thus includes the terms with $E_{p}=E_{q}$. Taking into account a possible continuous part in the spectrum of $H$, the spectral function can be written as a sum
\be\label{rhosum}\rho(\omega) = \rs(\omega) + \rd(\omega) + \rn(\omega)\, ,\ee
where $\rho_{\text s}$ is a smooth function associated with the continuous spectrum, $\rho_{\text d}$ is a sum of $\delta$-function contributions centered at non-zero frequencies associated with the discrete spectrum and $\rn(\omega)=\beta n_{0}\omega\delta(\omega)$ is the zero-frequency contribution.

\noindent\emph{Remark}: our subsequent discussion does not depend on reality conditions on the operators $A$ and $B$. However, let us note that, in the typical case $B=A^{\dagger}$, the representation \eqref{rewrite} implies that $\rho$ is a real function, positive for $\omega>0$ and negative for $\omega<0$.

\subsection{Real-time correlators}

We define
\be\label{Cdef} C(t) = \bigl\langle A(t) B\bigr\rangle_{\beta} = \frac{1}{2\pi}\int_{-\infty}^{+\infty}\! \tilde C(\omega) e^{-i\omega t}\,\d\omega\, .\ee
By inserting a complete set of states in the above definition, one can straightforwardly derive the following spectral decomposition,
\be\label{Crhorel} \tilde C(\omega) = \frac{2\pi}{Z}\sum_{p,q}e^{-\beta E_{p}}A_{pq}B_{qp}\delta(\omega+E_{p}-E_{q})\, .\ee
Using \eqref{rewrite}, this yields
\be\label{Crhorel2} \tilde C(\omega) = \frac{2\pi}{1-e^{-\beta\omega}}\,\rho(\omega)\, .\ee
Note that terms with $E_{p}=E_{q}$ a priori contribute in the sum \eqref{Crhorel}. Accordingly, the zero-frequency piece in $\rho$ can contribute in an essential way to $\tilde C$. Similarly, other real-time two-point functions can be studied,
\begin{align}
\label{xidef} \xi(t)&=\frac{1}{2}\bigl\langle [A(t), B]\bigr\rangle_{\beta} = \frac{1}{2\pi}\int_{-\infty}^{+\infty}\! \tilde \xi(\omega) e^{-i\omega t}\,\d\omega\, ,\\
\label{Sdef} S(t)&=\frac{1}{2}\bigl\langle \{A(t), B\}\bigr\rangle_{\beta} = \frac{1}{2\pi}\int_{-\infty}^{+\infty}\! \tilde S(\omega) e^{-i\omega t}\,\d\omega\, ,\\
\label{chirdef} \chi_{\text r}(t) &= 2i\theta(t)\xi(t) = \frac{1}{2\pi}\int_{-\infty}^{+\infty}\! \tilde \chi_{\text r}(\omega) e^{-i\omega t}\,\d\omega\, ,\\
\label{chiadef} \chi_{\text a}(t) &= -2i\theta(-t)\xi(t) = \frac{1}{2\pi}\int_{-\infty}^{+\infty}\! \tilde \chi_{\text a}(\omega) e^{-i\omega t}\,\d\omega\, ,\\
\label{DFeyndef} D(t) & = \bigl\langle \text T A(t) B\bigr\rangle_{\beta}
 = \frac{1}{2\pi}\int_{-\infty}^{+\infty}\! \tilde D(\omega) e^{-i\omega t}\,\d\omega\, ,
\end{align}
where T is the usual time-ordering. It is straightforward to check that all these two-point functions can be expressed in terms of the spectral function, which thus contains all the relevant information,\footnote{As usual, $\epsilon$ is an infinitesimal strictly positive parameter.}
\begin{align}
\label{xirhorel} \tilde\xi(\omega) &= \pi\rho(\omega)\, ,\\
\label{Srhorel} \tilde S(\omega) & = \pi\frac{e^{\beta\omega}+1}{e^{\beta\omega}-1}\,\rho(\omega)\, ,\\
\label{chirrhorel} \tilde\chi_{\text r}(\omega) &= -\int_{-\infty}^{+\infty}\!\frac{\rho(\omega')}{\omega-\omega'+i\epsilon}\,\d\omega'\, ,\\
\label{chiarhorel} \tilde\chi_{\text a}(\omega) &= -\int_{-\infty}^{+\infty}\!\frac{\rho(\omega')}{\omega-\omega'-i\epsilon}\,\d\omega'\, ,\\
\label{Drhorel} \tilde D(\omega) &= i\int_{-\infty}^{+\infty}\frac{\rho(\omega')}{e^{\beta\omega'}-1}\biggl[\frac{e^{\beta\omega'}}{\omega-\omega'+i\epsilon}-\frac{1}{\omega-\omega'-i\epsilon}\biggr]\d\omega'\, .
\end{align}
Evaluating $\xi(0)$ and $S(0)$ from the above relations yields the following important sum rules,
\begin{align}\label{sr1} &\int_{-\infty}^{+\infty}\!\rho(\omega)\,\d\omega = \bigl\langle [A,B]\bigr\rangle_{\beta}\, ,\\\label{sr2}&
\int_{-\infty}^{+\infty}\!\frac{e^{\beta\omega}+1}{e^{\beta\omega}-1}\,\rho(\omega)\,\d\omega = \bigl\langle \{A,B\}\bigr\rangle_{\beta}\, .
\end{align}
In particular, taking into account the fact that $\rd$ picks contributions only at non-zero frequencies and that the integral on the left-hand side of \eqref{sr2} must converge, we find
\be\label{rscond}\rd(\omega=0)=0\, ,\quad \rs(\omega=0) = 0\, .\ee
\subsection{Euclidean-time correlator}

We define
\be\label{Gdef} G(\tau) = \bigl\langle \text{T} A_{\text E}(\tau) B\bigr\rangle_{\beta}\quad\text{for}\ -\beta<\tau<0\ \text{or}\ 0<\tau<\beta\, .\ee
The standard KMS condition reads 
\be\label{KMS} G(\tau) = G(\tau+\beta)\ee
if $-\beta<\tau<0$. We can thus expand $G$ in Fourier series as in \eqref{GFourier} and use this expansion to extend the definition of $G$ for all values of $\tau$ that are not multiples of $\beta$. Note that if $A$ and $B$ do not commute, $G$ is discontinuous at $\tau= k\beta$, $k\in\mathbb Z$, with
\be\label{Gdisc} G(0^{+}) - G(0^{-})  = \bigl\langle [A,B]\bigr\rangle_{\beta}\, .\ee
Moreover, from Dirichlet theorem, we get
\be\label{GDirich}G(0^{+}) + G(0^{-}) = \bigl\langle \{A,B\}\bigr\rangle_{\beta}=\frac{2}{\beta}\Bigl(G_{0}+\sum_{k=1}^{\infty}\bigl(G_{k}+G_{-k}\bigr)\Bigr)\, .\ee

The Fourier coefficients $G_{k}$ admit the following spectral decomposition,
\be\label{GkSD} G_{k} = \frac{1}{Z}\sum_{\substack{p,q\\E_{p}\not = E_{q}}}\frac{e^{-\beta E_{p}}-e^{-\beta E_{q}}}{E_{q}-E_{p}-i\nu_{k}}\, A_{pq} B_{qp} + \beta n_{0}\delta_{k,0}\, .\ee
It is important to note that, in general, the zero mode $G_{0}$ picks a  contribution proportional to $n_{0}$ defined in \eqref{n0def}.

Even more generally, one can consider the two-point function for complex time $t-i\tau$, $(t,\tau)\in\mathbb R^{2}$, defined by
\be\label{Gcomdef}\mathscr G(t-i\tau) = \bigl\langle\text{T}_{\tau}A(t-i\tau)B\bigr\rangle_{\beta}\quad\text{for}\ -\beta<\tau<0\ \text{or}\ 0<\tau<\beta\, .\ee
The symbol $\text{T}_{\tau}$ denotes the time-ordering with respect to the Euclidean time $\tau$. The KMS condition reads $\mathscr G(t-i\tau) = \mathscr G(t-i\tau-i\beta)$ for $-\beta<\tau<0$. $\mathscr G$ is then extended to the whole complex time plane by $\beta$-periodicity in $\tau$. It is analytic in the strips $k\beta<\tau<(k+1)\beta$, $k\in\mathbb Z$ and possibly discontinuous for $\tau = k\beta$ with $\mathscr G(t-i0^{+})-\mathscr G(t-i0^{-}) = 2\xi(t)$. Moreover, $\mathscr G(-i\tau)=G(\tau)$. The function $\mathscr G$ can be expressed in terms of the spectral density $\rho$. If we expand
\be\label{Gcurlexp} \mathscr G(t-i\tau) = \frac{1}{2\pi\beta}\sum_{k\in\mathbb Z}\int_{-\infty}^{+\infty}\!\tilde{\mathscr G_{k}}(\omega)\, e^{-i\omega t - i\nu_{k}\tau}\,\d\omega\, ,\ee
it is straightforward to check that
\be\label{Gtildekom} \tilde{\mathscr G_{k}}(\omega) = \frac{2\pi}{\omega-i\nu_{k}}\rho(\omega)\ee
or, equivalently, that
\be\label{Gcurlform} \mathscr G(t-i\tau) = \int_{-\infty}^{+\infty}\!\frac{e^{-i\omega(t-i\tau)}}{1-e^{-\beta\omega}}\,\rho(\omega)\d\omega\quad\text{for}\ 0<\tau<\beta\, .\ee
\subsection{The resolvent}

The resolvent is defined by
\be\label{resdef} R(z) = \int_{-\infty}^{+\infty}\!\frac{\rho(\omega)}{z-\omega}\,\d\omega\ee
for any complex $z$ with $\im z\not = 0$. This is equivalent to 
\be\label{resdefBIS} R(z) = \left\{
\begin{aligned} & -\int_{0}^{\infty}\!\chi_{\text{r}}(t)e^{izt}\,\d t\quad\text{if}\ \im z>0\\
& -\int_{-\infty}^{0}\!\chi_{\text{a}}(t)e^{izt}\,\d t\quad\text{if}\ \im z<0\, .
\end{aligned}\right.\ee

This is a very useful object and it is going to play a central role in our subsequent discussion. The spectral decomposition of $R$ reads
\be\label{resSD} R(z) = \frac{1}{Z}\sum_{\substack{p,q\\E_{p}\not = E_{q}}}\frac{e^{-\beta E_{p}}-e^{-\beta E_{q}}}{z+E_{p}-E_{q}}A_{pq}B_{qp}\, .\ee
Note that the zero-frequency piece $\rho_{0}$ in $\rho$ does not contribute to $R$; equivalently, states with $E_{p}=E_{q}$ do not contribute in \eqref{resSD}.

The resolvent has the following set of fundamental properties:

\noindent\ i) It is holomorphic in the half-planes $\im z>0$ and $\im z<0$.

\noindent\ ii) For any $\eta>0$ and $|\im z|\geq\eta$, $R$ has a simple large $|z|$ asymptotic expansion
\be\label{Rexp} R(z) = \frac{\langle [A,B]\rangle_{\beta}}{z} + O(1/z^{2})\, .\ee
This follows from the definition \eqref{resdef} and the sum rule \eqref{sr1}.

\noindent\ iii) When the spectrum of the Hamiltonian is discrete, the only singularities of $R$ are simple poles on the real axis at the Bohr frequencies $E_{q}-E_{p}\not = 0$.

\noindent\ iv) More generally, when the spectral function is decomposed as in \eqref{rhosum}, the resolvent is discontinuous accross the support of $\rs+\rd$, with 
\be\label{resdisc}\rs(\omega)+\rd(\omega) = \frac{i}{2\pi}\bigl( R(\omega+i\epsilon) - R(\omega-i\epsilon)\bigr)\, .\ee
In particular, taking into account \eqref{rscond}, we see that $R(0)$ is well-defined.

\noindent\ v) The real-time correlators $\xi(t)$, $\chi_{\text r}(t)$ and $\chi_{\text a}(t)$, that do not depend on the zero-frequency piece in the spectral function, can be obtained from the knowledge of $R$ alone. This is a direct consequence of the relation \eqref{resdisc}. In particular,
\be\label{chiRrel} \tilde\chi_{\text r}(\omega) = -R(\omega+i\epsilon)\, ,\quad \tilde\chi_{\text a}(\omega) = -R(\omega-i\epsilon)\, .\ee
On the other hand, the real-time correlators $C(t)$, $S(t)$ and $D(t)$ are given in terms of $R$ up to a time-independent piece given by $n_{0}$ in each case.

\noindent\ vi) The function $-R$ yields the analytic continuations to complex frequencies of $\tilde\chi_{\text r}$ and $\tilde\chi_{\text a}$ for $\im\omega>0$ and $\im\omega<0$ respectively. This is a direct consequence of \eqref{chiRrel}. 

\noindent\ vii) The Euclidean-time correlator $G(\tau)$ can be obtained from  $R$ up to the time-independent piece $n_{0}$. Indeed, the Fourier coefficients $G_{k}$ are given by 
\be\label{GkRrel} G_{k}=-R(i\nu_{k}) + \beta n_{0}\delta_{k,0}\, .\ee
This is a direct consequence of the spectral representations \eqref{GkSD} and \eqref{resSD}.

\subsection{The Carlson's theorem}

From the knowledge of the real-time correlator $C(t)$, we get the spectral density, including the zero-frequency piece proportional to $n_{0}$, by using \eqref{Crhorel2}. We then get the Euclidean-time correlator from \eqref{Gcurlform}. The analytic continuation from real-time to Euclidean-time is thus rather straightforward.

At non-zero temperature, the converse is much more subtle. The Euclidean-time physics is coded in the set of Fourier coefficients $G_{k}$ associated with the discrete Matsubara frequencies $\nu_{k}=2\pi k/\beta$, whereas the real-time physics is determined by the spectral function $\rho$ defined for all real frequencies $\omega$. To go from the Euclidean time to the real time, one must thus convert a discrete set of data into a continuous set of data. The fact that this can be done in a unique way is ensured by the famous Carlson's theorem. From the holomorphicity of the resolvent $R$ in the half-planes $\im z>0$ and $\im z<0$ and the asymptotic behaviour \eqref{Rexp}, the theorem implies that $R$ is uniquely determined on the upper half-plane by the values $R(i\nu_{k})=-G_{k}$ for $k\geq 1$ and on the lower half-plane by the values $R(i\nu_{k})=-G_{k}$ for $k\leq -1$. One can then obtain the full spectral density from the $G_{k}$s: the smooth and discrete pieces are derived from \eqref{resdisc} and the zero frequency piece is derived from \eqref{GkRrel} at $k=0$, $n_{0}=(G_{0}+R(0))/\beta$.

The Carlson's theorem thus implies that real-time and Euclidean-time data are equivalent. However, it does not provide a constructive way to obtain $\rho(\omega)$ from the $G_{k}$s. One application of our results, presented in the next section, is to obtain an infinite set of equivalent explicit reconstruction procedures, generalizing the results of Cuniberti et al.\ \cite{Cuniberti}.

\section{\label{ARGSec} The Analytic Renormalization Group}
\subsection{\label{BISec} Basic idea}

\begin{figure}
\centerline{\includegraphics[width=5in]{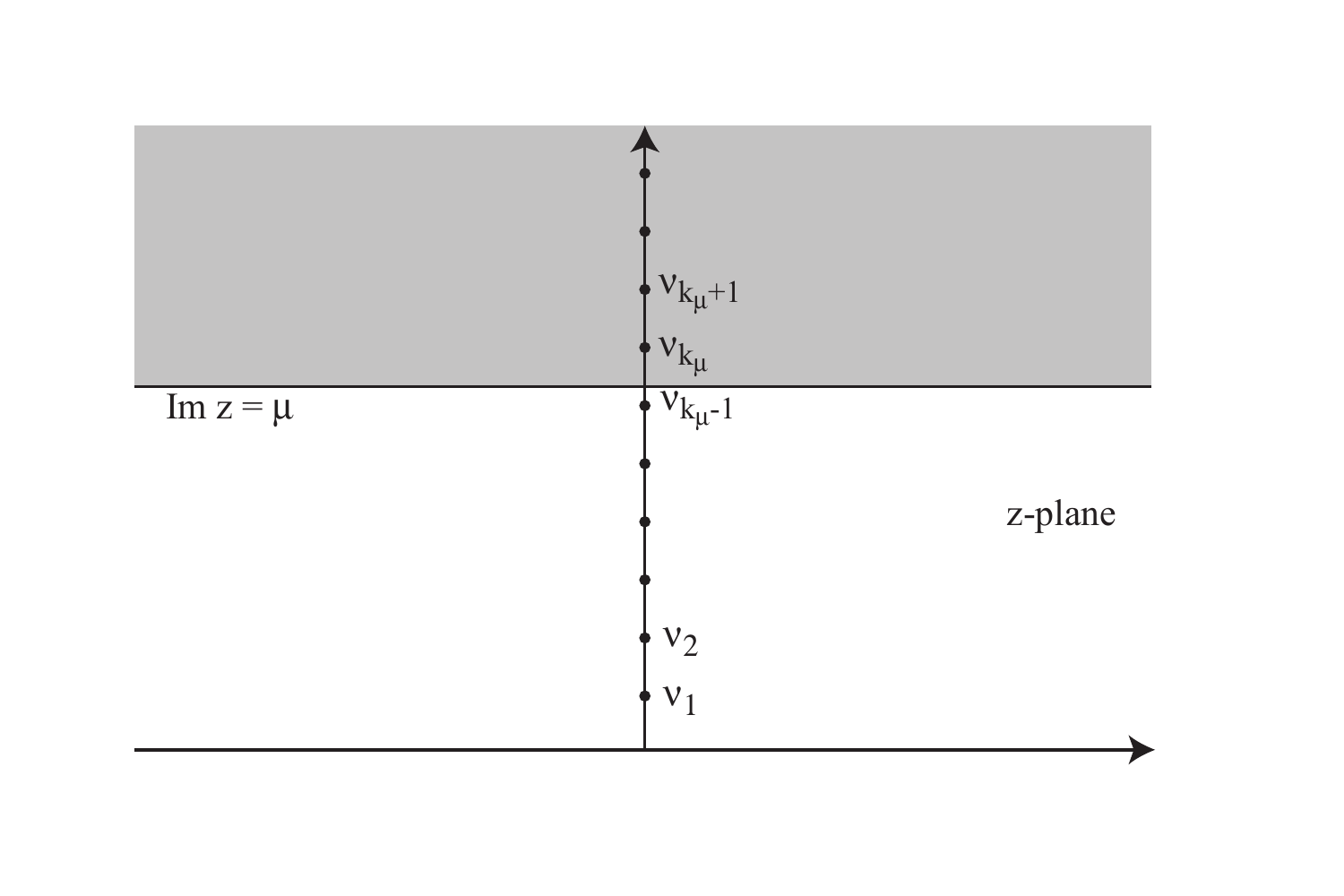}}
\caption{The upper half $z$-plane, $\im z>0$. The Fourier coefficients $G_{k}$ in the UV region above the cut-off $\mu$ ($k\geq\km$, gray area), determine $R$ for any $z$ in this region and thus, by analytic continuation, for any $z$ in the upper half-plane. As a consequence, the Fourier coefficients $G_{k}=-R(i\nu_{k})$ in the IR region below the cut-off $\mu$ ($0< k <\km$, white area), together with the spectral function and all two-point correlators, are fixed in terms of the Fourier coefficients in the UV region. This is the ARG map.
\label{fig1}}
\end{figure}

Let us pick an arbitrary RG scale $\mu>0$ and let $\km\in\mathbb N^{*}$ be defined as in \eqref{kmdef}. The existence of the ARG map relies on a very simple idea, which is illustrated on Fig.\ \ref{fig1}. We consider the resolvent $R$ in the domain $\im z>\nu_{\km -1}$. By applying the standard Carlson's theorem to the function $R_{\mu}(z)=R(z-\nu_{\km -1})$, we find that $R$ for $\im z>\nu_{\km -1}$ is uniquely determined by the Fourier coefficients $G_{k}$ for $k\geq\km$. But since $R$ is holomorphic for $\im z>0$, the principle of analytic continuation implies that the knowledge of $R$ for $\im z>\nu_{\km -1}$ uniquely fixes $R$ on the whole upper half-plane $\im z>0$. In particular, all the low energy Fourier coefficients $G_{k}=-R(i\nu_{k})$ for $1\leq k\leq\km-1$ are then fixed in terms of the high energy Fourier coefficients $G_{k}$ for $k\geq\km$. In other words, there must exist a map $\mathsf{A}_{\km}^{+}$ such that
\be\label{alpplus} (G_{1},G_{2},\ldots,G_{\km -1}) = \mathsf{A}_{\km}^{+}(G_{\km},G_{\km + 1},\ldots)\, .\ee
Similarly, there must exist a map $\mathsf{A}_{\km}^{-}$ such that
\be\label{alpminus} (G_{-1},G_{-2},\ldots,G_{-\km +1}) = \mathsf{A}_{\km}^{-}(G_{-\km},G_{-\km - 1},\ldots)\, .\ee
Moreover, the maps $\mathsf{A}_{\km}^{\pm}$ are linear, being the composition of the linear maps between the $G_{k}$ for $|k|\geq\km$ and $R$, and between $R$ and the $G_{k}$ for $|k|<\km$, $k\not = 0$.

One can actually further refine the above reasoning. The analytic function $R_{\mu}$ is also completely fixed by its values at $z=i\nu_{\km+\delta k}$, for any strictly positive integer $\delta$ and $k\geq 0$. This yields the general ARG maps of ``index $\delta$,'' $\mathsf{A}_{\km,\delta}^{\pm}$ of \eqref{ARGmapi} and \eqref{ARGmapii}. Similarly, we also get the maps $\mathsf{A}_{\km,\delta}^{\pm}(\omega,\Omega)$ of \eqref{ARGmapi2} and \eqref{ARGmapii2}, since the advanced and retarded correlators can be obtained from the resolvent.\footnote{An even stronger result can be derived \cite{Rubel}. Let $\sigma\subset\mathbb N^{*}$. Let $\sigma(k)$ be the number of elements in $\sigma$ that are less than or equal to $k$. Then, if $\limsup_{k\rightarrow\infty} \sigma(k)/k = 1$, there exists a linear ARG map $\smash{\mathsf{A}_{\mu,\delta,\sigma}^{+}}(\omega,\Omega)$ acting on the coefficients $G_{\km + \delta k}$ for $k\in \sigma$. A similar map $\smash{\mathsf{A}_{\mu,\delta,\sigma}^{-}}(\omega,\Omega)$ also exists.} Our goal, in the remaining of this section, is to find explicit expressions for these maps.

\subsection{\label{MathSec} On Laguerre and Pollaczek functions}

We now briefly review some useful results on Laguerre and Pollaczek polynomials.

For any real $a>-1$ and integer $n\geq 0$, generalized Laguerre polynomials can be defined in terms of Kummer's confluent hypergeometric function by
\be\label{Lagdef} L_{n}^{(a)}(t) = \frac{(a +1)_{n}}{n!} {}_{1}F_{1}(-n,a+1,t)\, ,\ee
where
\be\label{Pochdef} (u)_{n} = u (u+1)\cdots (u+n-1)=\frac{\Gamma(u+n)}{\Gamma (u)}\ee
denotes the usual Pochhammer symbol. The $L_{n}^{(a)}$ are degree $n$ polynomials, with
\be\label{Lagatzero} L_{n}^{(a)}(0) = \frac{(a +1)_{n}}{n!}\,\cdotp\ee
The associated Laguerre functions
\begin{align}\label{Lagfuncdef} \lag_{n}^{(a)}(t) &= \sqrt{\frac{n!}{\Gamma(n+1+a)}}\, t^{a/2}e^{-t/2}L_{n}^{(a)}(t)\\
& = \sqrt{\frac{\Gamma(n+1+a)}{n!}}\, \frac{t^{a/2}e^{-t/2}}{\Gamma(a+1)}\, {}_{1}F_{1}(-n,a+1,t)
\, ,\quad n\geq 0\, ,
\end{align}
form a real complete orthonormal basis of the Hilbert space $L^{2}(\mathbb R^{+})$,
\begin{align}\label{OrthLag} \int_{0}^{\infty}\!\lag_{n}^{(a)}(t)\lag_{m}^{(a)}(t)\, \d t &= \delta_{n,m}\, \\\label{CompLag} \sum_{n= 0}^{\infty}\lag_{n}^{(a)}(t)\lag_{n}^{(a)}(t') & = \delta(t-t')\, .
\end{align}

Similarly, for any real $\alpha>0$ and integer $n\geq 0$, we define the generalized Pollaczek polynomials in terms of the ordinary hypergeometric function $F={}_{2}F_{1}$ by
\be\label{Polldef} P_{n}^{(\alpha)}(x) = i^{n}\frac{(2\alpha)_{n}}{n!} F(-n,\alpha+i x,2\alpha,2)\, .\ee
The $P_{n}^{(\alpha)}$ are degree $n$ polynomials. The factor $i^{n}$ is inserted to make them real. It will be useful to know that
\be\label{Polparity} P_{n}^{(\alpha)}(-x) = (-1)^{n}P_{n}^{(\alpha)}(x)\ee
and that
\be\label{Polatzero} P_{2m}^{(\alpha)}(0) = (-1)^{m}\frac{(\alpha)_{m}}{m!}\,\cdotp\ee
The associated Pollaczek functions
\begin{align}\label{Polfuncdef} \pol_{n}^{(\alpha)}(x) &= 2^{\alpha}\sqrt{\frac{n!}{2\pi\Gamma(n+2\alpha)}}\,\Gamma(\alpha + i x) P_{n}^{(\alpha)}(x)\\
& = i^{n}2^{\alpha}\sqrt{\frac{\Gamma(n+2\alpha)}{2\pi n!}}\,\frac{\Gamma(\alpha + i x)}{\Gamma(2\alpha)} F(-n,\alpha+ix,2\alpha,2)
\end{align}
form a complete orthonormal basis of the Hilbert space $L^{2}(\mathbb R)$,
\begin{align}\label{OrthPol} \int_{-\infty}^{\infty}\!\pol_{n}^{(\alpha)*}(x)\pol_{m}^{(\alpha)}(x)\, \d x &= \delta_{n,m}\, \\\label{CompPol} \sum_{n= 0}^{\infty}\pol_{n}^{(\alpha)*}(x)\pol_{n}^{(\alpha)}(x') & = \delta(x-x')\, .
\end{align}

There exists a very natural relation between the Laguerre and the Pollaczek functions. Let us consider the linear map $U: L^{2}(\mathbb R^{+}) \rightarrow L^{2}(\mathbb R)$ defined by
\be\label{Udef} U(f)(x) = \frac{1}{\sqrt{2\pi}}\int_{0}^{\infty}\! t^{-\frac{1}{2}+i x}f(t)\, \d t\, .\ee
A straightforward calculation shows that $U$ is a unitary operator,
\be\label{normU} \int_{0}^{\infty}\! \bigl|f(t)\bigr|^{2}\,\d t = \int_{-\infty}^{+\infty}\! \bigl| U(f)(x)   \bigr|^{2}\,\d x\, .\ee
Morevoer, its inverse is given by the Mellin inversion theorem,
\be\label{Uinvdef} U^{-1}(\phi)(t) = \frac{1}{\sqrt{2\pi}}\int_{-\infty}^{+\infty}\! t^{-\frac{1}{2}-i x}\phi(x)\, \d x\, .\ee
Using the identity
\be\label{Batemanid} \int_{0}^{\infty}\! e^{-u t}t^{\beta}L_{n}^{(a)}(t)\,\d t = 
\frac{\Gamma(\beta+1)\Gamma (a+n+1)}{n!\,\Gamma (a+1)}u^{-\beta-1}{}_{2}F_{1}(-n,\beta+1,a+1,1/u)\, ,\ee
which is valid when $\re\beta >-1$ and $\re u>0$ \cite{Bateman}, we find that, up to a phase, the image under $U$ of the orthonormal basis of $L^{2}(\mathbb R^{+})$ given by the Laguerre functions \eqref{Lagfuncdef} is the orthonormal basis of $L^{2}(\mathbb R)$ give by the Pollaczek functions \eqref{Polfuncdef},
\be\label{LagPolrel} U\bigl(\lag_{n}^{(a)}\bigr)(x) = i^{-n}2^{ix}\pol_{n}^{(\frac{a+1}{2})}(x)\, .\ee
\subsection{\label{gARGSec} The general ARG map}
\begin{figure}
\centerline{\includegraphics[width=6in]{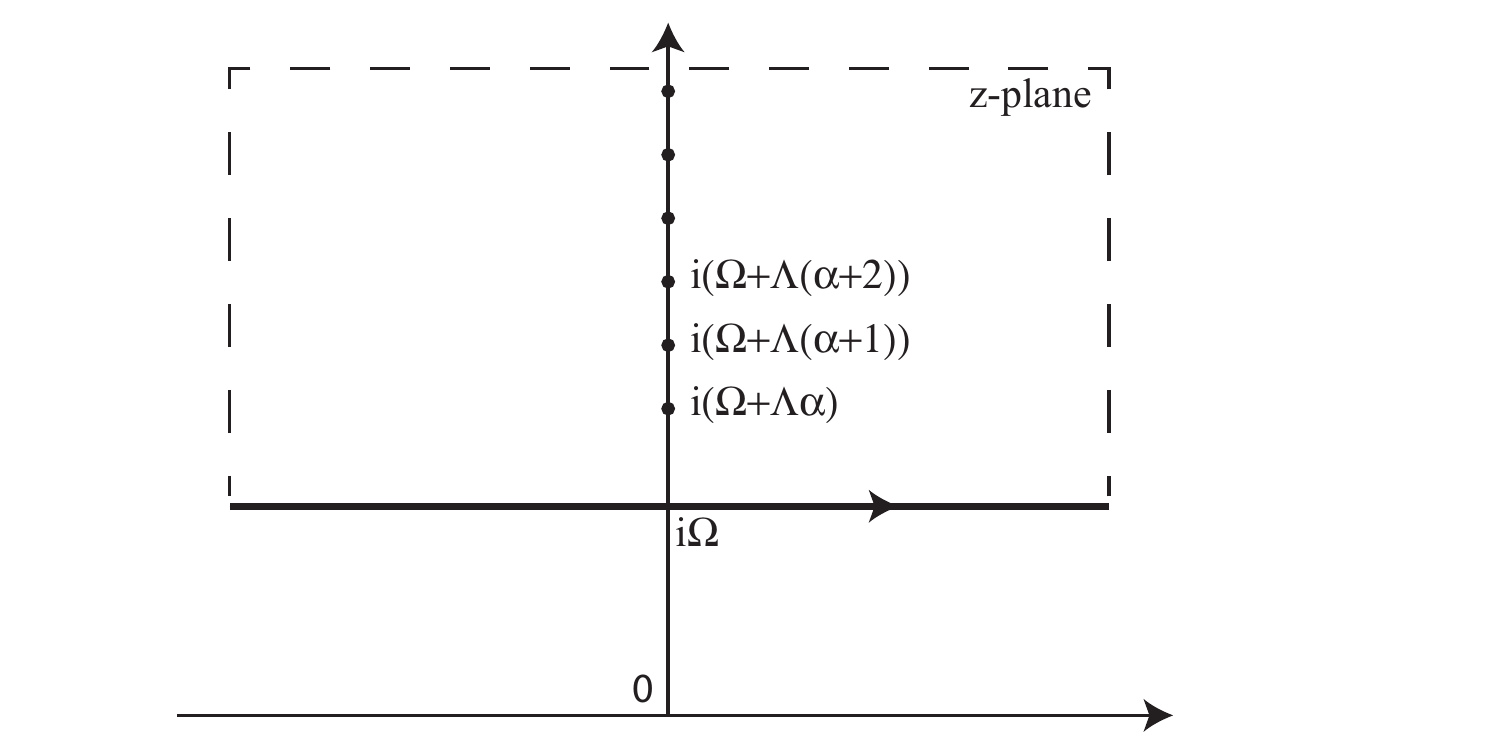}}
\caption{The contour of integration $]i\Omega-\infty,i\Omega-\infty[$ (thick line) used in the integral \eqref{cnf2}. The contour can be closed by an infinite semi-rectangle from above (dashed line) and the integral is given by an infinite sum associated with the poles of the $\Gamma$ function (black dots).\label{fig2}}
\end{figure}

Let us consider
\be\label{rdef} r_{\La,\Omega}(x) = R(-\La x + i \Omega)\, .\ee
The scales $\La>0$ and $\Omega>0$ are arbitrary, the sign in front of $\La x$ being chosen for future convenience. From \eqref{Rexp}, it is clear that $r_{\La,\Omega}\in L^{2}(\mathbb R)$. We can thus expand on a basis of Pollaczek functions $(\mathscr P_{n}^{(\alpha)})$, for any choice of $\alpha>0$:
\be\label{rexp} r_{\La,\Omega}(x) = \sum_{n=0}^{\infty} c_{n,\alpha}(\La,\Omega)\mathscr P_{n}^{(\alpha)}(x)\, ,\ee
with
\be\label{cnf1} c_{n,\alpha}(\La,\Omega) = \int_{-\infty}^{+\infty}\!\mathscr P_{n}^{(\alpha)*}(x) r_{\La,\Omega}(x)\, \d x\, .\ee
It is a bit more convenient to rewrite the integral in the variable $z=-\La x + i\Omega$. Explicitly, we get
\be\label{cnf2} c_{n,\alpha}(\La,\Omega) = - 2^{\alpha}\sqrt{\frac{n!}{2\pi\Gamma(n+2\alpha)}}\frac{1}{\La}\int_{i\Omega-\infty}^{i\Omega+\infty}\!
\Gamma\Bigl(\alpha+\frac{\Omega + i z}{\La}\Bigr) P_{n}^{(\alpha)}\Bigl(\frac{i\Omega - z}{\La}\Bigr)R(z)\, \d z\, .\ee
The contour of integration in the complex $z$-plane is depicted on Fig.\ \ref{fig2}. Using \eqref{Rexp} and the good asymptotic behaviour of the $\Gamma$ function given by Stirling formula, it is easy to show that the integral can be computed by closing the integration contour from above by an infinite semi-rectangle. In the region encircled by the rectangle, $R$ is holomorphic. The only poles we pick come from the $\Gamma$ function at the non-positive integer values of its argument. This yields
\be\label{cnf3} c_{n,\alpha}(\La,\Omega) = 2^{\alpha}\sqrt{\frac{2\pi n!}{\Gamma(n+2\alpha)}}\sum_{k=0}^{\infty}\frac{(-1)^{k}}{k!} P_{n}^{(\alpha)}\bigl(-i(\alpha+k)\bigr)R\Bigl(i\bigl(\Omega + \La(\alpha + k)\bigr)\Bigr)\, .\ee
We now see the magic of using the Pollaczek functions basis: due to the presence of the $\Gamma$ function, the coefficients of the expansion depend only on the values of $R$ at the discrete set of points $i\bigl(\Omega+\La(\alpha + k)\bigr)$, $k\in\mathbb N$, on the imaginary axis.

Since the only data we want to use are the Fourier coefficients $G_{k}=-R(i\nu_{k})$, we choose the scale
\be\label{Lachoice} \La = \frac{2\pi\delta}{\beta}\,\cvp\ee
where the ``index'' $\delta$ is an arbitrary strictly positive integer. We then choose a cut-off scale $\mu\geq\Omega$, associate to it the integer $\km$ as in \eqref{kmdef} and set
\be\label{alphachoice} \alpha = \alpha_{\km,\delta}(\Omega)= \frac{1}{\delta}\Bigl(\km-\frac{\beta\Omega}{2\pi}\Bigr)\, .\ee
Note that $\mu\geq\Omega$ implies $\alpha_{\mu,\delta}(\Omega)>0$, as required. With these choices, Eq.\ \eqref{cnf3} and \eqref{rexp} yield
\be\label{cnf4}\left\{
\begin{aligned}c_{n,\alpha}(\Lambda,\Omega) & =- 2^{\alpha}\sqrt{\frac{2\pi n!}{\Gamma(n+2\alpha)}}\sum_{k=0}^{\infty}\frac{(-1)^{k}}{k!} P_{n}^{(\alpha)}\bigl(-i(\alpha+k)\bigr)G_{\km+\delta k}\, ,\\
R(\omega+i\Omega) & = \sum_{n=0}^{\infty}c_{n,\alpha}(\Lambda,\Omega)\mathscr P_{n}^{(\alpha)}\Bigl(-\frac{\beta\omega}{2\pi\delta}\Bigr)\, .
\end{aligned}\right.\ee
Of course, the same reasoning as above can be repeated on the lower half-plane $\im z<0$.

To write down the final result in a convenient way, we introduce the coefficients
\be\label{chidef} \boxed{\chi_{n}^{\pm}(\km,\delta,\alpha) = \sum_{k=0}^{\infty}\frac{(-1)^{k}}{k!} F(-n,2\alpha+k,2\alpha,2) G_{\pm(\km + \delta k)}}\, .\ee
The Eq.\ \eqref{cnf4}, together with the similar formula valid in the lower half-plane, is then equivalent to
\be\label{ARGG}\boxed{
\begin{aligned} R(\omega\pm i\Omega) & = -\frac{2^{2\alpha}}{\Gamma(2\alpha)^{2}}\Gamma\Bigl(\alpha \mp \frac{i\beta\omega}{2\pi\delta}\Bigr)\\&\hskip 1.15cm
\sum_{n=0}^{\infty}(-1)^{n}\frac{\Gamma(2\alpha+n)}{n!}F\Bigl(-n,\alpha\mp\frac{i\beta\omega}{2\pi\delta},2\alpha,2\Bigr)\,\chi_{n}^{\pm}(\km,\delta,\alpha)
\\
\alpha& =\alpha_{\km,\delta}(\Omega)= \frac{1}{\delta}\Bigl(\km-\frac{\beta\Omega}{2\pi}\Bigr)\, ,\quad 0<\Omega\leq\mu\, .
\end{aligned}}\ee
This is the fundamental formula of the ARG, from which everything else can be derived. Taking into account \eqref{chiRrel}, it provides in particular the explicit form of the general ARG maps $\mathsf A^{\pm}_{\km,\delta}(\omega,\Omega)$ introduced in \eqref{ARGmapi2} and \eqref{ARGmapii2}.

\noindent\emph{Important remarks}:

\noindent i) The formula \eqref{chidef} is manifestly linear in the Fourier coefficients $G_{k}$ and thus the ARG map \eqref{ARGG} is linear as well, as expected.

\noindent ii) At large $k$, the general term of the series defining the coefficients $\chi_{n}^{\pm}$ is equivalent to
\be\label{largekser} \frac{(-2)^{n}}{(2\alpha)_{n}}\frac{(-1)^{k}}{k!}k^{n}G_{\pm (\km+\delta k)} = O\Bigl(k^{n-3/2}e^{-k(\ln k -1 )}\Bigr) \ee
and thus the sum over $k$ in \eqref{chidef} converges rapidly.

\noindent iii) The sum over $k$ in \eqref{chidef} must be performed first and the sum over $n$ in \eqref{ARGG} second. Indeed, if one makes the sum over $n$ first, one gets infinity. This is a very important qualitative property of the ARG maps, to be discussed further in Section \ref{ExSec}. 

\noindent iv) In Eq.\ \eqref{ARGG}, we have complete freedom in choosing the index $\delta\geq 1$ and the cut-off $\km$, as long as $\mu\geq\Omega>0$.\footnote{Let us note that the construction in \cite{Cuniberti} corresponds to the special values $\Omega = \pi/\beta$, $\km=1$ and $\delta=1$ (and thus $\alpha=1/2$).}  Of course, $R(\omega+i\Omega)$ does not depend on these arbitrary choices. This automatically yields highly non-trivial Analytic Renormalization Group equations, which take the form of \emph{universal linear relations constraining any admissible set of coefficients $G_{k}$.}

\subsection{\label{AmapSec} The maps $\mathsf{A}_{\km,\delta}^{\pm}$}

The construction of the ARG maps $\mathsf{A}_{\km,\delta}^{\pm}$ is now completely straightforward. We simply set $\omega=0$ and $\Omega=\nu_{k}$ in \eqref{ARGG}. The formula simplifies because
\be\label{Fsimform} F(-n,\alpha,2\alpha,2) = \left\{
\begin{aligned} & 0\quad\text{if}\ n\ \text{is odd.}\\
& \frac{(2m)!}{m!}\frac{(\alpha)_{m}}{(2\alpha)_{2m}}\quad\text{if}\ n=2m\, ,
\end{aligned}\right.\ee
which is equivalent to \eqref{Polparity} and \eqref{Polatzero}. We get
\be\label{ARGmap1}\boxed{G_{k} = \beta n_{0}\delta_{k,0}+ \sum_{m=0}^{\infty}\sum_{p=0}^{\infty}\Acoef G_{\pm(\km + \delta p)}}\ee
with
\be\label{ARGmap2}\boxed{
\begin{aligned} \Acoef &= 
\frac{2^{2\alpha}}{\Gamma (2\alpha)}\frac{\Gamma (m+\alpha)}{m!} \frac{(-1)^{p}}{p!}F(-2m,2\alpha + p, 2\alpha ,2)\, ,\\
\alpha &= \frac{\km-k}{\delta}
\end{aligned}}\ee
and the sign $\pm$ on the right-hand side of \eqref{ARGmap1} is chosen according to the sign of $k$.

\noindent\emph{Remarks}

\noindent i) As in \eqref{ARGG}, the order of the sums in \eqref{ARGmap1} is essential. For example, the case $p=0$ involves $F(-2m,2\alpha,2\alpha,2)=1$ and the series $\sum_{m\geq 0}\frac{\Gamma(m+\alpha)}{m!}$ clearly diverges.\footnote{It is funny to note that, by using Euler identity, one can easily prove $\smash{\sum_{m=0}^{\infty}\frac{\Gamma(m+\alpha)}{m!}F(-2m,b,c,2)} =\smash{\frac{\Gamma(c)\Gamma(\alpha)\Gamma(b-\alpha)\Gamma(c-b-\alpha)}{2^{2\alpha}\Gamma(b)\Gamma(c-b)\Gamma(c-2\alpha)}}$ if $b>\alpha>0$ and $c>b+\alpha$. However, these conditions are not met in our case.}

\noindent ii) One can immediately write down ARG equations, for example
\be\label{ARGEq2} \sum_{m=0}^{\infty}\sum_{p=0}^{\infty}\mathsf{A}_{\km+1,\delta}^{m,k,p}G_{\pm(\km + 1+\delta p)}= \sum_{m=0}^{\infty}\sum_{p=0}^{\infty}\Acoef G_{\pm(\km + \delta p)}\, ,\ee
for all $0\leq k<\km$, and also similar identities obtained by varying $\delta$. 

\noindent iii) The relations \eqref{ARGmap1}, for $k\not = 0$, are universal, model-independent linear constraints on the Fourier coefficients. From this point of view, they are not different from \eqref{ARGEq2} and for this reason we also call them ``ARG equations.'' More generally, any universally valid linear relation between the Fourier coefficients is called an ARG equation in the present paper.

\subsection{\label{realSec} Real-time physics from Euclidean data}

\subsubsection{The spectral function}

By using \eqref{resdisc}, the general ARG map \eqref{ARGG}, applied for $\Omega=\epsilon$, allows to reconstruct explicitly the spectral function $\rho(\omega)$ from the Euclidean Fourier coefficients $G_{k}$. This provides a full solution to the problem of reconstructing the real-time two-point functions in terms of the Euclidean data, using \eqref{xirhorel}--\eqref{Drhorel}. Actually, we have obtained an infinite set of equivalent reconstruction formulas, each associated with a choice of cut-off $\km$ and index $\delta$, using only subsets of Fourier coefficients $G_{\pm\km\pm \delta k}$ for $k\geq 0$ (as usual, we also need $G_{0}$ to get the zero-frequency piece in the spectral function, if $n_{0}\not = 0$). 

\subsubsection{\label{rtacSec} The real-time retarded and advanced correlators}

Explicit formulas can be obtained for the retarded and advanced functions $\chi_{\text r}(t)$ and $\chi_{\text a}(t)$. The most general formulas are actually obtained by considering $e^{-\Omega t}\chi_{\text r}(t)$ and $e^{\Omega t}\chi_{\text a}(t)$, for any $\Omega\geq 0$. For example, using the analyticity of the resolvent $R$ on the upper half-plane, we get, from \eqref{chirdef} and \eqref{chiRrel},
\be\label{chi1} e^{-\Omega t}\chi_{\text r}(t) = -\frac{1}{2\pi}\int_{-\infty}^{+\infty}\! R(\omega + i \Omega+ i\epsilon)e^{-i\omega t}\,\d\omega\, .\ee
From \eqref{cnf4}, we see that to evaluate this integral we need to know the Fourier transform of the Pollaczek functions. But the Fourier transform of an arbitrary function $\phi$ is directly given in terms of the unitary operator $U^{-1}$ defined in Section \ref{MathSec}. Indeed, Eq.\ \eqref{Uinvdef} is equivalent to
\be\label{FtransU} \frac{1}{2\pi}\int_{-\infty}^{+\infty}\!\phi(\omega)e^{-i\omega t}\,\d\omega = \frac{1}{\sqrt{2\pi}}e^{t/2}U^{-1}(\phi)(e^{t})\, .\ee
The result \eqref{LagPolrel} thus tells us that the Fourier transform of the Pollaczek functions can be expressed in terms of the Laguerre functions. Putting all the factors together we get in this way
%
\be\label{chir3}\boxed{
\begin{aligned} &\chi_{\text r}(t) = \frac{2\pi\delta}{\beta}\frac{2^{2\alpha}}{\Gamma(2\alpha)} e^{-\frac{2\pi}{\beta}(2\delta\alpha-\km)t}e^{-e^{-\frac{2\pi\delta t}{\beta}}}
\sum_{n=0}^{\infty}(-1)^{n}L_{n}^{(2\alpha-1)}\Bigl(2e^{-\frac{2\pi\delta t}{\beta}}\Bigr)
\chi_{n}^{+}(\km,\delta,\alpha)\, ,\\
 &\chi_{\text a}(t) = \frac{2\pi\delta}{\beta}\frac{2^{2\alpha}}{\Gamma(2\alpha)} e^{\frac{2\pi}{\beta}(2\delta\alpha-\km)t}e^{-e^{\frac{2\pi\delta t}{\beta}}}
\sum_{n=0}^{\infty}(-1)^{n}L_{n}^{(2\alpha-1)}\Bigl(2e^{\frac{2\pi\delta t}{\beta}}\Bigr)
\chi_{n}^{-}(\km,\delta,\alpha)\, ,\\
& 0<\alpha<\frac{\km}{\delta}\,\cdotp
\end{aligned}}\ee
As usual, these equations are valid for any choice of strictly positive integers $\km$ and $\delta$. Moreover, the parameter $\alpha$, being related to the arbitrary $\Omega$ that we have introduced in \eqref{chi1} by the equation \eqref{alphachoice}, can be chosen at will in the interval $]0,\frac{\km}{\delta}[$.

\subsubsection{A very general form of the ARG equations}

Using the fact that the left-hand sides of \eqref{chir3} do not depend on $\km$, $\delta$ or $\alpha$, we immediately get many ARG equations. Moreover, causality immediately implies
\be\label{argeqGEN}\boxed{
\begin{aligned}
&\sum_{n=0}^{\infty}(-1)^{n}L_{n}^{(2\alpha-1)}(u)
\chi_{n}^{\pm}(\km,\delta,\alpha)=0\, ,\\
&\text{for any}\ (\km,\delta)\in\mathbb N^{*2}\, ,\ 0<\alpha<\frac{\km}{\delta}\ \text{and}\ u>2\, ,
\end{aligned}}\ee
since $\chi_{\text r}(t)=0$ if $t<0$ and $\chi_{\text a}(t)=0$ is $t>0$.

\subsubsection{The long time behaviour}

Of particular interest is the long time behaviour of the correlation functions. In particular, linear response theory implies that the retarded correlator $\chi_{\text r}(t)$ governs the response of the operator $A$ to a small perturbation of the system by the operator $B$. If the system thermalizes, we thus have $\lim_{t\rightarrow\infty}\chi_{\text r}(t)=0$. In many interesting examples, $\chi_{\text r}(t)$ decays exponentially,
\be\label{typicallarget} \chi_{\text r}(t) \underset{t\rightarrow\infty}{\propto} e^{-\gamma t}\, ,\ee
where $1/\gamma >0$ is the thermalization time scale. The behaviour \eqref{typicallarget} occurs when the analytic continuation of the resolvent $R(z)$ from the upper half-plane to the lower half-plane admits poles for $\im z<0$. If $z_{0}$ is the pole closest to the real axis, then $\gamma = -\im z_{0}$. 

The representation \eqref{chir3} allows to study quite efficiently the large time behaviour of $\chi_{\text r}$. For example, if we choose $\alpha = \frac{\km}{2\delta}$, which is equivalent to $\Omega = \frac{\pi\km}{\beta}$, and use \eqref{Lagatzero}, we get
\be\label{larget1} \lim_{t\rightarrow\infty}\chi_{\text r}(t) = \frac{2\pi\delta}{\beta}\frac{2^{\km/\delta}}{\Gamma(\km/\delta)^{2}}
\sum_{n=0}^{\infty}(-1)^{n}\frac{\Gamma(n+\km/\delta)}{n!}
\chi_{n}^{+}\bigl(\km,\delta,\frac{\km}{2\delta}\bigr)\, .\ee
More generally, let us assume that the large time behaviour is of the form \eqref{typicallarget}. Let us then pick a $\tilde\gamma>0$ and choose $\km>\frac{\beta\tilde\gamma}{2\pi}$. If we examine the $t\rightarrow\infty$ limit of \eqref{chir3} for
\be\label{alpch23} \alpha = \frac{1}{2\delta}\Bigl(\km+\frac{\beta\tilde\gamma}{2\pi}\Bigr)\, ,\ee
we find that
\be\label{critgamma}
\sum_{n=0}^{\infty}(-1)^{n}\frac{\Gamma(2\alpha +n)}{n!}\chi_{n}^{+}(\km,\delta,\alpha) =\left\{
\begin{aligned}
& 0\quad\text{if}\ \tilde\gamma <\gamma\, ,\\
& \infty\quad\text{if}\ \tilde\gamma >\gamma\, .
\end{aligned}\right.\ee
This provides a sharp criterion to compute the thermalization time scale $\gamma$ from the Euclidean data.

\subsection{\label{MFSec} The ARG equations and the space of two-point functions}

As announced in Section \ref{IntroSec}, we have shown that analyticity implies an infinite set of linear constraints on the Fourier coefficients $G_{k}$, the ARG equations. In other words, the space $\mathscr M$ of two-point functions is a subspace of infinite codimension of the space $\mathscr F$ of Fourier coefficients.

It is not too difficult to understand that the full set of ARG equations is enough to characterize $\mathscr M$: if a set of Fourier coefficients $(G_{k})_{k\in\mathbb Z}$ satisfy all the ARG equations mentioned above, then it belongs to $\mathscr M$. This is equivalent to saying that there exists a resolvent $R$, with the analyticity properties discussed above, such that $G_{k}=-R(i\nu_{k})$ for $k\not = 0$. The argument to show this goes as follows.

One starts with the full set of ARG equations \eqref{argeqGEN}, together with the equations ensuring that the right-hand sides of \eqref{chir3} do not depend on the choice of $\km$, $\delta$ and $\smash{\alpha\in ]0,\frac{\km}{\delta}[}$. One then uses Eq.\ \eqref{chir3} to define $\chi_{\text r}$ and $\chi_{\text a}$ and Eq.\ \eqref{resdefBIS} to define $R$. Thank's to \eqref{argeqGEN}, $R$ is automatically analytic in the upper and lower half-planes. Moreover, evaluating explicitly the integrals in \eqref{resdefBIS} starting from \eqref{chir3} amounts to doing the inverse of the Fourier transform performed in Section \ref{rtacSec}. This obviously yields the formula \eqref{ARGG} for $R$. We can then evaluate $R(i\nu_{k})$ by using the ARG equations \eqref{ARGmap1}, which eventually yields $G_{k}=-R(i\nu_{k})$ as was to be shown.

A more difficult question is to find a minimal set of ARG equations that fully characterize $\mathscr M$. This is non-trivial, because non-trivial linear relations between the ARG equations do exist, see Section \ref{useARGSec}. A detailed discussion of this issue if beyond the scope of the present paper. One may conjecture that the equations \eqref{argeqGEN}, for an arbitrary but unique choice of parameters $\km$, $\delta$ and $\alpha\in ]0,\frac{\km}{\delta}[$, but for all $u>2$, form a complete set of linearly independent ARG equations.

\section{\label{ExSec} Numerical analysis and simple applications}

We are now going to explain how the formalism of the previous section can be implemented numerically and used in practice. Our aim is to get more intuition on how the ARG actually works and to illustrate the ARG maps and equations on simple explicit examples.

\subsection{\label{GrSec} General remarks}

Let us start by discussing three qualitatively important properties of the numerical implementation of the ARG.

\subsubsection{Finite precision and the matrix form of the ARG}

The ARG maps, as well as the ARG equations, all entail a sum over $n\geq 0$ involving the coefficients $\chi_{n}^{\pm}$ defined in \eqref{chidef}. To obtain a numerical approximation, we keep only a finite number of terms in this sum, restricting the integer $n$ to the interval $0\leq n\leq N$. Obviously, the more terms we keep, the better precision we get. For this reason, we call $N$ the ``precision'' of the numerical implementation. Of course, the actual numerical precision achieved for a given choice of $N$ depends on the particular example under study. Equalities at precision $N$ will be denoted by $\eN$.

For a given finite precision $N$, the sum over $n$ and the sum \eqref{chidef} over $k$ can be permuted. Unlike the exact maps, the finite-precision linear ARG maps can thus be written in a familiar finite-dimensional matrix form. For example, the general ARG map \eqref{ARGG} at precision $N$ is given by
\be\label{argg1} R(\omega\pm i\Omega) \underset{N}{=}\sum_{p=0}^{\infty} \mathsf A^{\pm}_{\km,\delta}(N;\omega,\Omega,p)\, G_{\pm(\km + \delta p)}\ee
where the matrix elements $\mathsf A^{\pm}_{\km,\delta}(N;\omega,\Omega,p)$ are universal numbers given by 
\be\label{argg2}\left\{
\begin{aligned} &\mathsf A^{\pm}_{\km,\delta}(N;\omega,\Omega,p) = 
-\frac{2^{2\alpha}}{\Gamma(2\alpha)^{2}}\Gamma\Bigl(\alpha\mp\frac{i\beta\omega}{2\pi\delta}\Bigr)\frac{(-1)^{p}}{p!}\\ &\hskip 1.75cm\sum_{n=0}^{N}
(-1)^{n}\frac{\Gamma(2\alpha + n)}{n!} F\Bigl(-n,\alpha\mp\frac{i\beta\omega}{2\pi\delta},2\alpha,2\Bigr) F\bigl(-n,2\alpha+p,2\alpha,2\bigr)\, ,\\&
\alpha = \frac{1}{\delta}\Bigl(\km - \frac{\beta\Omega}{2\pi}\Bigr)\, ,\quad 0<\Omega\leq\mu\, .
\end{aligned}\right.\ee
Similarly, recalling that $m=2n$ and denoting by $[N/2]$ the integer part of $N/2$, \eqref{ARGmap1} and \eqref{ARGmap2} are rewritten as
\be\label{argg3} G_{k} \underset{N}{=} \beta n_{0}\delta_{k,0}+\sum_{p=0}^{\infty}\mathsf A_{\km,\delta}(N;k,p)\, G_{\pm(\km+\delta p)}\, ,\ee
the sign $\pm$ being fixed by the sign of $k$ and
\be\label{argg4}\left\{\begin{aligned} & \mathsf A_{\km,\delta}(N;k,p)   = \frac{2^{2\alpha}}{\Gamma(2\alpha)}\frac{(-1)^{p}}{p!}\sum_{m=0}^{[N/2]}\frac{\Gamma(m+\alpha)}{m!} F(-2m,2\alpha+p,2\alpha,2)\, ,\\
& \alpha = \frac{\km-k}{\delta}\,\cdotp
\end{aligned}\right.\ee
And, finally, \eqref{chir3} corresponds to 
\be\label{argg5} \chi_{\text{r,a}}(t) \underset{N}{=}
\frac{2\pi\delta}{\beta}\frac{2^{2\alpha}}{\Gamma(2\alpha)} e^{\mp\frac{2\pi}{\beta}(2\delta\alpha-\km)t}e^{-e^{\mp\frac{2\pi\delta t}{\beta}}}
\sum_{p=0}^{\infty}
\mathsf A_{\alpha}\bigl(N;2e^{\mp\frac{2\pi\delta t}{\beta}},p\bigr)\, G_{\pm(\km + \delta p)}\ee
where the matrix elements $\mathsf A_{\alpha}(N;u,p)$ are given by
\be\label{argg6} \mathsf A_{\alpha}(N;u,p) = \frac{(-1)^{p}}{p!}\sum_{n=0}^{N}(-1)^{n}L_{n}^{(2\alpha -1)}(u) F(-n,2\alpha + p,2\alpha,2)\, ,\ee
for any $0<\alpha<\km/\delta$ and $u>0$. In particular, the ARG equations \eqref{argeqGEN} take the form
\be\label{argg7} \sum_{p=0}^{\infty}\mathsf A_{\alpha}(N;u,p)G_{\pm (\km + \delta p)} \underset{N}{=} 0\, ,\ee
for any $0<\alpha<\km/\delta$ and $u>2$.

\subsubsection{\label{UVSec} Decoupling of the UV}

Using
\be\label{Fasymp} F(-n,2\alpha+p,2\alpha,2)\underset{p\rightarrow\infty}{\sim}\frac{\Gamma(2\alpha)}{\Gamma(2\alpha + n)}\, (-2p)^{n}\, ,\ee
we see that the matrix elements in \eqref{argg2}, \eqref{argg4} or \eqref{argg6} are proportional to $p^{N}/p!$ at large $p$. They are thus decreasing very quickly when $p\rightarrow\infty$. This implies that the Fourier coefficients $G_{k}$ above some Euclidean UV cut-off $K$, i.e.\ for $|k|> K$, are totally irrelevant to evaluate the ARG maps. Of course, the UV cut-off must be much larger than the RG scale, $K\gg\km$.
Moreover, if we increase the precision $N$ or the index $\delta$, $K$ must also be increased accordingly. In practice, working with $K\sim \delta N \gg \km$ is more than enough, see the examples below. 

The conclusion is that we can always use a finite dimensional data set $(G_{k})_{|k|\leq K}$ in numerical calculations, the UV cut-off $K\gg\km$ being chosen according to the precision goal.

This phenomenon of decoupling of the UV physics is of course one of the most important consequence of the usual RG ideas. Here we obtain a mathematically rigorous and universal version of this decoupling, as a consequence of analyticity. Moreover, the $p$-dependence of the matrix elements of the ARG maps (see e.g.\ Fig.\ \ref{fig3} and \ref{fig4}) quantifies in a very precise way how the low energy physics can be influenced by the data above the RG scale, as a function of energy.

\subsubsection{\label{extSec} Extreme sensitivity on the data set}

On top of their large $p$ behaviour and the associated decoupling of the UV that we have just mentioned, the ARG maps matrix elements have another remarkable feature: in the range of energy where they are not infinitesimally small (i.e.\ for $p\ll K$), they are typically huge. This is due to the fact that the sums over $n$ (or $m$) in \eqref{argg2}, \eqref{argg4} and \eqref{argg6} diverge, as already emphasized in Section \ref{ARGSec}. This property implies an \emph{extreme} sensibility, which increases with the precision $N$, of the ARG maps on the values of the Fourier coefficients $G_{k}$ in the relevant energy range.

\subsubsection{\label{EEnumSec} Illustrations}

\begin{figure}
\centerline{\includegraphics[width=3in]{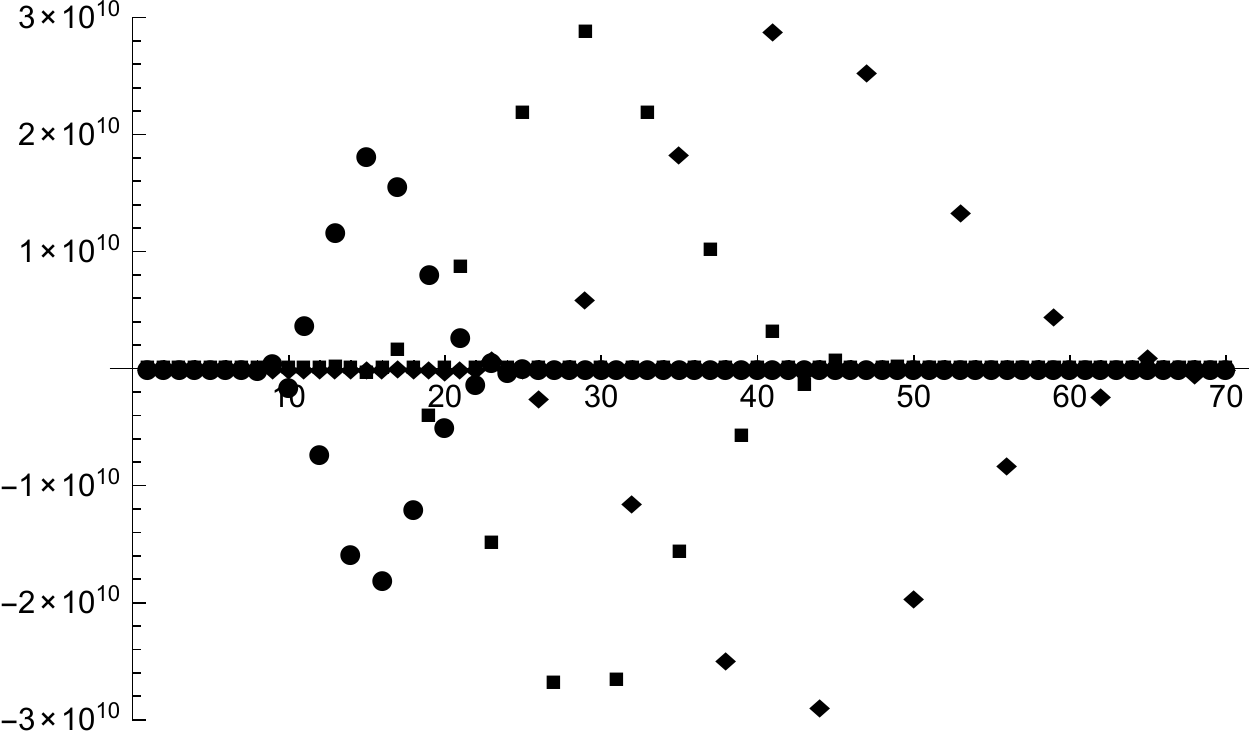}\includegraphics[width=3in]{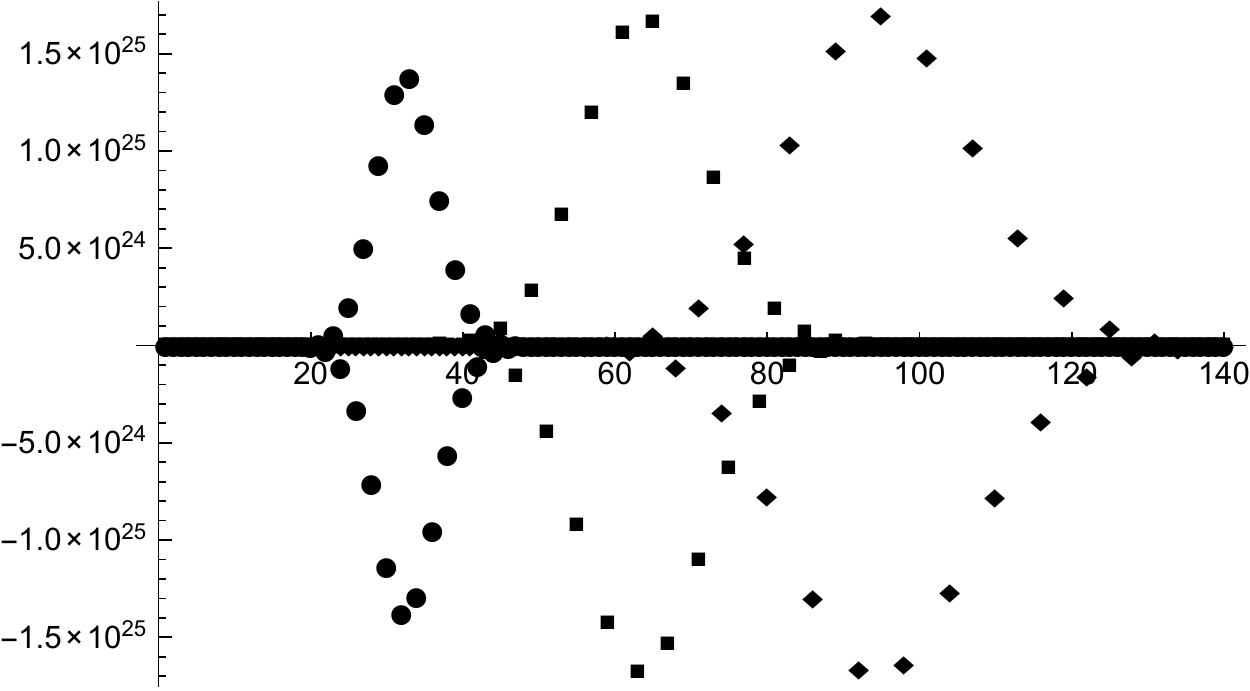}}
\caption{The matrix elements $\mathsf A_{\km=5,\delta}(N;k=1,p)$ for $\delta=1,2,3$ (dots, squares and diamonds) and $N=100,500$ (left inset, right inset), as a function of the ``energy'' $\km+\delta p$. The peaks on the plots correspond to energy ranges for which the matrix elements are huge. This implies an extreme sensibility of the ARG maps at these energies.\label{fig3}}

\vskip .25cm

\centerline{\includegraphics[width=3in]{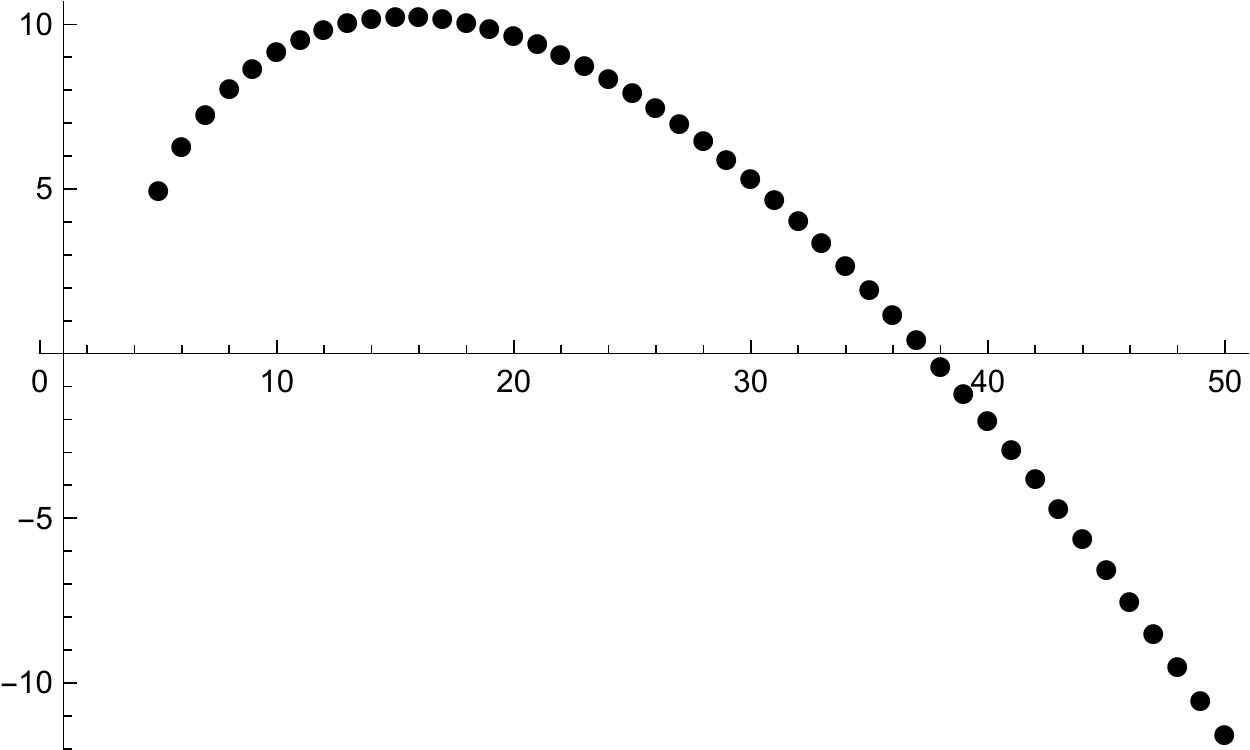}\includegraphics[width=3in]{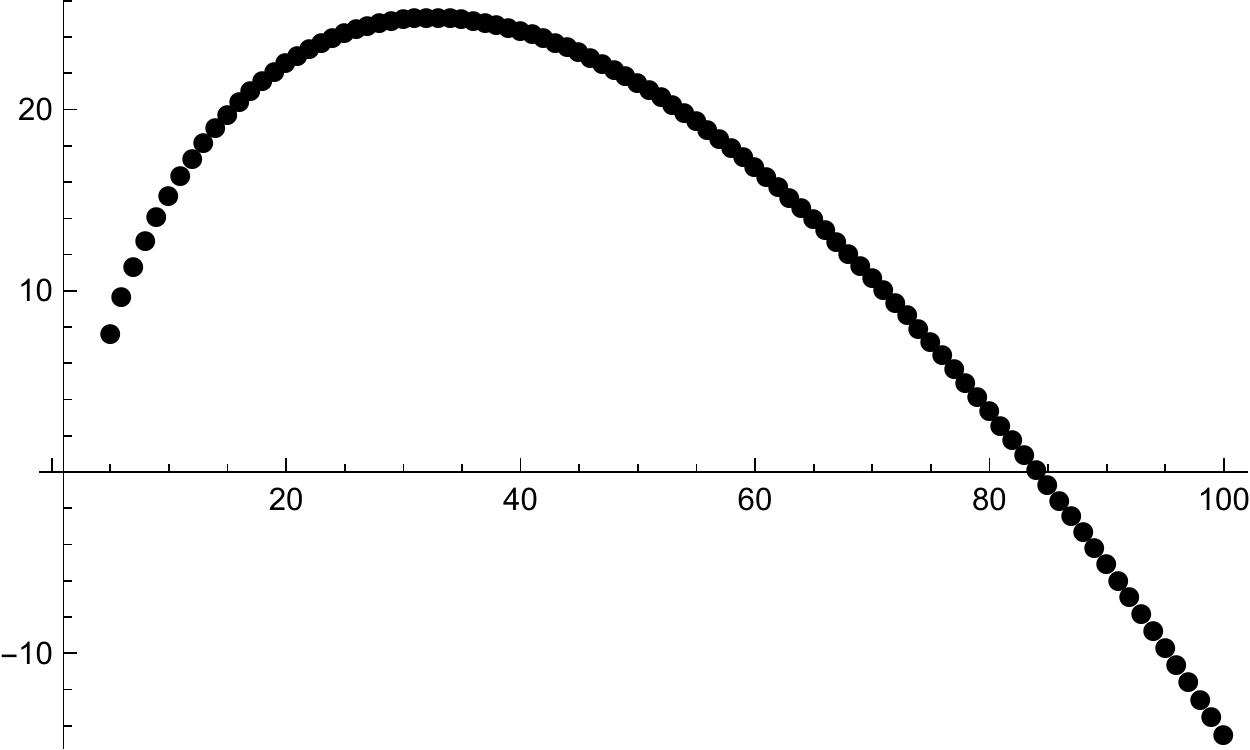}}
\caption{The logarithms of the matrix elements, $\log_{10} \mathsf A_{\km=5,\delta=1}(N;k=1,p)$, for $N=100,500$ (left inset, right inset), as a function of the ``energy'' $\km+\delta p$. The decoupling of the UV physics is clearly seen.\label{fig4}}

\vskip .25cm

\centerline{\includegraphics[width=3in]{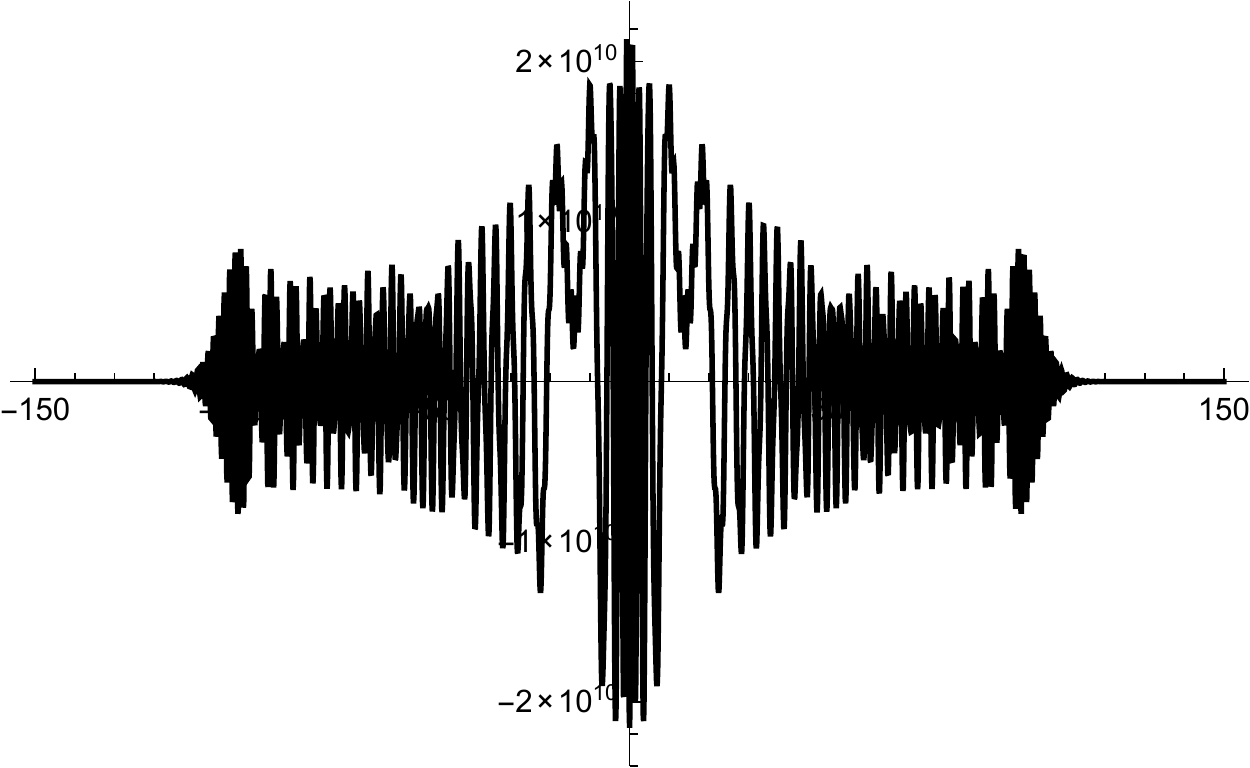}\includegraphics[width=3in]{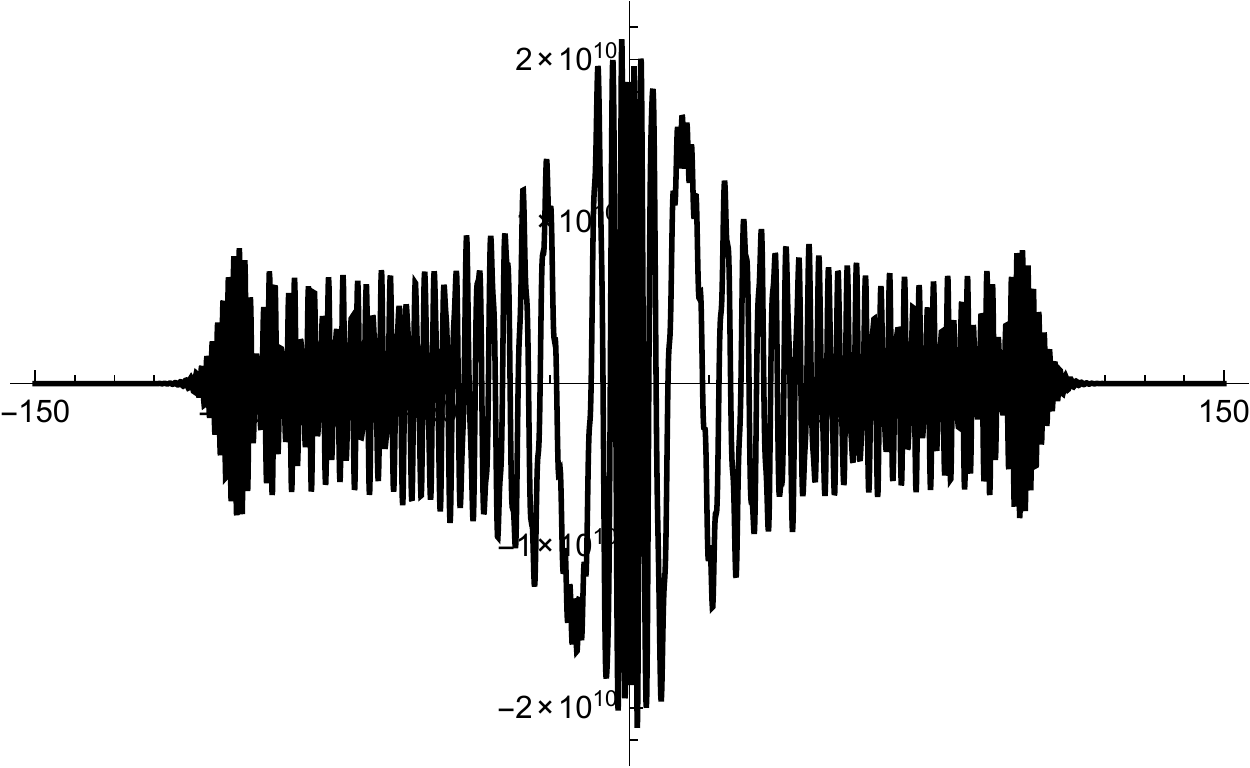}}
\caption{The real and imaginary parts (left inset, right inset) of the matrix elements $\mathsf A^{+}_{2,1}(100;\omega,0,11)$ as a function of the real-time frequency $\omega$ and for $\beta=2\pi$. As Eq.\ \eqref{argg2} shows, these matrix elements are suppressed at large $|\omega|$, \smash{$\mathsf A_{\km,\delta}^{\pm}(N;\omega,\Omega,p)\propto |\omega|^{N+\alpha-1/2}e^{-\frac{\beta |\omega|}{4\delta}}$.}
\label{fig5}}
\end{figure}

There is nothing better than a few plots to illustrate the properties listed above. On Fig.\ \ref{fig3} and \ref{fig4}, we have depicted the values of some matrix elements \eqref{argg4} at $\km=5$ and $k=1$, for $\delta=1,2,3$ and precisions $N=100,500$. Very similar plots are obtained for different values of $\km$ and $k$, or for matrix elements $\mathsf A^{\pm}_{\km,\delta}(N;\omega,\Omega,p)$ and $\mathsf A_{\alpha}(N;u,p)$. We see that:

\noindent i) On Fig.\ \ref{fig3}, the dots that are visibly above or below the abscissa axis on the plots correspond to matrix elements that are huge in some range of the energy $\km+\delta p$, of order $10^{10}$ for $N=100$ and $10^{25}$ for $N=500$! For example, $\mathsf A_{5,1}(100;1,11)\simeq 1.82\, 10^{10}$. This means that a tiny error in the Fourier coeffients $G_{16}$ that multiplies this huge number in the ARG map \eqref{argg3}, let's say of order $10^{-5}$, would yield a huge error in the coefficient $G_{1}$ given by the ARG map \eqref{argg3}, of order $10^{5}$!

\noindent ii) On the logarithmic plots of Fig.\ \ref{fig4}, which both correspond to $\delta=1$, we clearly see that the matrix elements have a maximum for some energy and that they remain sizeable below this energy. For example, $\mathsf A_{5,1}(100;1,0)\simeq 9.64\, 10^{4}$. This property is of course completely generic and in particular remains valid for the other values of $\delta$.

\noindent iii) Fig.\ \ref{fig4} clearly shows that the UV physics decouple. The UV cut-off can be taken to be $K\sim 50$ for the case $\km=5,\delta=1,N=100$ (for example, we find $\mathsf A_{5,1}(100;1,50) = 3.1\, 10^{-11}$) and $K\sim 100$ for $\km=5,\delta=1,N=500$. Similarly, by imposing that the matrix elements are $\sim 10^{-10}$ or smaller above the UV cut-off, we get $K\sim 100$ for $\km=5,\delta=2,N=100$, $K\sim 150$ for $\km=5,\delta=3,N=100$, $K\sim 200$ for $\km=5,\delta=2,N=500$ and $K\sim 300$ for $\km=5,\delta=3,N=500$.

In Fig.\ \ref{fig5}, we have depicted the real-time frequency dependence of a typical large matrix element of the general ARG map \eqref{argg2}. This dependence is very complicated, but is eventually tamed for large frequencies. This property is true for all the matrix elements.

\subsection{\label{NSecE} Using the ARG with an exact data set}

We are now going to illustrate explicitly how the ARG works, by using mainly (but not exclusively) the simple example of the damped harmonic oscillator. This case captures well many basic qualitative features of more realistic models. We shall give examples of the maps $\mathsf A^{\pm}_{\km,\delta}$ and of the reconstruction of the spectral function and of the real-time functions from the Euclidean data. We fix the overall energy scale by setting the temperature to
\be\label{betaN} \beta = 2\pi\, .\ee
\subsubsection{The damped harmonic oscillator}

For an oscillator frequency $m>0$ and damping coefficient $\Gamma>0$, we consider the resolvent function
\be\label{resDO} R(z) = \left\{
\begin{aligned} &\frac{1}{z^{2}-m^{2}+2i\Gamma z}\quad\text{if}\ \im z>0\\
&\frac{1}{z^{2}-m^{2}-2i\Gamma z}\quad\text{if}\ \im z<0\, .
\end{aligned}\right.\ee
The associated Euclidean Fourier coefficients are
\be\label{GkDO} G_{k} = \frac{1}{m^{2}+\nu_{k}^{2}+2\Gamma |\nu_{k}|}
\ee
and the spectral function is given by
\be\label{rhoDO} \rho(\omega) = \frac{2}{\pi}\frac{\Gamma\omega}{(\omega^{2}-m^{2})^{2}+4\Gamma^{2}\omega^{2}}\,\cdotp\ee
The spectral function is smooth and has no discrete or zero-frequency piece. The retarded two-point function is given by
\be\label{chirDO} \tilde\chi_{\text r}(z) = \frac{1}{m^{2}-z^{2}-2i\Gamma z}\ee
on the complex frequency plane and by
\be\label{chirDO2} \chi_{\text r}(t) = \left\{
\begin{aligned}
& \frac{\theta(t)}{\sqrt{m^{2}-\Gamma^{2}}}\, e^{-\Gamma t}\sin\bigl(\sqrt{m^{2}-\Gamma^{2}}\, t\bigr)\quad\text{if}\ m>\Gamma\\
& \theta(t)\, t e^{-\Gamma t}\quad\text{if}\ m=\Gamma\\
&
\frac{\theta(t)}{2\sqrt{\Gamma^{2}-m^{2}}}\, \Bigl(e^{-(\Gamma-\sqrt{\Gamma^{2}-m^{2}})\,t} - e^{-(\Gamma+\sqrt{\Gamma^{2}-m^{2}})\,t}\Bigr)\quad\text{if}\ m<\Gamma\, .
\end{aligned}\right.\ee
in real time. The case $m>\Gamma$ corresponds to mild damping, the poles of $\tilde\chi_{\text r}(z)$ on the lower half-plane having a non-zero real part, whereas $\Gamma>m$ corresponds to strong damping, with poles on the imaginary axis. Similar formulas give the advanced function too.

\begin{figure}
\centerline{\includegraphics[width=2.9in]{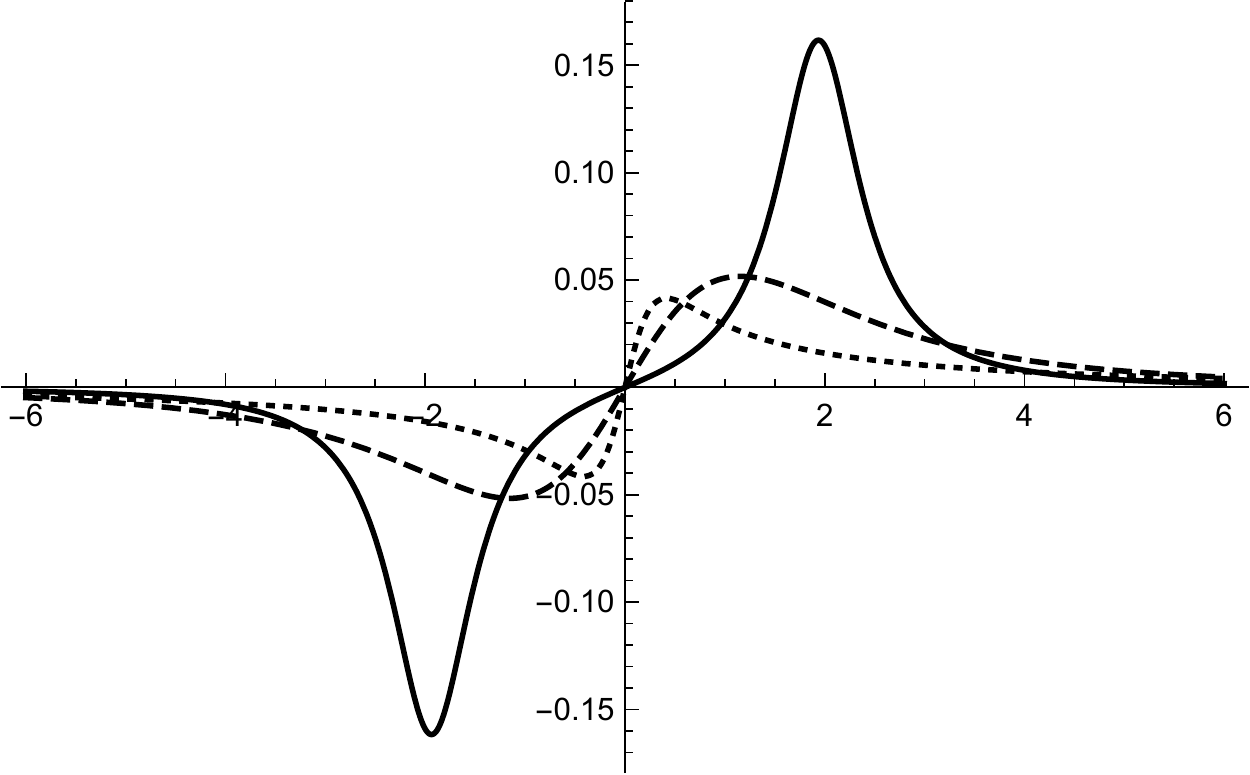}\hskip .2in\includegraphics[width=2.9in]{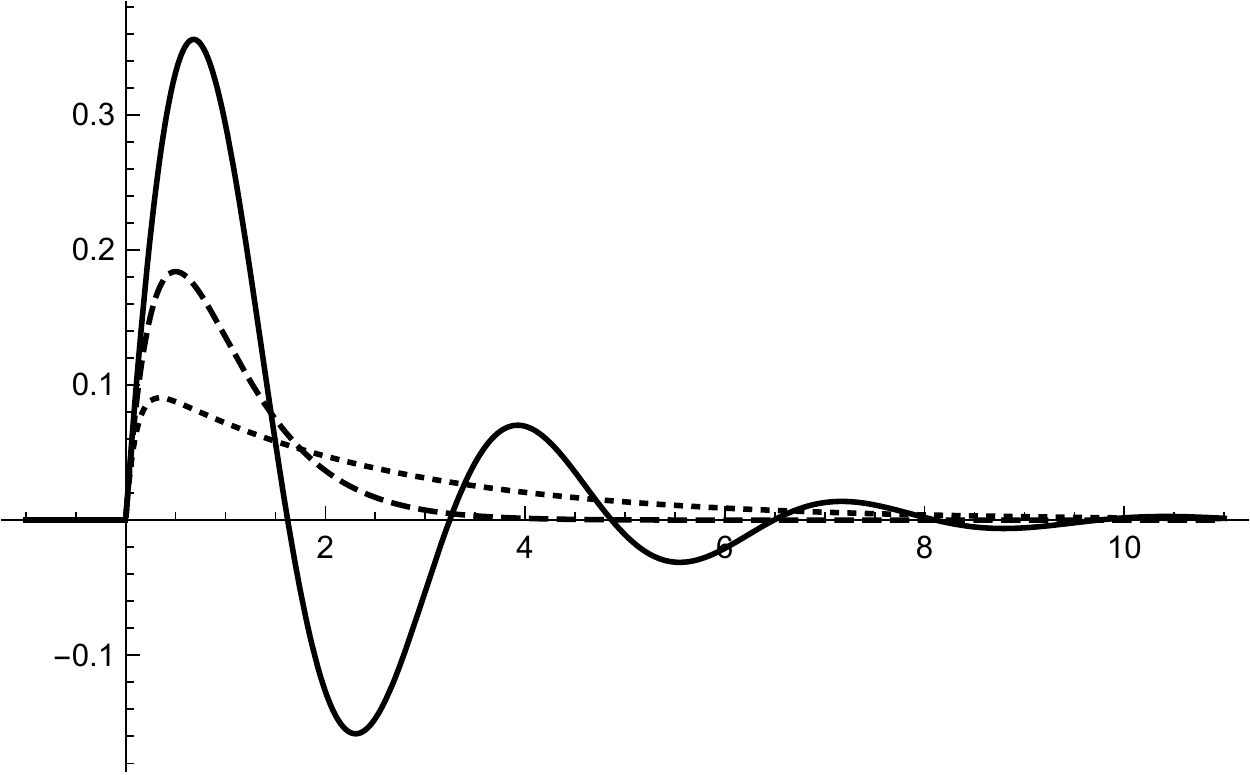}}
\caption{The spectral density $\rho(\omega)$ (left inset) and the retarded function $\chi_{\text r}(t)$ (right inset) of the damped harmonic oscillator for $(m,\Gamma)=(2,0.5), (2,2), (2,5)$ (plain, dashed and dotted lines respectively). \label{fig6}}
\end{figure}

We shall focus on the three representative cases $(m,\Gamma) = (2,0.5)$ (mild damping), $(m,\Gamma) = (2,2)$ (limit case) and $(m,\Gamma) = (2,5)$ (strong damping), see Fig.\ \ref{fig6}.

\subsubsection{\label{DOap1} The ARG maps $\mathsf A^{+}_{\km,\delta}$}

We focus on the maps $\mathsf A^{+}_{\km,\delta}$. The maps $\mathsf A^{-}_{\km,\delta}$ work in a similar way (note that, moreover, $G_{k}=G_{-k}$ for the damped harmonic oscillator). 

\begin{figure}
\centerline{\includegraphics[width=2.9in]{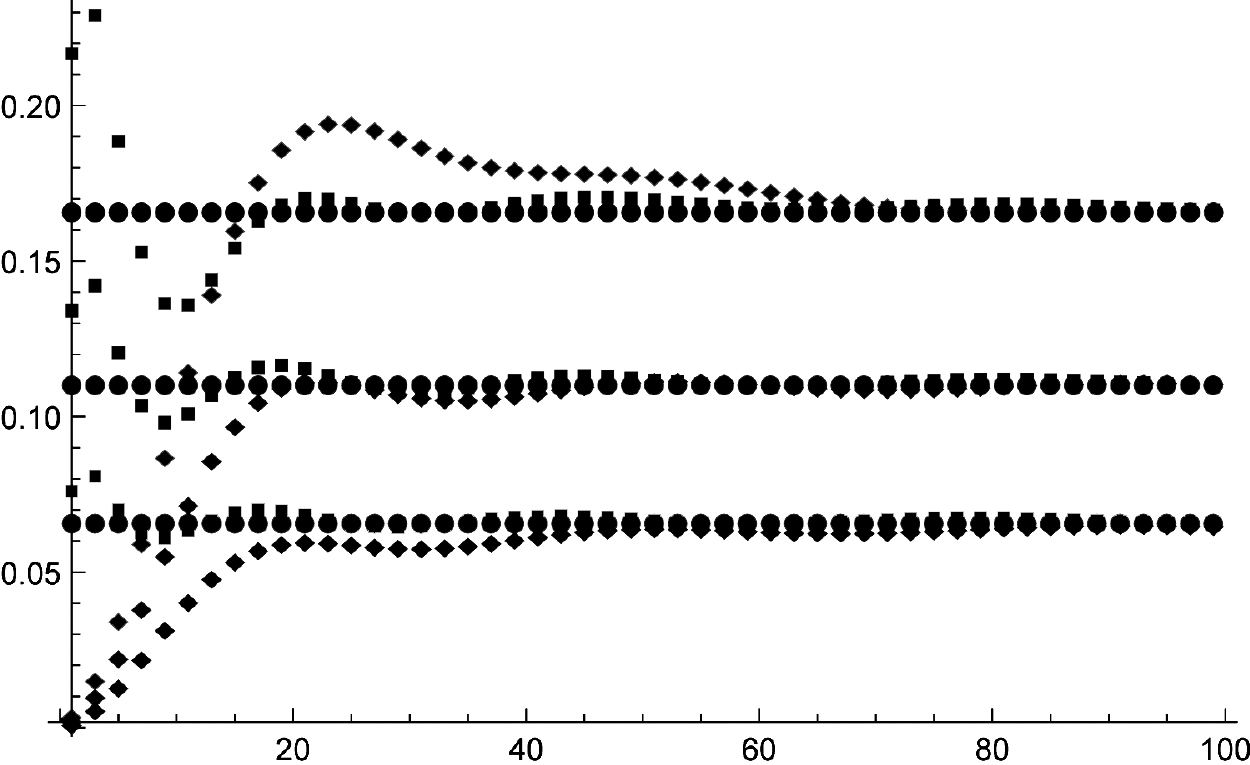}\hskip .2in\includegraphics[width=2.9in]{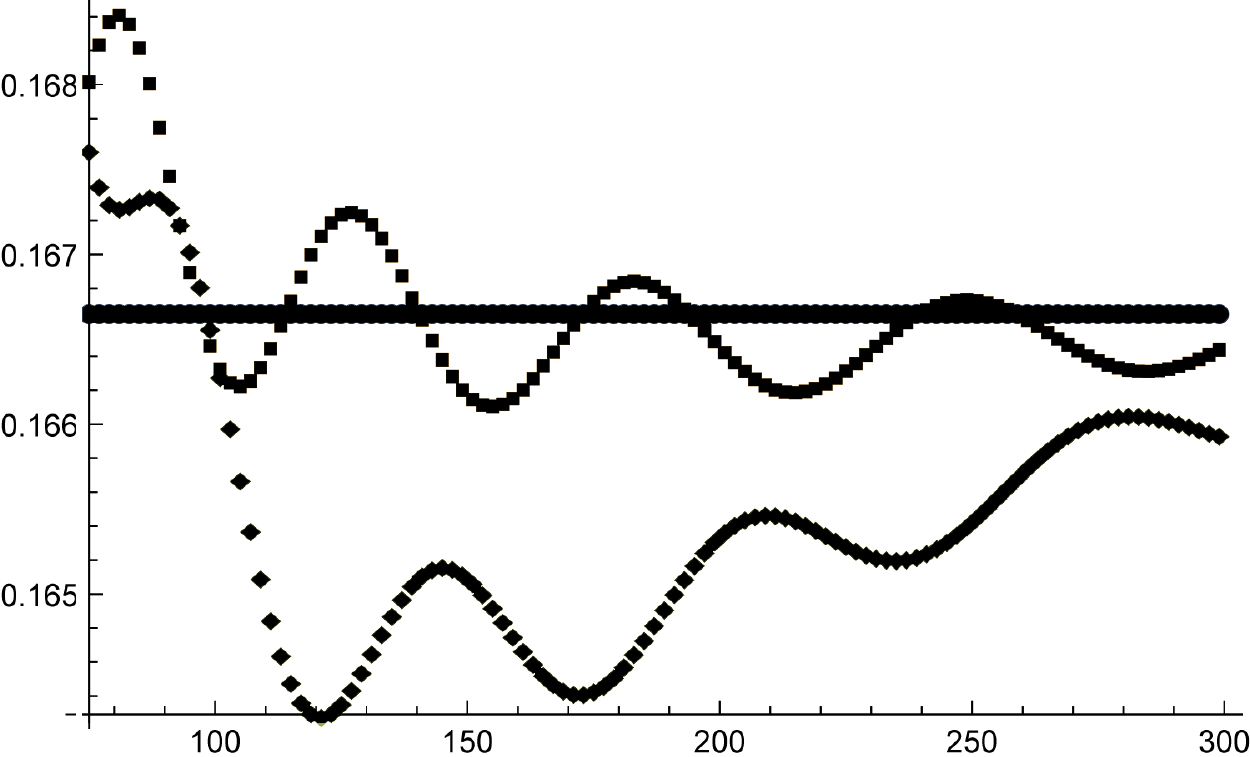}}
\caption{Left inset: the Fourier coefficient $G_{1}$ reconstructed from the ARG maps $\mathsf A^{+}_{2,1}$ (squares) and $\mathsf A^{+}_{5,1}$ (diamonds) as a function of the precision $N$, in the cases $(m,\Gamma)=(2,0.5), (2,2), (2,5)$ (top, middle and bottom). The flat lines (dots) represent the exact values. Right inset: high precision reconstruction of $G_{1}$ from $\mathsf A^{+}_{2,1}$ (squares) and $\mathsf A^{+}_{5,1}$ (diamonds) in the case $(m,\Gamma)=(2,0.5)$. \label{fig7}}

\vskip .5 cm

\centerline{\includegraphics[width=2.9in]{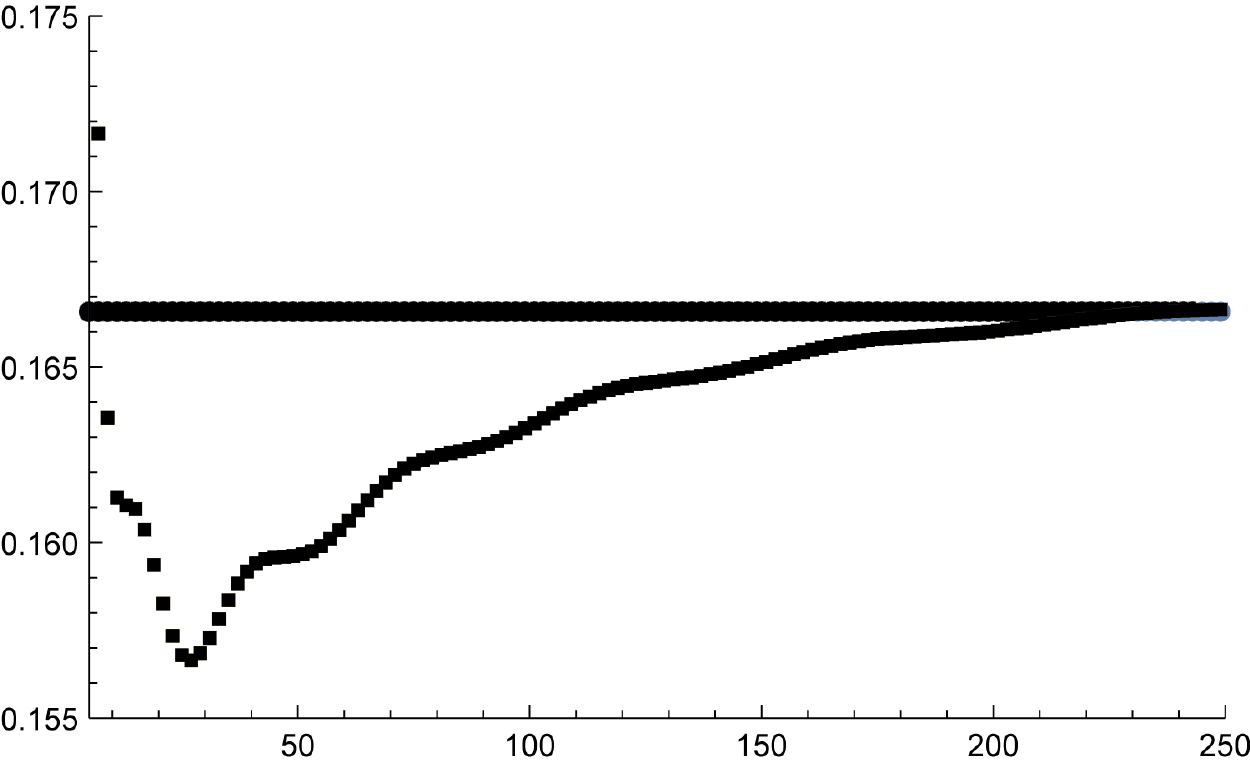}\hskip .2in\includegraphics[width=2.9in]{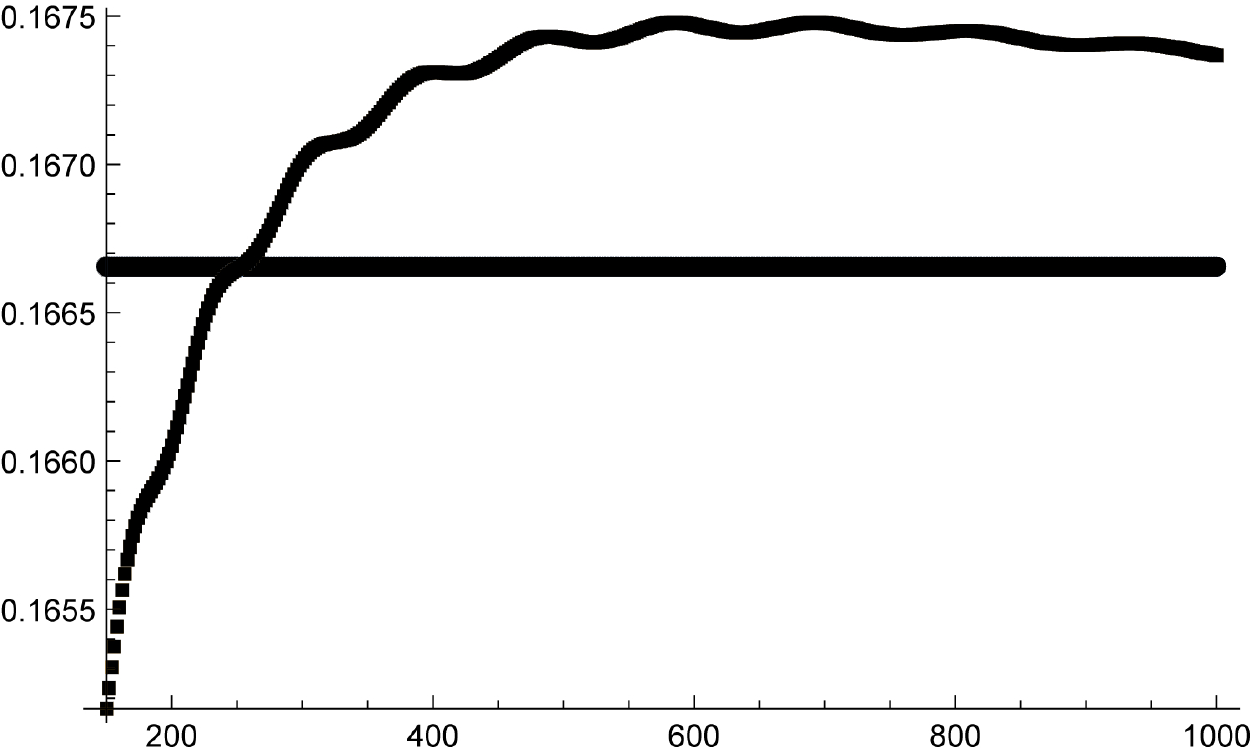}}
\caption{The Fourier coefficient $G_{1}$ reconstructed from the ARG maps $\mathsf A^{+}_{2,2}$ as a function of the precision $N$ in the case $(m,\Gamma)=(2,0.5)$. The flat line represents the exact values. \label{fig8}}

\end{figure}

On Fig.\ \ref{fig7}, we have depicted the Fourier coefficient $G_{1}$ obtained from the maps $\mathsf A^{+}_{\km=2,\delta=1}$ and $\mathsf A^{+}_{\km=5,\delta=2}$, as a function of the precision $N$ (the UV cut-off being adjusted according to the precision). Similar plots are obtained for the reconstruction of other Fourier coefficients and for other values of the RG scale $\km$ and the index $\delta$; convergence is slower when $\km$ and/or $\delta$ are increased. An example with $\delta=2$ is provided on Fig.\ \ref{fig8}.

\subsubsection{\label{DOap2} The spectral functions from the Euclidean data}
\begin{figure}
\centerline{\includegraphics[width=2.9in]{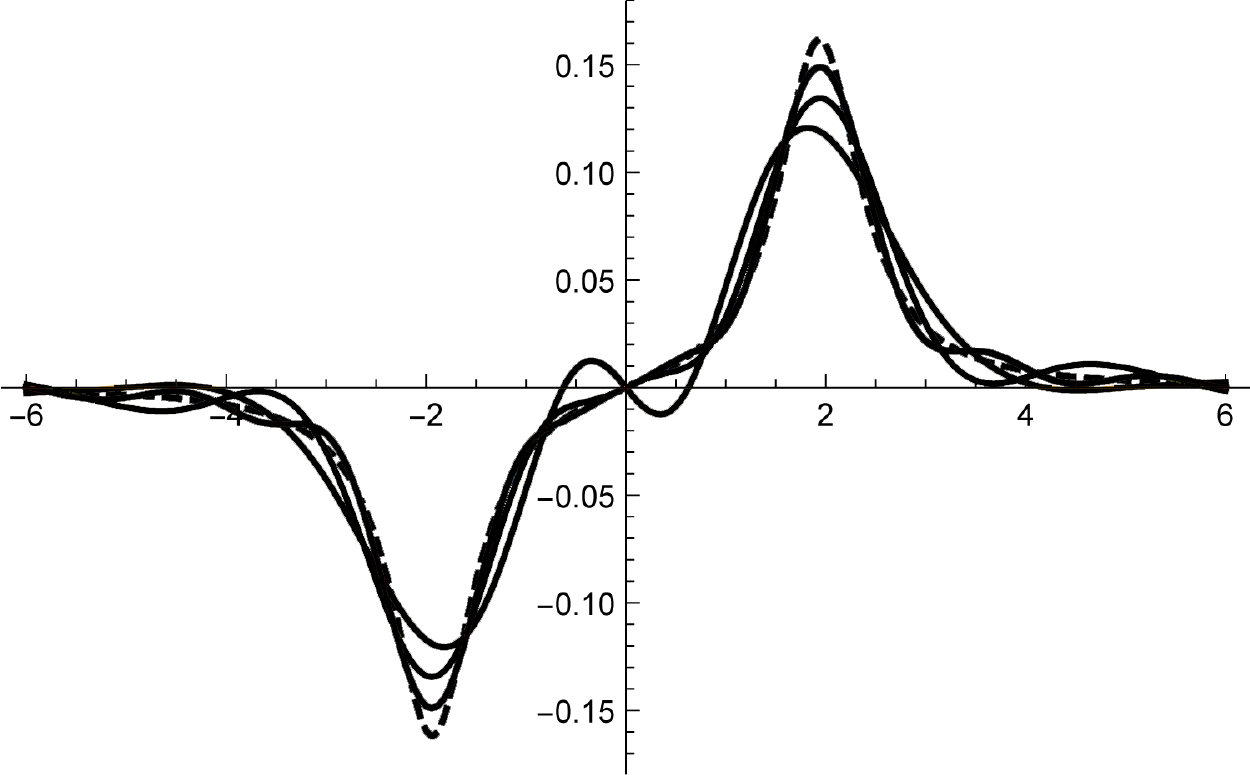}\hskip .2in\includegraphics[width=2.9in]{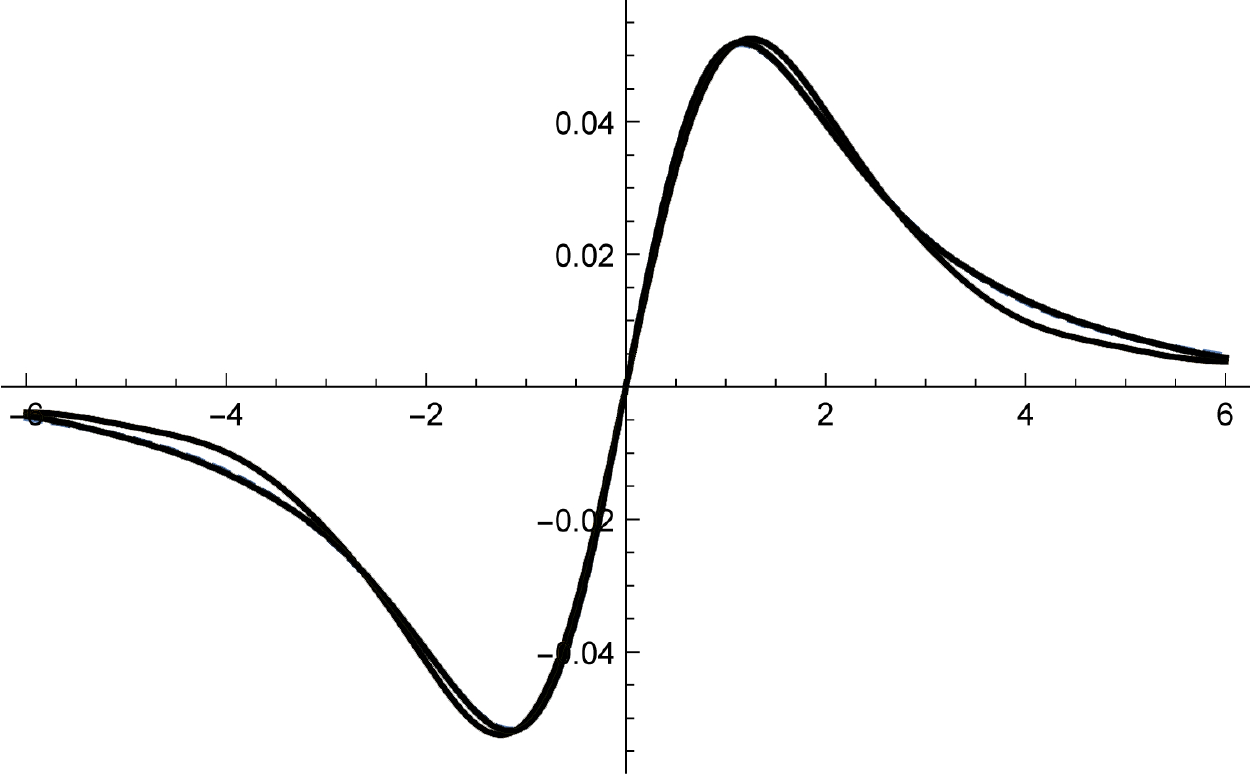}}

\vskip .5cm

\centerline{\includegraphics[width=2.9in]{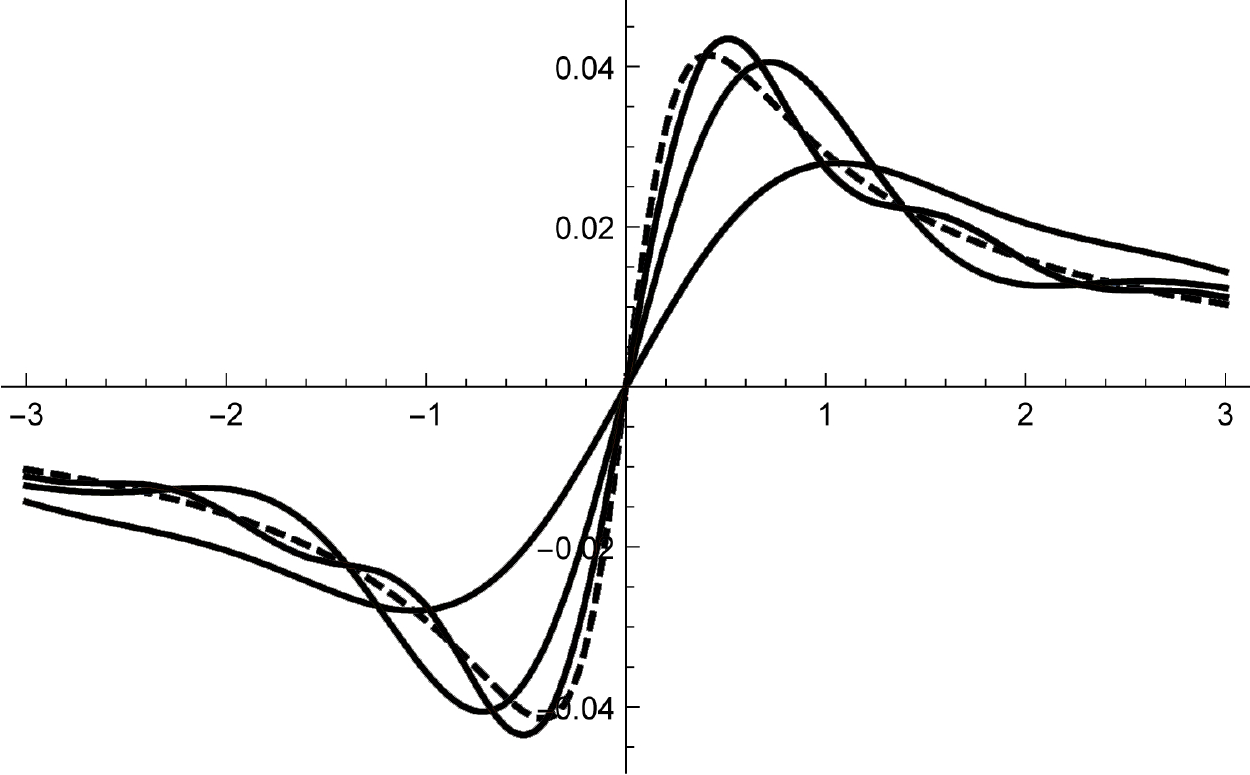}\hskip .2in\includegraphics[width=2.9in]{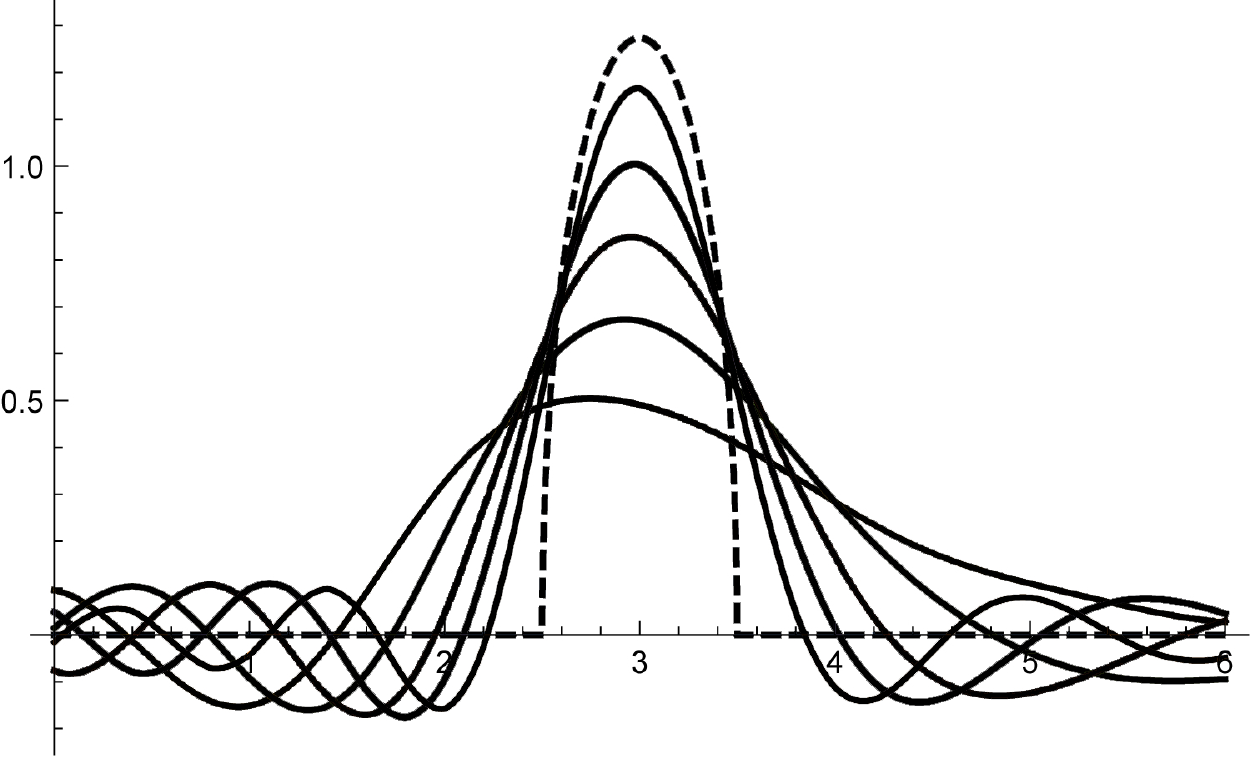}}

\caption{\label{fig9} Reconstruction of the spectral function from the Euclidean data. In all cases, the dashed line represents the exact result. Upper-left inset: $(m,\Gamma)=(2,0.5)$ for $N=50, 100, 500$; upper-right inset: $(m,\Gamma) = (2,2)$ for $N=15, 50$; lower-left inset: $(m,\Gamma)$ for $N=5,50,500$; lower-right inset: Wigner semi-circle law for $a=2.5$ and $b=3.5$ reconstructed for $N=25,50,100,200,500$.}
\end{figure}

In the case of the damped harmonic oscillator \eqref{resDO}, Eq.\ \eqref{resdisc} yields
\be\label{DOrho1} \rho(\omega) = -\frac{1}{\pi}\im R(\omega + i\epsilon)\, .\ee
On Fig.\ \ref{fig9}, we have depicted the reconstruction of the spectral function $\rho$ from \eqref{DOrho1} using the general ARG map \eqref{argg1} and \eqref{argg2}. In some cases, an excellent result is obtained using a small precision (e.g.\ the spectral density obtained for $N=15$ in the case $(m,\Gamma)=(2,2)$ is already excellent), whereas in other cases a much higher precision is needed. Quite generally, a reliable reconstruction of sharp peaks requires a high value of $N$. 

We have also included an example for the Wigner's semi-circle law. For any choices of $b>a$, it corresponds to the Fourier coefficients
\be\label{FGkWigner} G_{k} = \frac{8}{(b-a)^{2}}\Bigl[ ik - \frac{a+b}{2} - \sqrt{(ik-a)(ik-b)}\Bigr]\ee
and the spectral function
\be\label{rhoWigner} \rho(\omega) = \left\{
\begin{aligned} 
& \frac{8}{\pi (b-a)^{2}}\sqrt{(b-\omega)(\omega - a)}\quad\text{for}\ \omega\in [a,b]\\
& 0 \quad\text{for}\ \omega\not\in [a,b]\, .\end{aligned}\right.
\ee
This example has some qualitative difference with the damped harmonic oscillator: the spectral function has a compact support, the resolvent has square root branch cuts and the real-time two-point functions has a power-law decay at large time instead of an exponential decay.

\subsubsection{\label{DOap3} The real-time retarded function from the Euclidean data}

\begin{figure}
\centerline{\includegraphics[width=2.9in]{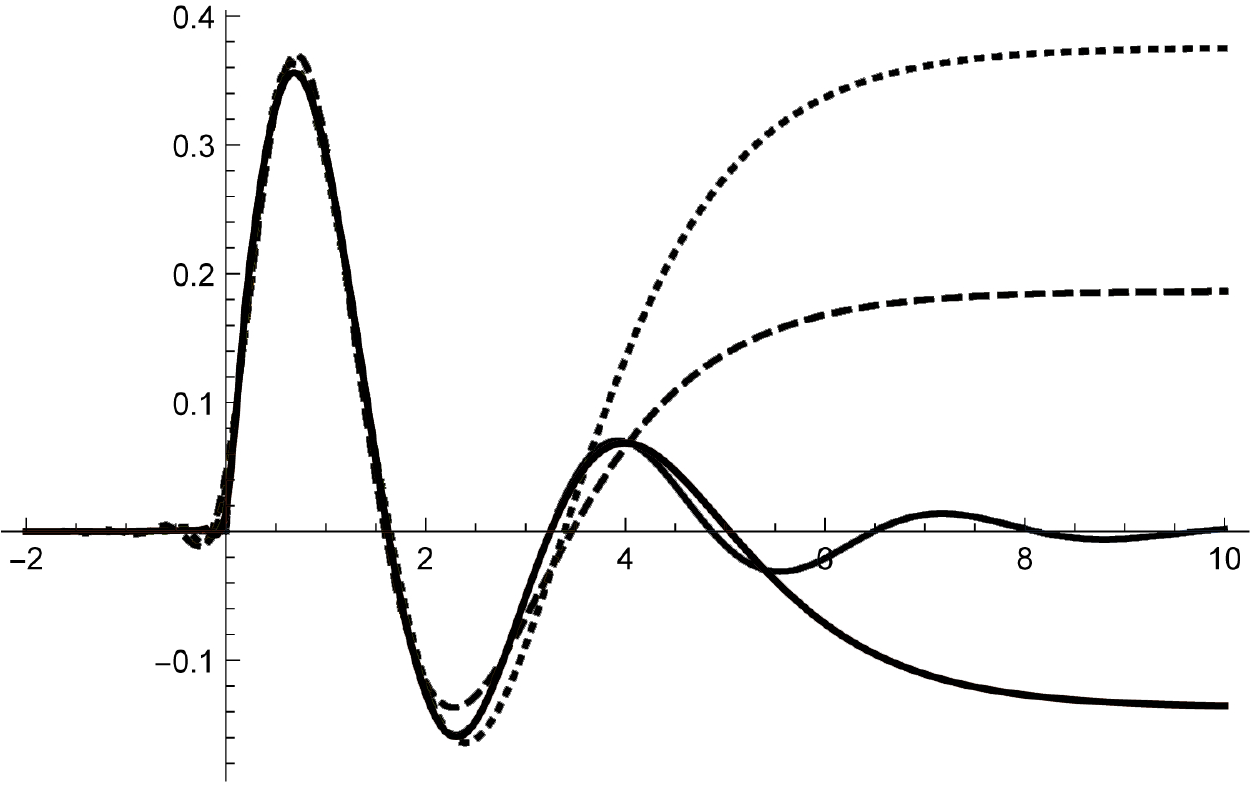}\hskip .2in\includegraphics[width=2.9in]{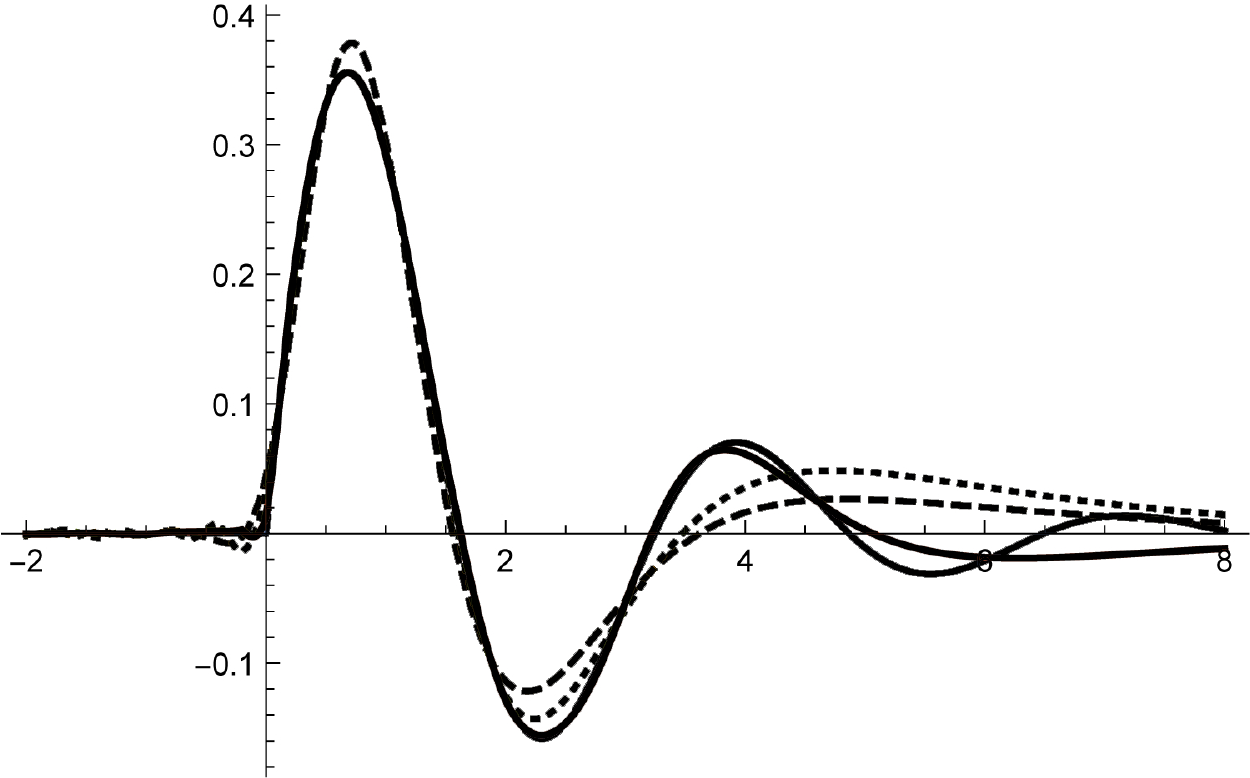}}

\vskip .5cm

\centerline{\includegraphics[width=2.9in]{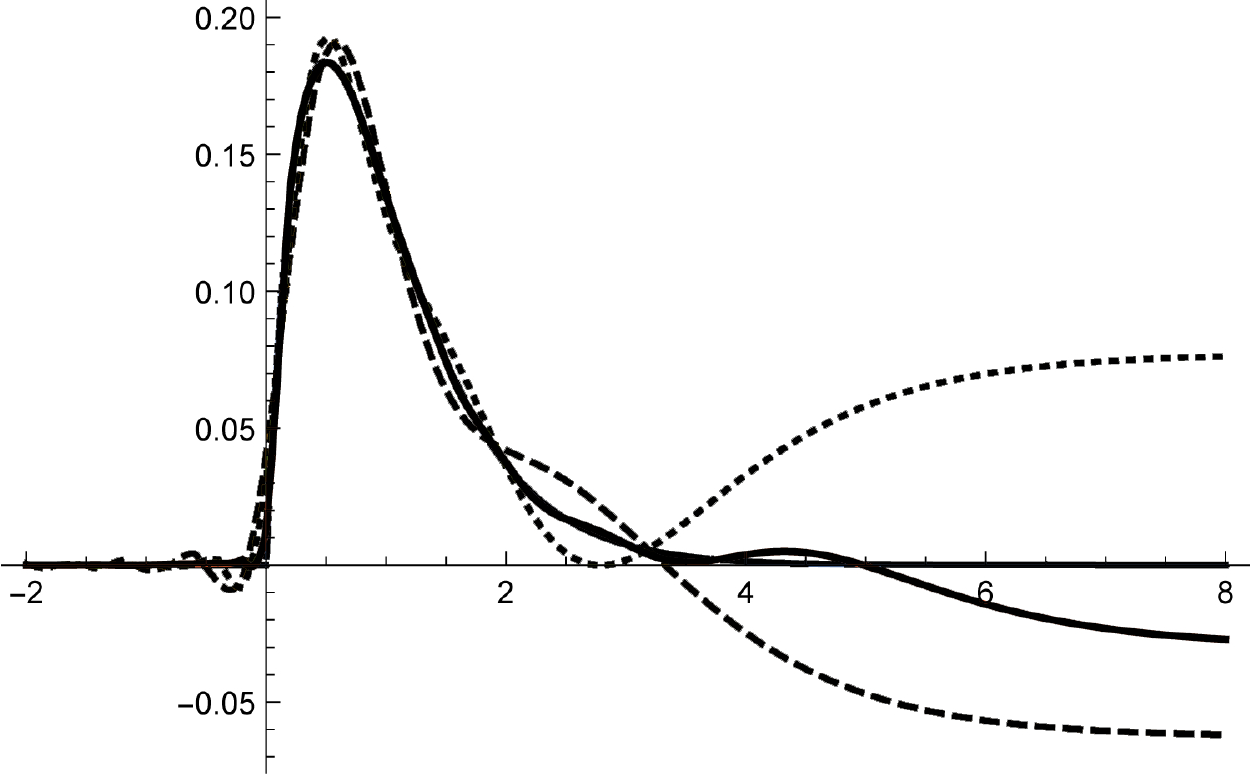}\hskip .2in\includegraphics[width=2.9in]{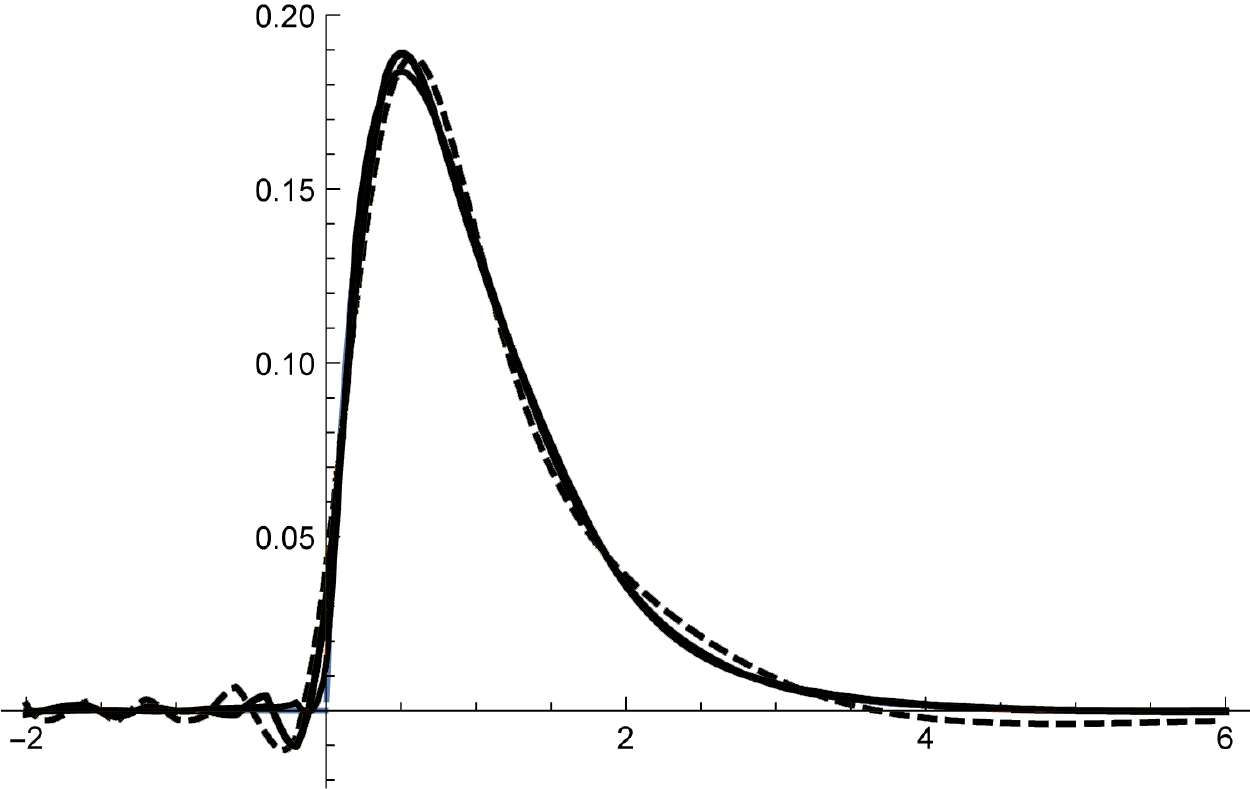}}

\vskip .5cm

\centerline{\includegraphics[width=2.9in]{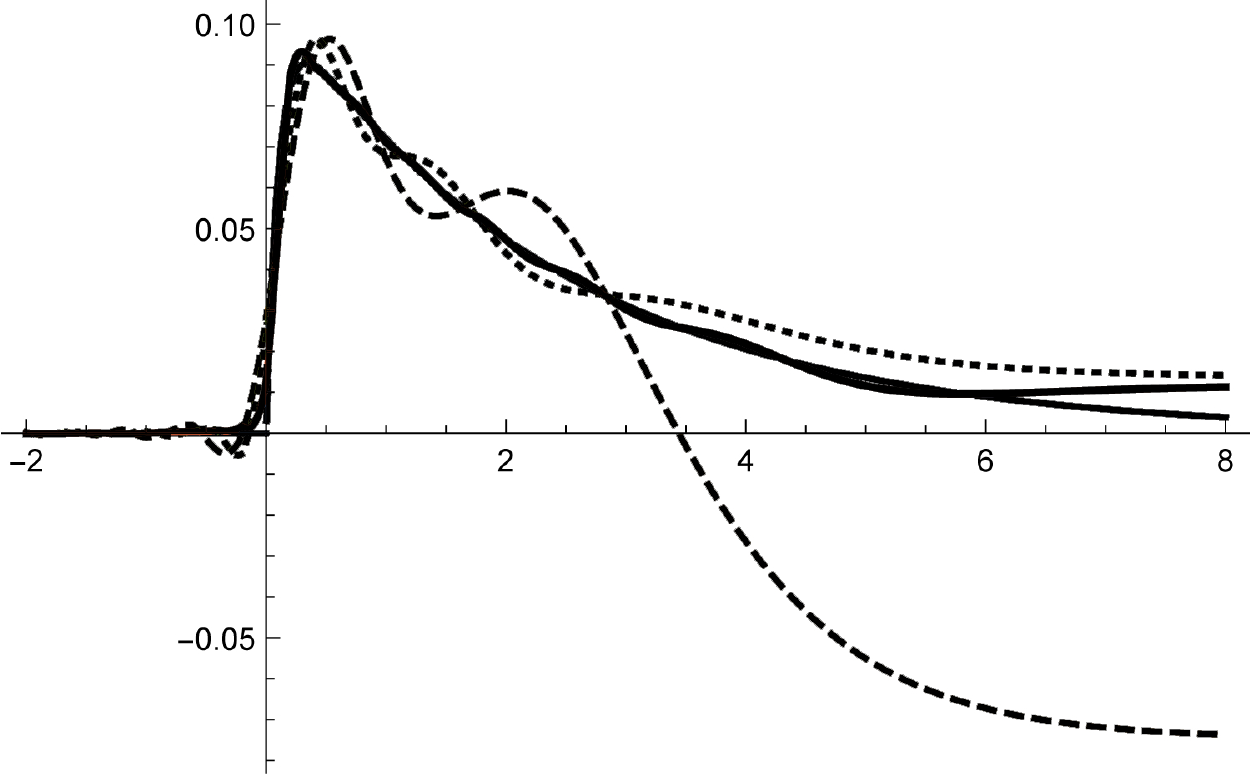}\hskip .2in\includegraphics[width=2.9in]{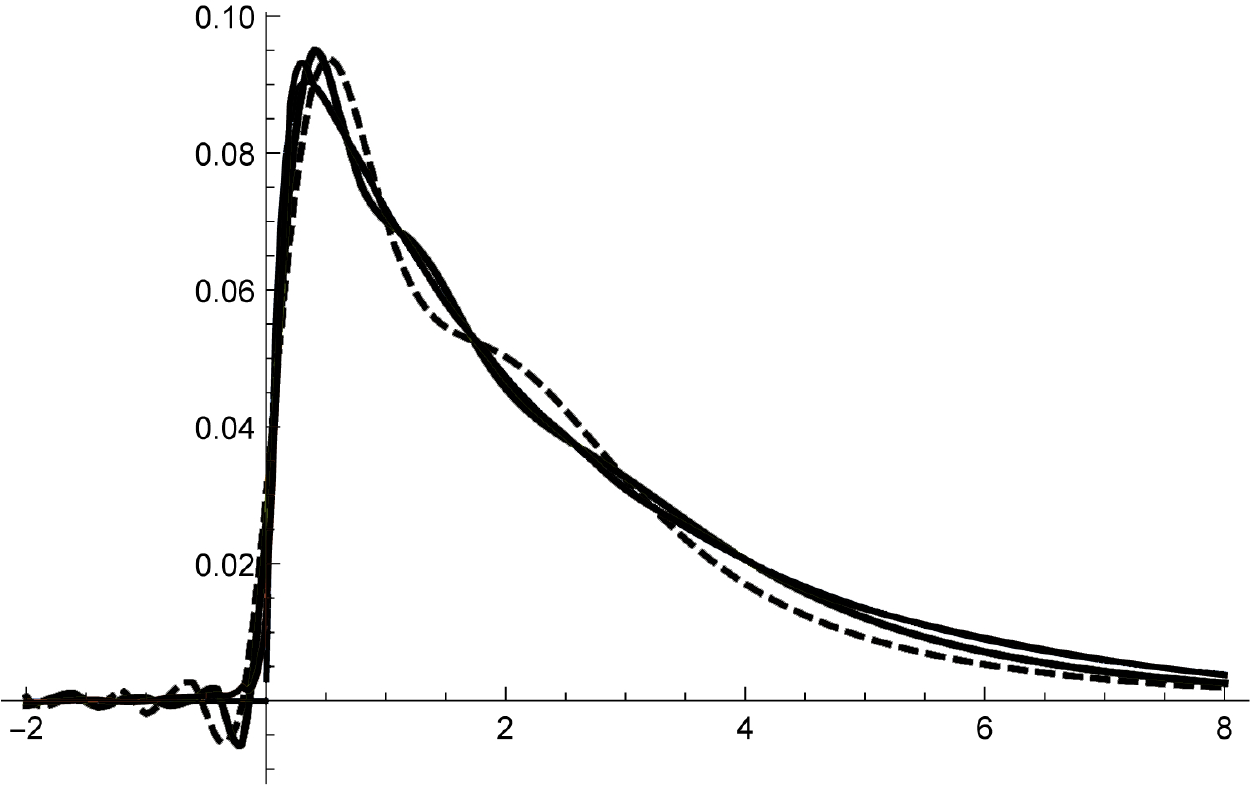}}

\caption{\label{fig10} The retarded two-point function $\chi_{\text r}(t)$ reconstructed from the Euclidean data by using the ARG maps \eqref{chir3} with $(\km,\delta,\alpha)=(2,1,1)$ (left insets) and $(\km,\delta,\alpha) = (1,1,3/4)$ (right insets), in the cases $(m,\Gamma)=(2,0.5)$ (upper plots), $(m,\Gamma)=(2,2)$ (center plots) and $(m,\Gamma)=(2,5)$ (lower plots). The normal lines correspond to the exact solutions whereas the dashed, dotted and thick lines correspond to $N=25,50,300$ respectively.}
\end{figure}

Fig.\ \ref{fig10} illustrates the direct reconstruction of the real-time retarded two-point functions $\chi_{\text r}(t)$ from the Euclidean data, using \eqref{argg5} and \eqref{argg6}. We are using two different ARG maps: one with $(\km,\delta,\alpha)=(2,1,1)$, for which the exponential pre-factor in \eqref{argg5} vanishes; and one with $(\km,\delta,\alpha)=(1,1,3/4)$, for which there is a non-vanishing exponential pre-factor $e^{-t/2}$. 

For relatively short times, an excellent reconstruction can be obtained using modest values for the precision. However, to get the long-time behaviour right requires higher and higher precisions. Using values of $(\km,\delta,\alpha)$ for which there is an explicit exponential damping factor in \eqref{chir3} of course helps in this respect, since the exact result goes to zero at large time. This is true even if the rate of damping associated with the chosen values of $\km$, $\delta$ and $\alpha$ doesn't match the exact result, as for the choice $(\km,\delta,\alpha)=(1,1,3/4)$ in Fig.\ \ref{fig10}. 

\subsubsection{\label{DOap4} The long-time behaviour from the Euclidean data}

\begin{figure}
\centerline{\includegraphics[width=2.9in]{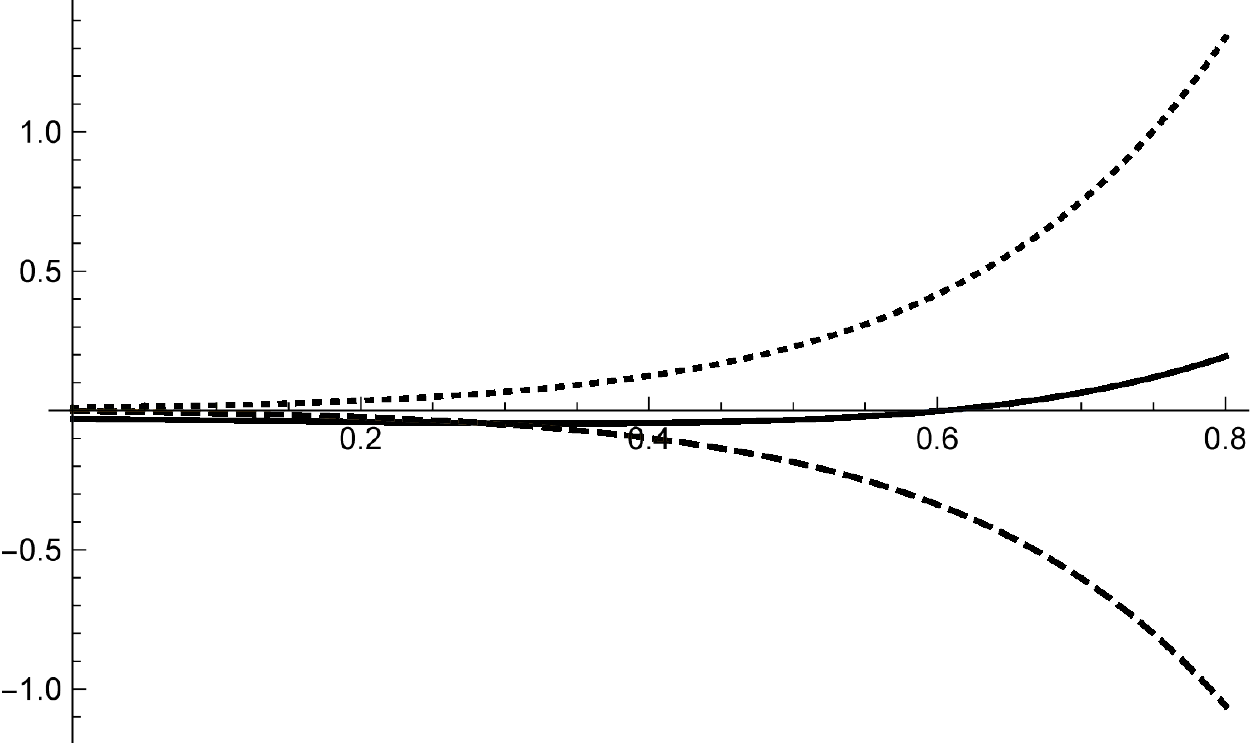}\hskip .2in\includegraphics[width=2.9in]{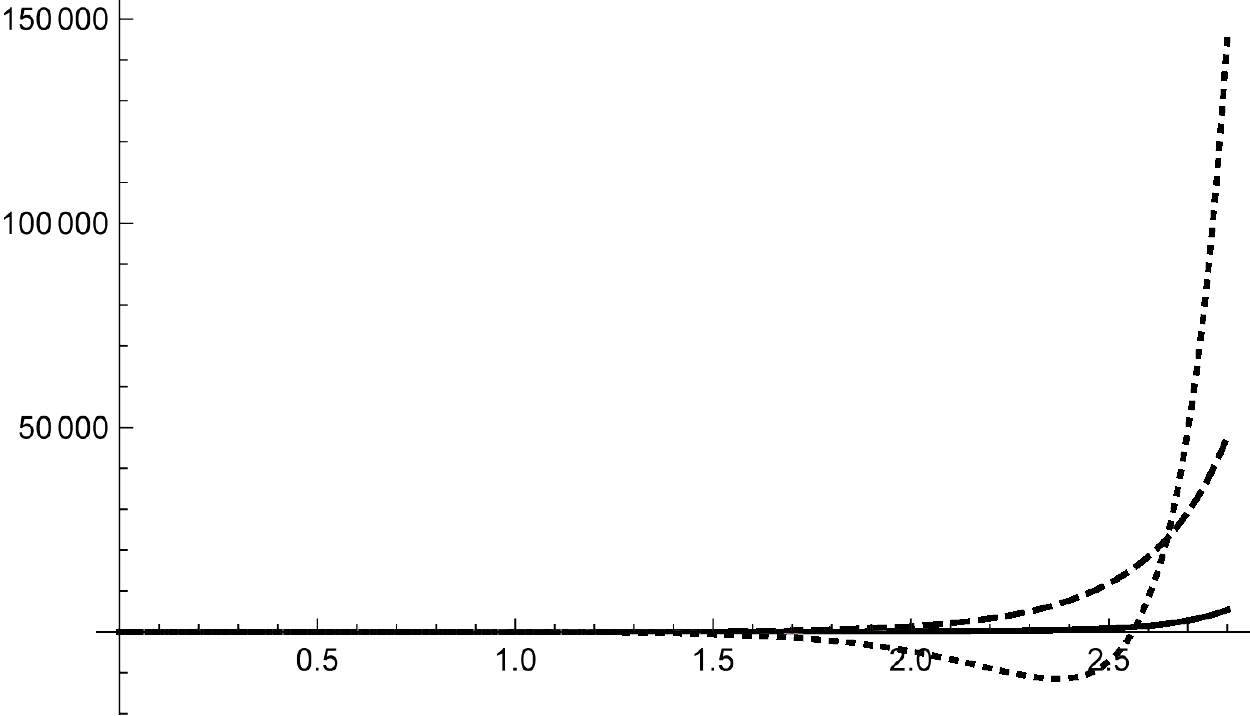}}
\caption{\label{fig11} Plots of the left-hand side of \eqref{critgamma} as a function of $\tilde\gamma$, for precisions $N=100, 300, 1000$ (plain, dashed and dotted lines). Left inset: case $(m,\Gamma)=(2,0.5)$, for which $\Gamma=0.5$, for the choice $\km=\delta=1$. Right inset: case $(m,\Gamma)=(2,2)$, for which $\gamma=2$, for the choice $\km=3$, $\delta=1$. The case $(m,\Gamma)=(2,5)$, for which $\gamma=5-\sqrt{21}\simeq 0.417$ is similar to $(m,\Gamma)=(2,0.5)$.}
\end{figure}

Let us now illustrate the criterion \eqref{critgamma} for the computation of the thermalization time scale $\gamma^{-1}$, defined by \eqref{typicallarget}, from the Euclidean data. The results of the previous subsection showed that a reliable description of the long-time behaviour of $\chi_{\text r}$ requires a very high precision $N$ and thus we do not expect that the condition \eqref{critgamma} will be satisfied very sharply for moderate values of $N$. However, Fig.\ \ref{fig11} is rather suggestive. It clearly hints at the existence of two qualitatively disctinct regions for the left-hand side of \eqref{critgamma}, as a function of $\tilde\gamma$: one for which it is nearly zero and one for which it deviates from zero and tends to diverge. This is a convincing sign that the correlator decays exponentially when $t\rightarrow\infty$, but only a rough estimate of the corresponding thermalization time scale $\gamma^{-1}$ is obtained, even when one uses the high precision $N=1000$. 

\subsection{\label{IncomDataSec} Incomplete data set and analytic interpolation}

The ARG maps $\mathsf A_{\km,\delta}^{+}$ allow to reconstruct the low energy Fourier coefficients $G_{k}$ for $k<\km$ in terms of the high energy coefficients $G_{k}$ for $k\geq\km$. This amounts to performing an exact ``discrete'' analytic interpolation of the Fourier coefficients below some energy scale $\km$. If one knows the $G_{k}$ below some cut-off $K$ only, i.e.\ for $|k|\leq K$, one can still use the approximate ARG maps given by \eqref{argg3} and \eqref{argg4} to perform the analytic interpolation with some finite precision.

One can also consider more general analytic interpolation problems. For example, assuming that the Fourier coefficients are known for $1\leq k<k_{1}$ and for $k>k_{2}$, one could try to deduce the coefficients in the interval $k_{1}\leq k\leq k_{2}$. One obvious way to do this is to use an ARG map $\mathsf A_{\km,\delta}^{+}$ for some $\km>k_{2}$. This does not use the knowledge of the coefficients for $1\leq k<k_{1}$. In practice, using this knowledge allows to immensely improve the precision of the interpolation, thank's to the extreme sensitivity of the ARG maps on the data set discussed in \ref{extSec}.

To clearly understand this point, let us start with the simplest possible exercice: the reconstruction of a single unknown coefficient, say $G_{10}$, assuming that all the others are known. To do that, we may use \eqref{argg3} with, for instance, the choices $\km=2, \delta=1$, $k=1$ and some finite precision $N$. Due to the UV decoupling, the Fourier coefficients above a certain $N$-dependent UV cut-off $K$ are irrelevant. We thus get a single linear constraint with a finite number of terms,
\be\label{interex1} G_{1} \underset{N}{=} \sum_{p=0}^{K-2}\mathsf A_{2,1}(N;1,p) G_{2+p}\, ,\ee
which allows to obtain an approximate value for the single unknown $G_{10}$. The approximate value we obtain in this way is \emph{extremely} precise, because the matrix element $\mathsf A_{2,1}(N;1,10)$ multiplying the unknown coefficient is typically huge (see Fig.\ \ref{fig3} and \ref{fig4}), whereas the left-hand side of \eqref{interex1} involves the known coefficient $G_{1}$ multiplied by one! For example, $N=200$ yields the correct value for $G_{10}$ with a relative error of the order of $10^{-15}$.

More generally, it is convenient to use the ARG equations in the form \eqref{argg7}. If all the Fourier coefficients except $n$ are known, we can use $n$ equations \eqref{argg7}, obtained by choosing $n$ different values for the parameters $\alpha,\km,\delta,u$, to perform the analytic interpolation.

For instance, assume that all the coefficients $G_{k}$ are known, except for ten of them corresponding to $6\leq k\leq 15$. We choose $N=200$ (a cut-off $K=75$ is then amply enough), $\alpha=1/2$, $\km=1$, $\delta=1$ and we solve the linear equations
\be\label{lin2} \sum_{p=0}^{K}\mathsf A_{1/2}(N;u,p)G_{1+p} = 0\ee
for the ten values $u=3+j/5$, $0\leq j\leq 9$ to get the ten unknow Fourier coefficients. Let us denote by $\tilde G_{k}$, $6\leq k\leq 15$, the coefficients obtained in this way, whereas the notation $G_{k}$ is kept for the exact values. To evaluate the error, we compute
\be\label{err1} \sigma = \sqrt{\frac{1}{10}\sum_{k=6}^{15}\biggl(\frac{\tilde G_{k}-G_{k}}{G_{k}}\biggr)^{2}}\, .\ee
For the cases $(m,\Gamma)=(2,1/2)$, $(m,\Gamma)=(2,2)$ and $(m,\Gamma)=(2,5)$, we find, using this method, $\sigma \simeq 1.35\, 10^{-4}$, $\sigma \simeq 6.87\, 10^{-5}$ and $\sigma \simeq 1.47\, 10^{-4}$ respectively. 


%
\section{\label{useARGSec} ARG and data improvement}

In this last section, we show how to use the ARG equations to systematically improve random approximate Euclidean data sets obtained, for example, from Monte-Carlo simulations. The basic philosophy is very similar to the use of standard RG equations in field theory to improve perturbation theory: one relies on the fact that the RG equations are exact statements. By combining these exact statements with approximate data, it is not surprising that one can devise algorithms to improve the data.

The rather spectacular aspect of the method is that the ARG equations are completely universal, model-independent constraints. The very same algorithms can thus be applied in principle to improve Monte-Carlo data for problems as diverse as lattice QCD, strongly correlated electron systems or strongly coupled matrix quantum mechanical models of black holes, etc.

\subsection{\label{Approx1Sec} General principle}

As in the general presentation in Sec.\ \ref{IntroSec}, let $\mathscr F$ be the set of Fourier coefficients $(G_{k})_{k\in\mathbb Z}$ satisfying all the basic standard constraints (but not the ARG equations). We endow $\mathscr F$ with a scalar product, which induces a notion of distance $d$. The distance $d$ gives a measure of how much two sets of Fourier coefficients are physically close to each other. There may be several natural choices for $d$, see below.

Let us assume that we have at our disposal an imprecise data set $\Ga=(\Ga_{k})_{k\in\mathbb Z}$. This is of course a very common and important situation, since most interesting models cannot be solved exactly. The data $\Ga$ can be seen as a point in $\mathscr F$. It approximates an exact, but in principle unknown, set of Fourier coefficients $\Ge=(\Ge_{k})_{k\in\mathbb Z}$. The distance $d(\Ge,\Ga)$ measures the accuracy of the approximation.\footnote{One must not confuse the accuracy of the approximate data point $\Ga$ and the ``precision'' $N$ of the numerical analysis introduced in Sec. \ref{ExSec}. We shall keep using this terminology, accuracy of the data versus precision of the numerics, to avoid confusion.}

The point $\Ge$ belongs to the linear subspace $\mathscr M$ of $\mathscr F$ defined by the set of all the ARG equations, for example the equations \eqref{argeqGEN} for all the allowed choices of $\alpha$, $u$, $\km$ and $\delta$. The point $\Ga$, on the other hand, is a random point in $\mathscr F$ belonging to a certain ball centered on $\Ge$; the better the accuracy of the Monte-Carlo simulation, the smaller the radius of the ball. 

\begin{figure}
\centerline{\includegraphics[width=6in]{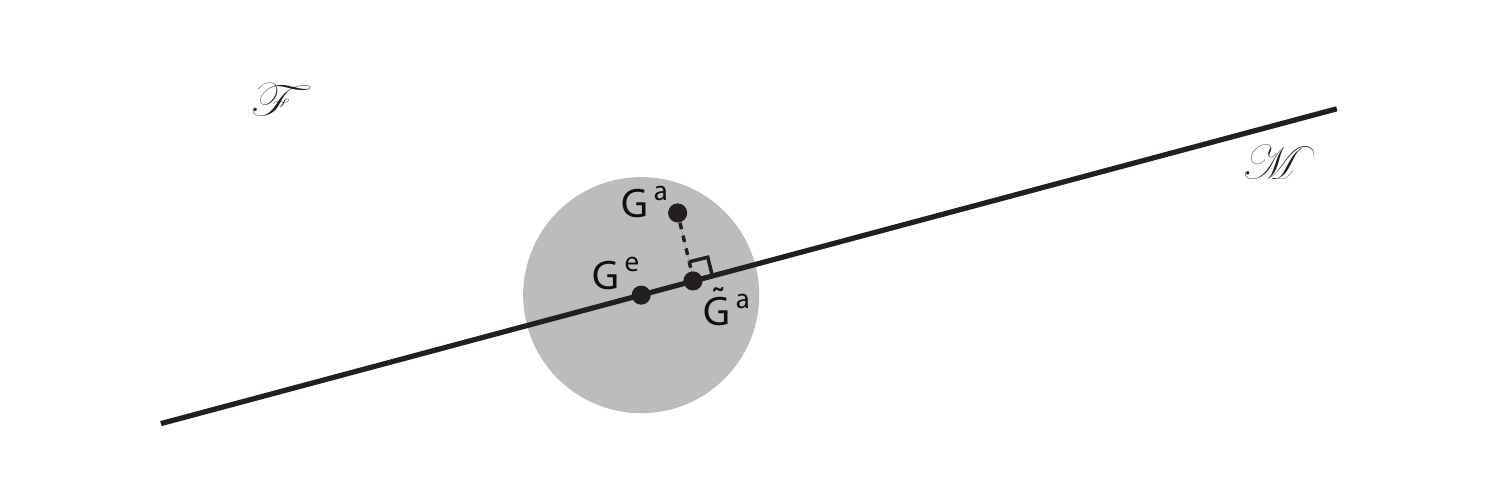}}
\caption{\label{fig12} A typical Monte-Carlo simulation yields an approximate data point $\Ga$ in the vicinity of the exact result $\Ge$. $\Ga$ can be seen as a random point belonging to a certain ball around $\Ge$ whose radius parameterizes the accuracy of the Monte-Carlo simulation. By projecting $\Ga$ onto the linear subspace $\mathscr M$ defined by the ARG equations, we obtain a new approximate data point $\tilde\Ga$. By construction, $\tilde\Ga$ is more accurate than $\Ga$, $d(\Ge,\tilde\Ga)\leq d(\Ge,\Ga)$.}
\end{figure}

One can then improve systematically the approximate data by using an extremely simple idea: we consider the orthogonal projection $\tilde\Ga$ of $\Ga$ onto $\mathscr M$. By the Pythagoras' theorem, $d(\Ge,\tilde\Ga)\leq d(\Ge,\Ga)$: \emph{the new data point $\tilde\Ga$ is automatically more accurate than the data point $\Ga$ we started with!} This simple method is illustrated on Fig.\ \ref{fig12}.\footnote{We do not discuss here the subtleties associated with the fact that the spaces $\mathscr F$ and $\mathscr M$ are infinite dimensional. Indeed, for all practical purposes, in numerical implementations, we work in finite dimension.}

\subsection{Important properties of the numerical implementation}

\paragraph{Finite dimensional space} 

In all numerical implementations, we work with a finite numerical precision $N$ and a finite cut-off $K$ (adjusted according to $N$). The vector space $\mathscr F$ is thus replaced by its finite dimensional version $\mathscr F_{K}$. A point in $\mathscr F_{K}$ is a set of Fourier coefficients $(G_{k})_{|k|\leq K}$. Moreover, it is convenient to separate the positive, $k>0$, zero, $k=0$, and negative, $k<0$, frequencies,
\be\label{Fdec} \mathscr F_{K} = \mathscr F_{K}^{+}\oplus\mathscr F_{0}\oplus\mathscr F_{K}^{-}\, ,\ee
since the ARG equations do not mix positive and negative frequencies.

\paragraph{Natural distance functions}

A priori, any scalar product on $\mathscr F_{K}$ can be used to define a distance. Since the approximate data set $\Ga$ plays a special role, a natural choice is
\be\label{ddef} d_{K}(G,G') = \sqrt{\frac{1}{2K+1}\sum_{k=-K}^{K}\biggl| \frac{G_{k}-G'_{k}}{\Ga_{k}}\biggr|^{2}}\, .\ee
In practice, we shall focus on $\mathscr F_{K}^{+}$, with distance function
\be\label{ddef2} d_{K}^{+}(G,G') = \sqrt{\frac{1}{K}\sum_{k=1}^{K}\biggl| \frac{G_{k}-G'_{k}}{\Ga_{k}}\biggr|^{2}}\, .\ee

\paragraph{A random data point never belongs to $\mathscr M$}

Let us note that a randomly chosen point $\Ga$ around $\Ge$ will virtually never belong to $\mathscr M$. Actually, even if the point $\Ga$ is very close to $\Ge$, that is to say, even if the accuracy of the approximation is excellent, the ARG equations will usually be violated by huge amounts. This is due to their extreme sensitivity on the data set, as explained in \ref{extSec}. In other words, the ARG equations constitute a very delicate set of constraints which allow to detect, with very high precision, whether a set of Fourier coefficients is consistent with analyticity or not. 

\begin{figure}
\centerline{\includegraphics[width=2.9in]{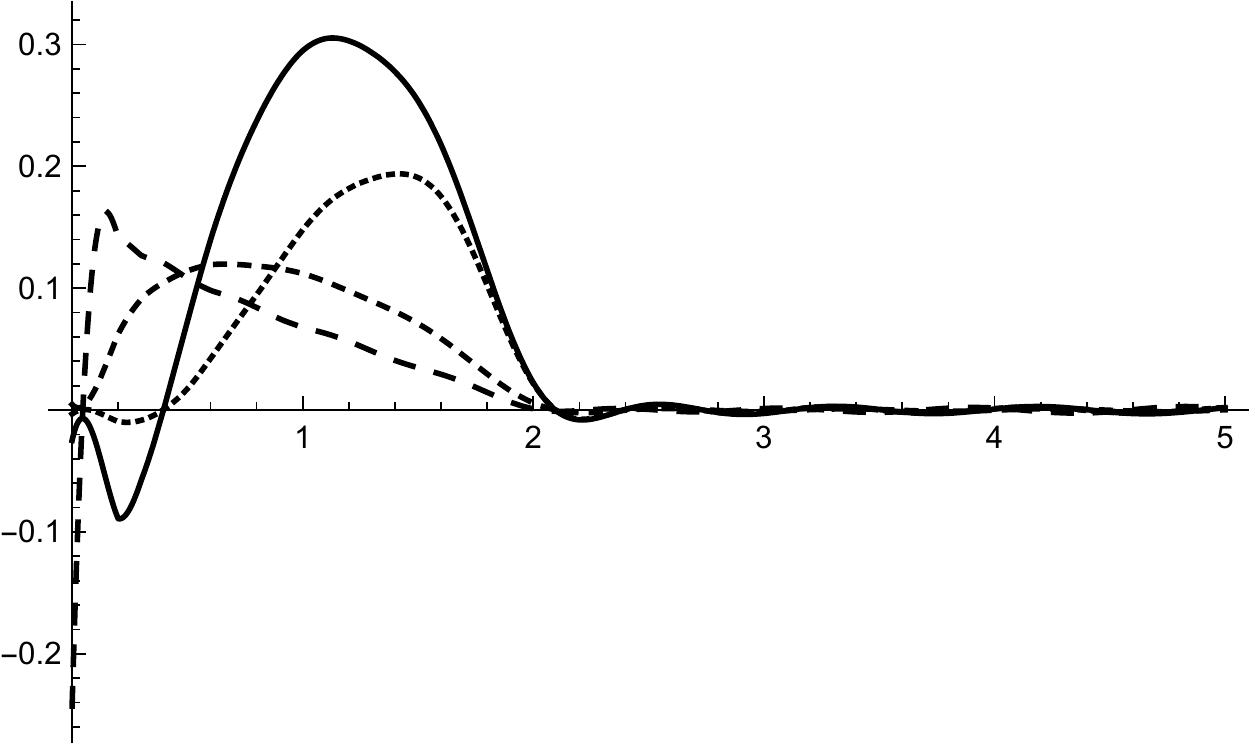}\hskip .2in\includegraphics[width=2.9in]{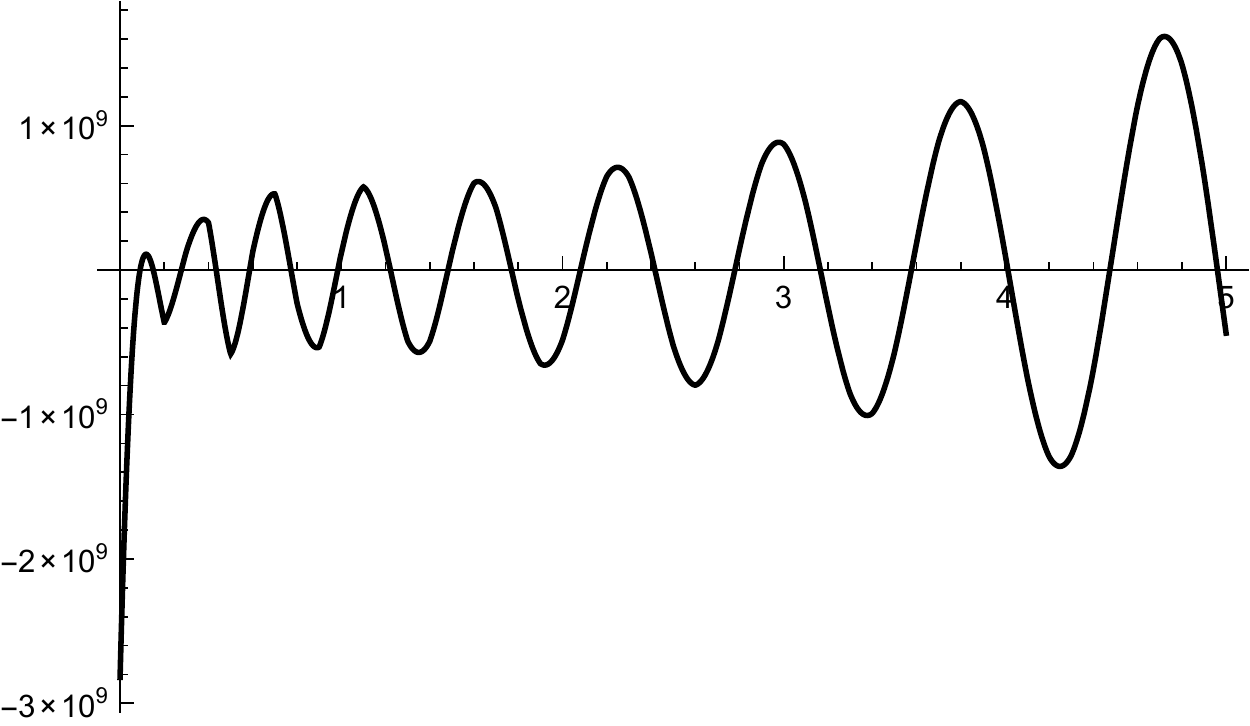}}
\caption{\label{fig13} Plot of the left-hand side of the ARG equations \eqref{argg7} as a function of $u$, for $N=200$ and $\alpha=1/2$, in the case of the damped harmonic oscillator with $(m,\Gamma)=(2,0.5)$. Left inset: the exact values for the coefficients $G_{k}$ are used, in the cases $(\km,\delta) = (1,1)$ (solid line), $(\km,\delta) = (2,1)$ (tiny dashed line), $(\km,\delta) = (2,2)$ (small dashed line) and $(\km,\delta) = (2,3)$ (medium dashed line). Right inset: the case $(\km,\delta) = (1,1)$, for which approximate values of the coefficients $G_{k}$ are used, the distance to the exact values being $\sim 10^{-5}$. Note the scale $\sim 10^{9}$ on the ordinate axis!}
\end{figure}

This property is illustrated on Fig.\ \ref{fig13}. On the left inset is plotted the left-hand side of \eqref{argg7}, as a function of $u$, for $N=200$,\footnote{For this value of the precision, a cut-off $K=75$ is amply enough.} $\alpha=1/2$ and various values of $\km$ and $\delta$, for the data set $\Ge$ corresponding to the damped harmonic oscillator \eqref{GkDO} at $(m,\Gamma)=(2,0.5)$. These functions all vanish (to a very good precision) for $u\geq 2$, as implied by the ARG equations \eqref{argg7}. On the right inset is plotted an instance of the same function, but using an approximate data set $\Ga$ instead of the exact values \eqref{GkDO}. The approximate data set is related to the exact data set by
\be\label{GaGerel} \Ga_{k} = \Ge_{k} \bigl( 1 + \varepsilon_{\sigma,k}\bigr)\, ,\ee
where the $\varepsilon_{\sigma,k}$ are independent Gaussian random variables of width $\sigma$, with probability density
\be\label{Gaussdens} f_{\varepsilon}(x) = \frac{1}{\sqrt{2\pi}\sigma} e^{-\frac{x^{2}}{2\sigma^{2}}}\, .\ee
On the plot we choose $\sigma = 10^{-5}$ which yields, on the particular realization we use, $d_{75}^{+}(\Ge,\Ga) \simeq 9.65\, 10^{-6}$. This means that $\Ga$ is a very good approximation to $\Ge$. In particular, the Euclidean correlators $G(\tau)$ computed from $\Ge$ and $\Ga$ are almost indistinguishable on a plot. Nevertheless, the graph on the right inset of Fig.\ \ref{fig13} clearly shows that the ARG equations are wildly violated. Actually, one must go to accuracies as good as $\sigma\sim 10^{-15}$ for the approximate plot to start looking like the exact plot! And this value of $\sigma$ would be even smaller if we were working at a higher precision $N$.

\paragraph{Singularity near the UV cut-off and the ``perturbative'' cut-off}

The region near the UV cut-off is, of course, singular. Indeed, for $|k|>K$ we set $G_{k}$ to zero artificially. This is manifestly inconsistent with the analyticity properties, except if $G_{k}=0$ for all $k$.

However, this is not a serious flaw. As we have explained previously, working with a finite cut-off has essentially no effect much below the cut-off. The only obvious limitation is that we cannot expect to improve significantly the data near the cut-off using the ARG. 

Moreover, let us note that, at high energies, a very reliable approximation to the coefficients $G_{k}$ can be obtained in many models of interest by using perturbation theory.\footnote{This includes all quantum mechanical models and the asymptotically free quantum field theories.} For a given accuracy goal, there exists a ``perturbative cut-off'' $K_{\text{p}}$ above which perturbation theory is enough to reach this accuracy goal. In practice, we thus choose $K$ sufficiently greater than $K_{\text{p}}$. For $|k|\leq K_{\text{p}}$ we use the ARG equations to improve the non-perturbative Monte-Carlo data. For $|k|>K_{\text p}$, we are satisfied with perturbation theory. 

\paragraph{Defining the space $\mathscr M$ in finite dimension}

\begin{figure}
\centerline{\includegraphics[width=3in]{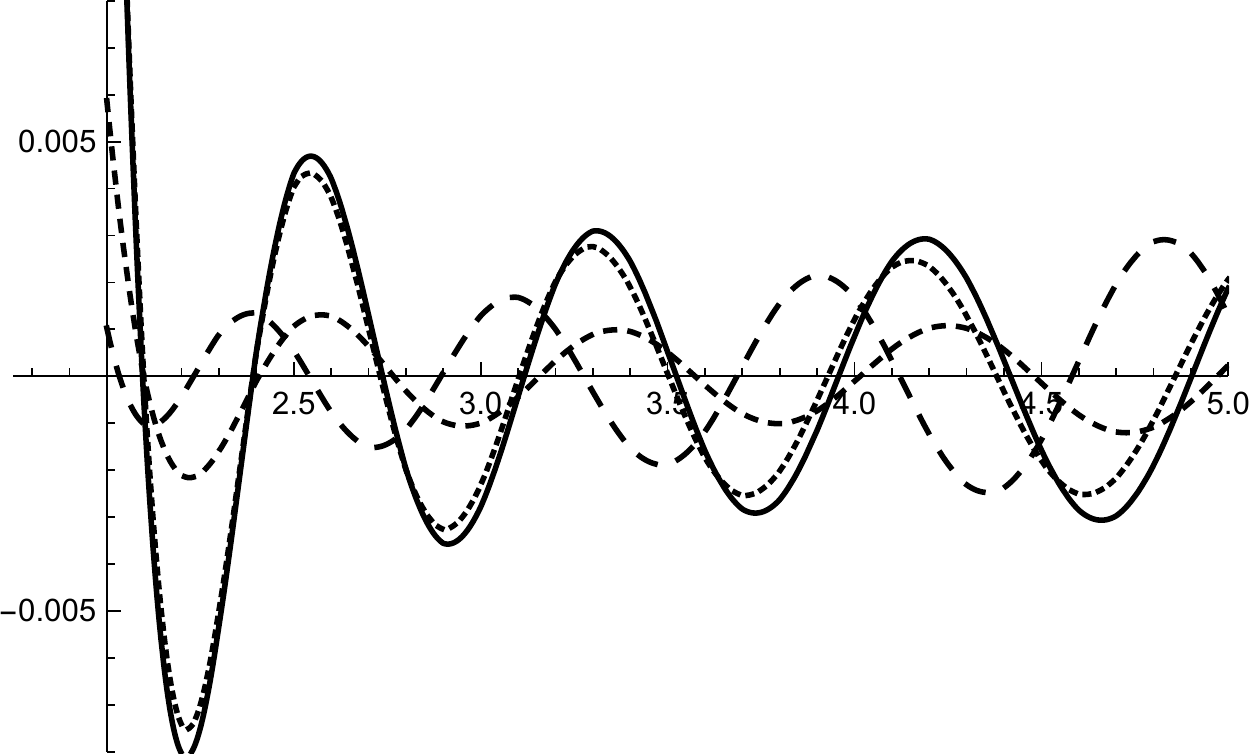}}
\caption{\label{fig14} Zoom of the plots on the left-hand side of Fig.\ \ref{fig13}. Working with a finite precision $N=200$, the ARG equations cannot be satisfied exactly.}
\end{figure}

Since we work with a finite dimensional space $\mathscr F_{K}$, it is clear that we cannot impose the infinite set of ARG equations on it. Indeed, if we used more than $K$ independent ARG equations, the only solution would be the trivial $G_{k}=0$ for all $k$. This is not surprising: working with a finite cut-off implies that we work with a finite precision $N$ and the ARG equations can be satisfied only approximately. For example, if we zoom the graph on the left-hand side of Fig.\ \ref{fig13}, we get the plots depicted on Fig.\ \ref{fig14}. This implies that there is no unique way to define a finite dimensional version $\mathscr M_{K}\subset\mathscr F_{K}$ of $\mathscr M$. This is an important feature that we have to deal with to implement in practice the general principles outlined in \ref{Approx1Sec}.

A possibility is to replace the ARG equations by linear inequalities. For example, we can replace \eqref{argg7} by
\be\label{argg8} -\varepsilon<\sum_{p=0}^{\infty}\mathsf A_{\alpha}(N;u,p)G_{\pm (\km + \delta p)} <\varepsilon\, ,\ee
for a suitable choice of $\varepsilon$, depending on the precision $N$. The advantage of this method is that we may use in principle as many values of $\km$, $\delta$, $\alpha$ and $u$ that we wish. However, the space $\mathscr M_{K}$ defined in this way is not a linear subspace of $\mathscr F_{K}$ and the resulting linear programming problem that we have to solve is rather complicated.

Instead, we are going to limit ourselves in the present paper to a much simpler approach. We define $\mathscr M_{K}$ by a finite number $\bar n$ of independent ARG equations of the form \eqref{argg7}, for certain choices of the parameters $\km$, $\delta$, $\alpha$ and $u$. There is an obvious ambiguity in these choices but we shall see that, to a large extent, this ambiguity is irrelevant. In particular, for a given $K$, it turns out that there is always a prefered order of magnitude for $\bar n$, that yields the codimension of $\mathscr M_{K}$. This codimension turns out to be largely independent of the precise set of ARG equations one chooses. Moreover, the accuracy of the improved data that we get also turns out to be largely independent of the choice of equations. The conclusion is that all reasonable choices seem to yield the construction of a subspace $\mathscr M_{K}$ which provides a good approximation to $\mathscr M$.

We are now going to illustrate very explicitly all the above-mentioned properties by implementing an explicit algorithm.

\subsection{\label{ExpalgSec} Explicit algorithm}

\paragraph{Step 1} We choose a precision $N$ and evaluate the associated cut-off $K$ as explained in Sec.\ \ref{GrSec}. Most of our explicit examples will correspond to $N=200$ and $K=75$, but we shall also use $N=1000$ and $K=150$.

\paragraph{Step 2} We build an approximate data set $(\Ga_{k})_{1\leq k\leq K}$ belonging to $\mathscr F_{K}^{+}$ from an exact data set $\Ge$ by using \eqref{GaGerel} for some $\sigma$. We will mostly use exact data sets corresponding to the formula \eqref{GkDO}, for $(m,\Gamma)=(2,0.5)$. We have studied several other values of $m$ and $\Gamma$ and they all yield very similar results; see also Sec.\ \ref{BHlastSec} for a very different example. 

\paragraph{Step 3} The perturbative, or high energy, expansion of \eqref{GkDO}, up to one loop, reads (recall that $\beta = 2\pi$)
\be\label{Gkexpert} G_{k} = \frac{1}{k^{2}} - \frac{2\Gamma}{k^{3}} + O(1/k^{4})\, .\ee
This formula is very poor for very low values of $k$, but the accuracy becomes excellent for large values of $k$. For example, for $(m,\Gamma)=(2,0.5)$, we get an accuracy better than $1\%$ for $|k|\geq 17$. Even if we use only the leading $1/k^{2}$ term in \eqref{Gkexpert}, we obtain an accuracy better than $5\%$ for $|k|\geq 25$. So, for this example, a reasonable value for the perturbative cut-off is $K_{\text{p}}\sim 25$. All the values of this order of magnitude yield similar results, see below.

We set
\be\label{Delzerodef} \Delta(0)=d_{K_{\text{p}}}^{+}(\Ge,\Ga)\, .\ee
This measures the accuracy of the non-per\-tur\-ba\-ti\-ve piece of the original approximate data set. It is this accuracy that we want to improve using the ARG equations.

\paragraph{Step 4} We now start the delicate discussion of how to get the best possible subspace $\mathscr M_{K}$. We focus on the positive frequency space $\mathscr M_{K}^{+}$ without loss of generality. This discussion will be continued in Step 6.

We define the subspaces $\mathscr M_{K,\bar n}^{+}$ by a set of equations
\be\label{Mkddef} \sum_{p=0}^{[\frac{K-\km}{\delta}]}\mathsf A_{\alpha}(N;u,p)G_{\km+\delta p} = 0\ee
for $n$ different values of $\{\km,\delta,\alpha,u\}$.\footnote{We could also use other ARG equations, like \eqref{argg3}, but this would not change our discussion or our results in any significant way.} We observe numerically that the linear equations \eqref{Mkddef} are not all independent in general, as was already suggested in Section \ref{MFSec}. We denote by $\bar n(n)\leq \min(n,K)$ the number of independent equations \eqref{Mkddef}, such that $\dim\mathscr M_{K,\bar n}^{+}=K-\bar n$.

Of course, the subspace $\mathscr M_{K,\bar n}^{+}$ depends on the precise values for $\{\km,\delta,\alpha,u\}$ that we use, and the choice of these values is a priori quite arbitrary. We have tested many possibilities. To be specific, we proceed as follows. We choose three lists, $l_{\delta}$, $l_{\alpha}$ and $l_{u}$, of possible values for $\delta$, $\alpha$ and $u$ that we want to use. Then, for a given value of $n$, we include all the possible $\{\km,\delta,\alpha,u\}$, in lexicographic order, starting from $\km=1$ and increasing. For example, if we pick $l_{\delta}=\{1,2\}$, $l_{\alpha}=\{3/4\}$, $l_{u}=\{3,5\}$ and $n=7$, we use
\begin{multline}
\nonumber\{\km,\delta,\alpha,u\}=\{1,1,3/4,3\}, \{1,1,3/4,5\}, \{2,1,3/4,3\}, \\\{2,1,3/4,5\}, \{2,2,3/4,3\}, \{2,2,3/4,5\},\{3,1,3/4,3\}\, .
\end{multline}
Note that $\{1,2,3/4,3\}$ and $\{1,2,3/4,5\}$ are not included because they do not satisfy the constraint $0<\alpha<\km/\delta$. We shall see that the various possible choices yield very similar results, but it seems to be always better to sample at least a few values of $\alpha$ and $u$.

\paragraph{Step 5} We construct the orthogonal projections $\tilde{G}^{\text a}(\bar n)$ of $\Ga$ onto the spaces $\mathscr M_{K,\bar n}^{+}$, associated with the distance function \eqref{ddef2}. We set 
\be\label{Deltadef1}\Delta(\bar n)=d_{K_{\text{p}}}^{+}\bigl(\Ge,\tilde{G}^{\text a}(\bar n)\bigr)\, .\ee

If the algorithm works, we expect that the function $\Delta(\bar n)$, which measures the accuracy of the improved data set $\tilde{G}^{\text a}(\bar n)$, will be a decreasing function of $\bar n$, up to some optimal value of $\bar n_{0}$ which, of course, must be less than $K$. Indeed, when $\bar n=K$, $\tilde{G}^{\text a}(\bar n)=0$ and $\Delta(K)\simeq 1$.

The question is, then, how to find this optimal value of $\bar n_{0}$ in general?

In the articifial situation where one actually knows the exact data set $\Ge$, the optimal value of $\bar n_{0}$ is, obviously, the one that minimizes $\Delta(\bar n)$. The \emph{accuracy gain} is then defined to be
\be\label{acg1} \bar w = \frac{\Delta(0)}{\Delta(\bar n_{0})}\, \cdotp\ee
An interesting observation is that, for sufficiently large $N$ and $K$, the optimal value $\bar n_{0}$ turns out to be always more or less the same, independently of the precise choice of the $\{\km,\delta,\alpha,u\}$ that defines $\mathscr M_{K,\bar n}$.

However, in a real-life calculation, the exact data set is unknown. One then needs a criterion to obtain an estimate $\tilde n_{0}$ of the optimal value $\bar n_{0}$. It is quite important for this estimate to be reliable: if the guess is over-evaluated, we are likely to get a totally non-sensical result like $\tilde{G}^{\text a}(\tilde n_{0})\simeq 0$; if it is under-evaluated, then we will get a data improvement significantly inferior to the maximal value the method can produce in principle. 

\paragraph{Step 6} We introduce the partial norm of the improved data, defined to be
\be\label{mudef}\mu(\bar n) = d_{K_{\text{p}}}^{+}\bigl(0,\tilde{\Ga}(\bar n)\bigr)\, .\ee
By construction, if $K_{\text{p}}$ is not too small, $\mu(\bar n)$ will be an almost always decreasing function of $\bar n$, with $\mu(0)=1$ and $\mu(K)=0$.\footnote{We may use the total norm $\nu(\bar n) = d_{K}^{+}\bigl(0,\tilde{\Ga}(\bar n)\bigr)$ which, by construction, is a strictly decreasing function of $\bar n$. Using $\nu$ instead of $\mu$ yields similar results.} 

The detailed behaviour of $\mu$ as a function of $\bar n$ is generically as follows (see the examples below). If $\bar n$ is below the optimal value $\bar n_{0}$, $\smash{\mathscr M_{K,\bar n}^{+}}$ yields a better and better approximation to the positive frequency space $\smash{\mathscr M^{+}}$ when $\bar n$ is increased. In this regime, the data point $\tilde{\Ga}(\bar n)$ will be slightly modified at each step $\bar n\mapsto\bar n+1$ and the function $\mu$ decreases mildly at each increment of $\bar n$. To the contrary, when $\bar n$ is above $\bar n_{0}$, the approximation of $\smash{\mathscr M^{+}}$ by $\smash{\mathscr M_{K,\bar n}^{+}}$ becomes inconsistent. The data point $\tilde{\Ga}(\bar n)$ then departs significantly from the correct value and tends to zero. We thus expect a rather sudden and sharp decrease of $\mu(\bar n)$ when $\bar n$ exceeds $\bar n_{0}$.

This sudden sharp decrease allows to ``detect'' $\bar n_{0}$. A very simple procedure is to set $\tilde n_{0}=[\frac{3}{4}\bar n_{1/2}]$, where $\bar n_{1/2}$ is the smallest value of $\bar n$ for which $\mu(\bar n) <1/2$ (the brackets denote the integer part). The \emph{effective accuracy gain} of the algorithm is then defined by
\be\label{acgeff} \tilde w = \frac{\Delta(0)}{\Delta(\tilde n_{0})}\, \cdotp\ee
The use of the factor $\frac{3}{4}$ in the definition of $\tilde n_{0}$ is of course a matter of choice, but it seems to be very reasonable. On the one hand, a greater value could jeopardize the whole scheme, by potentially producing, at least in some cases, an estimate beyond the value for which the approximation of $\mathscr M^{+}$ by $\mathscr M_{K,\bar n}^{+}$ makes sense.\footnote{Of course, using the factor of $\frac{3}{4}$ does not preclude this problem from happening on special cases. One could use a factor $\frac{1}{2}$ to be on the completely safe side.} On the other hand, $\frac{3}{4}$ is large enough to ensure that we are always not too far below the genuine optinal value $\bar n_{0}$ and thus that the effective accuracy gain $\tilde w$ is not much lower than its maximal possible value $\bar w$.

A finer procedure consists in estimating $\bar n_{0}$ by looking in more details at the shape of the curve representing $\mu(\bar n)$. This yields in general the best results, but, for our pruposes, the crude recipe proposed in the previous paragraph works well enough.

\subsection{Results}

\subsubsection{The algorithm on a specific case}

\begin{figure}
\centerline{\includegraphics[width=4in]{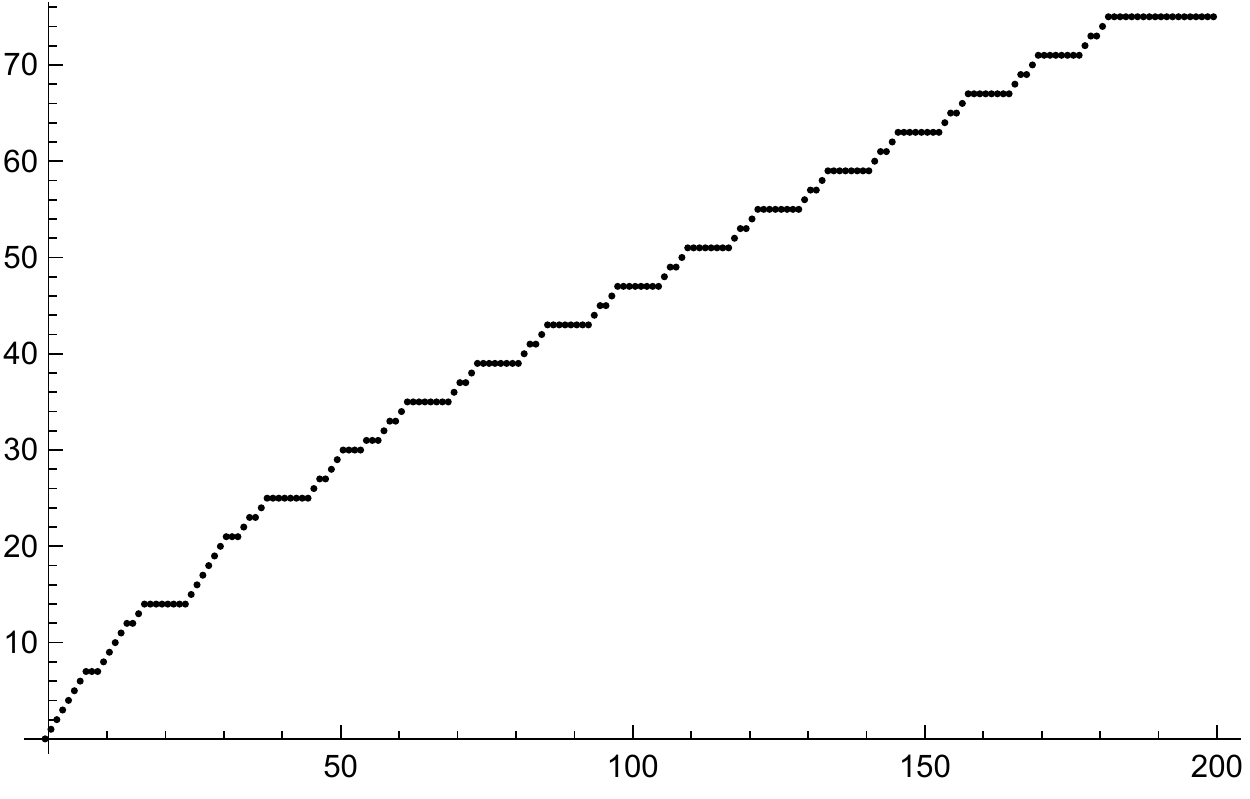}}
\caption{\label{fig15} The codimension $\bar n$ of the space $\mathscr M_{75,\bar n}^{+}$ as a function of the total number $n$ of ARG equations that we use to define it, for the choices $N=200$, $l_{\delta}=\{1,2\}$, $l_{\alpha}=\{1/4,1/2,3/4,1\}$, $l_{u}=\{3,5,7\}$.}
\end{figure}
\begin{figure}
\centerline{\includegraphics[width=4in]{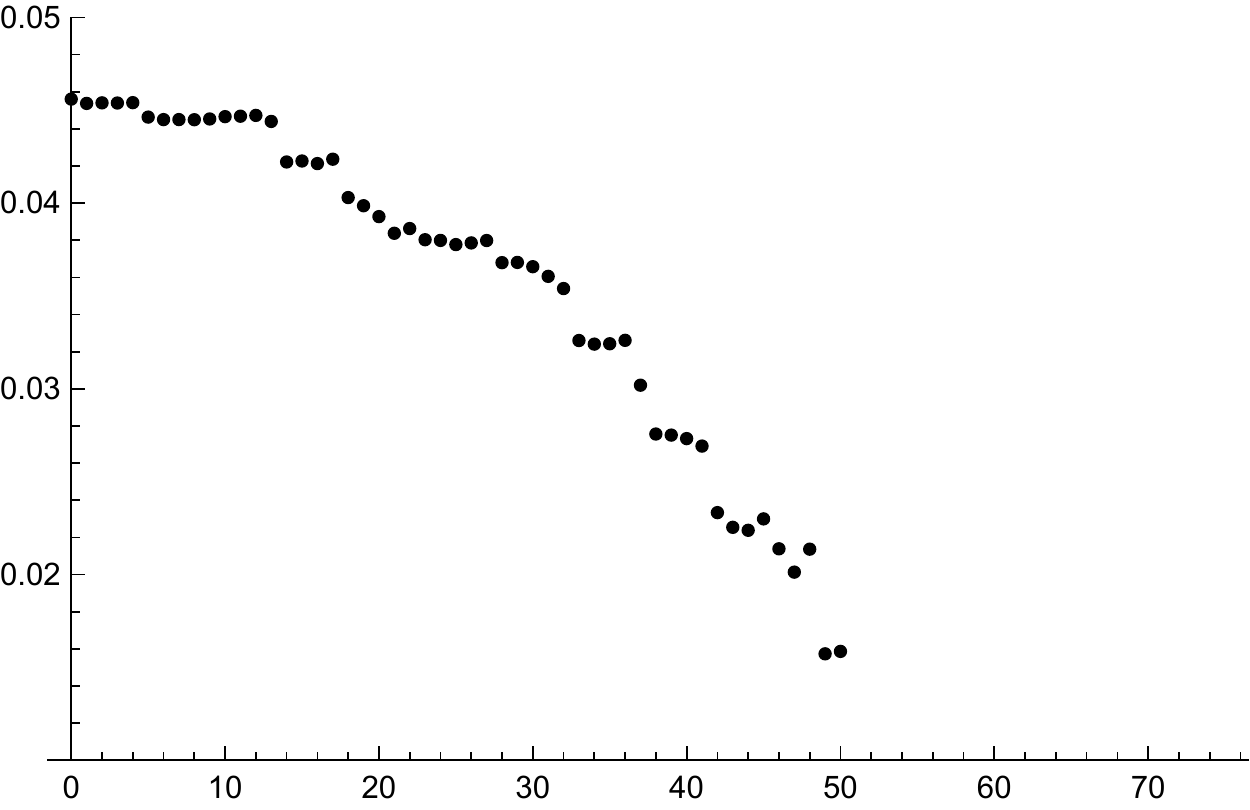}}
\caption{\label{fig16} The accuracy $\Delta(\bar n)$ of the improved data set produced by our algorithm, as a function of $\bar n$, starting from an approximate data set obtained from \eqref{GkDO} and \eqref{GaGerel} with $\sigma = 0.05$. We use the values $N=200$, $K=75$, $K_{\text{p}}=25$, $l_{\delta}=\{1,2\}$, $l_{\alpha}=\{1/4,1/2,3/4,1\}$, $l_{u}=\{3,5,7\}$. The algorithm yields an accuracy gain of about 2.9 in this case.}
\end{figure}
\begin{figure}
\centerline{\includegraphics[width=4in]{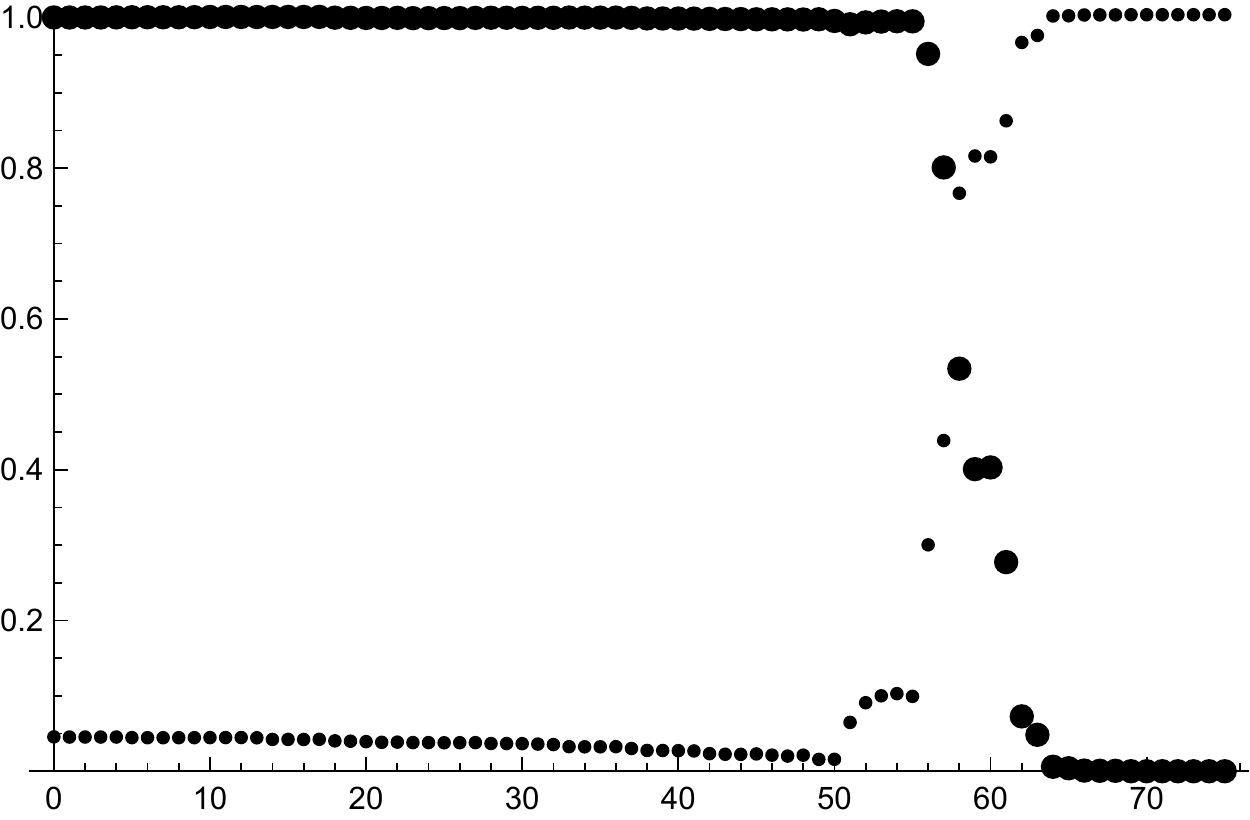}}
\caption{\label{fig17} Plot of the norm $\mu$ (thick dots) and of the accuracy $\Delta$ (thin dots) as a function of $\bar n$. We use the same data as in Fig.\ \ref{fig16}. The breakdown of the algorithm can be reliable detected by using the sudden sharp decrease of the norm.}
\end{figure}
\begin{figure}
\centerline{\includegraphics[width=2.9in]{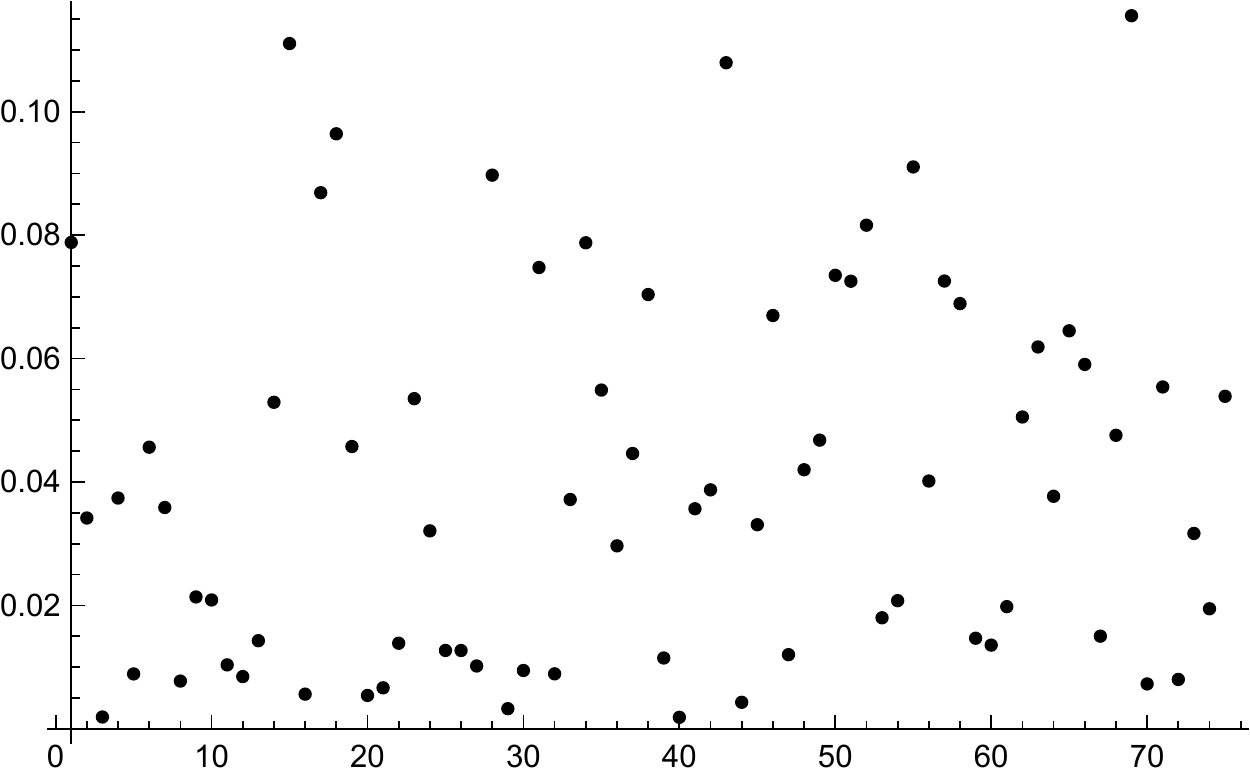}\hskip .2in\includegraphics[width=2.9in]{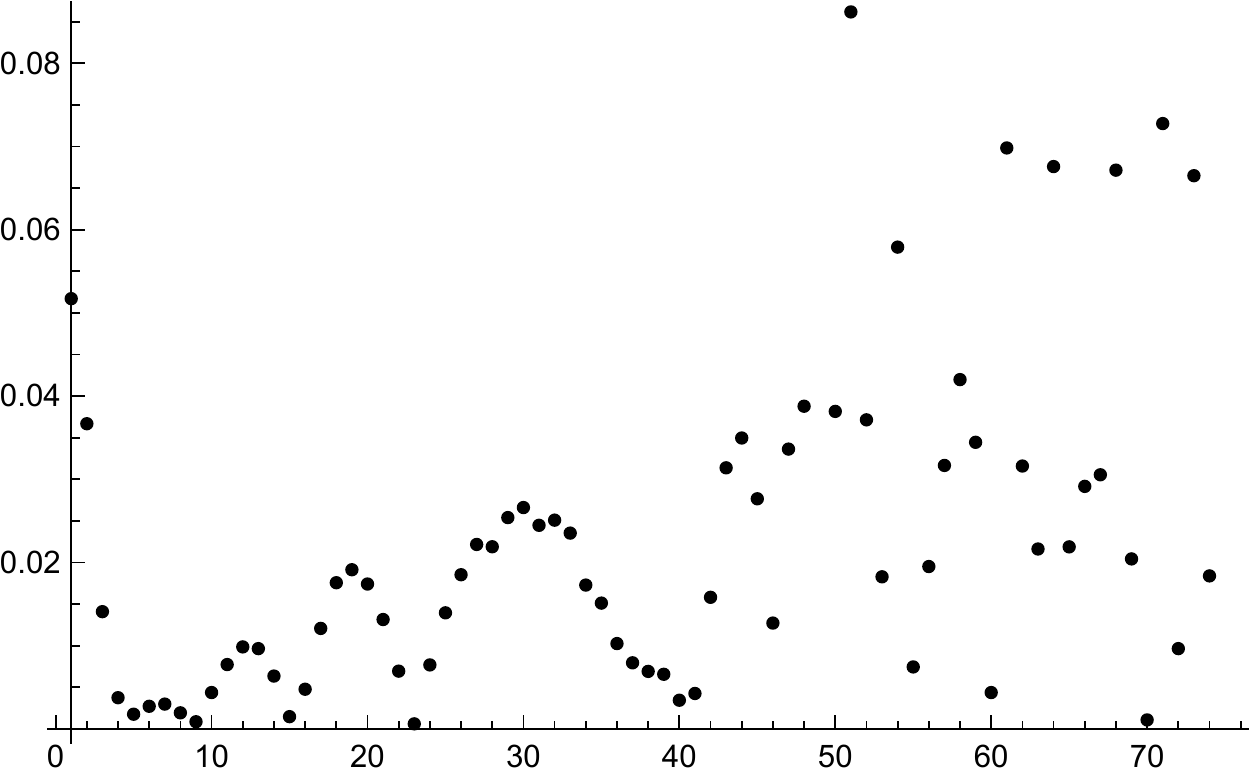}}
\caption{\label{fig18} Accuracy of the original data set (left inset) and of the best improved data set, corresponding to the point $\bar n=\bar n_{0}=49$ in Fig.\ \ref{fig16} (right inset), Fourier coefficients by Fourier coefficients. The accuracy of a given coefficient $G_{k}$ is defined to be $|(G_{k}-\Ge_{k})/\Ga_{k}|$, consistently with the definition \eqref{ddef2} of the distance.}
\end{figure}

Let us first illustrate all the basic properties of the algorithm on a specific typical example. We pick $N=200$, $K=75$, $(m,\Gamma)=(2,0.5)$ and build an approximate data set $\Ga$ from \eqref{GaGerel} with $\sigma = 0.05$. We also choose $K_{\text{p}}=25$. For the particular realization of $\Ga$ that we use, we find that $\Delta(0)\simeq 0.0456$. This means that our approximate data set has an accuracy of about $4.56\%$ for the first 25 Fourier coefficients. The goal is to improve these Fourier coefficients to get a better accuracy.

We define the spaces $\mathscr M_{K,\bar n}^{+}$ by using the lists of values $$l_{\delta}=\{1,2\}\, ,\ l_{\alpha}=\{1/4,1/2,3/4,1\} \, ,\ l_{u}=\{3,5,7\}\, ,$$ as explained below Eq.\ \eqref{Mkddef}. On Fig.\ \ref{fig15}, we plot the number $\bar n$ of independent ARG equations obtained in this way, as a function of the total number $n$ of equations that we use. We clearly see that all the equations are not independent. One needs $n=182$ equations to span the whole 75-dimensional space $\mathscr F_{75}^{+}$.

The accuracy function $\Delta$ is plotted on Fig.\ \ref{fig16}, as a function of $\bar n$. As expected, we observe that $\Delta(\bar n)$ decreases, down to the minimal value $\Delta(49)\simeq 0.0157$ obtained for $\bar n = \bar n_{0}=49$. The algorithm is thus able to produce an improved data set of accuracy $\sim 1.57\%$, starting from a sample of accuracy $\sim 4.56\%$. The accuracy gain is $\bar w\simeq 2.9$ in this case, which is quite good. When $\bar n>50$, the algorithm brutally breaks down. The corresponding data points $\tilde{\Ga}(\bar n)$ do not appear on the plot because they are off scale.

On Fig.\ \ref{fig17}, we have plotted both the norm $\mu(\bar n)$ and $\Delta(\bar n)$. The behaviour of $\mu$ is as described in Section \ref{ExpalgSec}. For $\bar n\leq\bar n_{0}$, it is a mildly decreasing function of $\bar n$. Then it sharply decreases, which means that the algorithm no longer provides a good approximation to $\mathscr M^{+}$. We can thus estimate the optimal value of $\bar n_{0}$ using the idea explained in the Step 6 of Section \ref{ExpalgSec}. We get in this way $\tilde n_{0}=44$, $\Delta(\tilde n_{0}) \simeq 0.0224$ and thus an effective accuracy gain of $\tilde w \sim 2$, a very decent result.

On Fig.\ \ref{fig18} is displayed the accuracy of the original data (left inset) versus the accuracy of the best improved data obtained at $\bar n=\bar n_{0}$ (right inset), Fourier coefficients by Fourier coefficients. This shows in great details how the algorithm acts on the data set to improve it. We see that our previous choice of $K_{\text{p}}=25$ was rather conservative, since the algorithm works pretty well up to $k\sim 40$.

\subsubsection{Varying parameters}

\begin{figure}
\centerline{\includegraphics[width=2.9in]{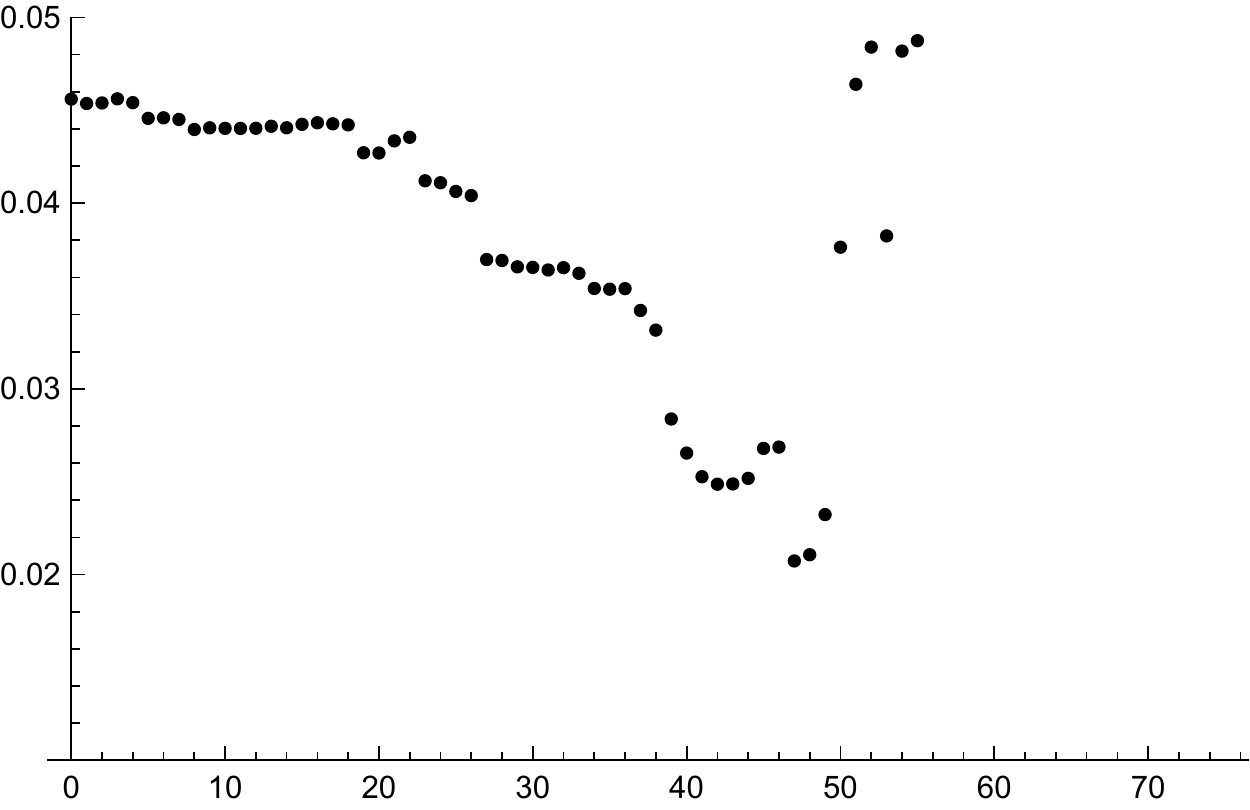}\hskip .2in\includegraphics[width=2.9in]{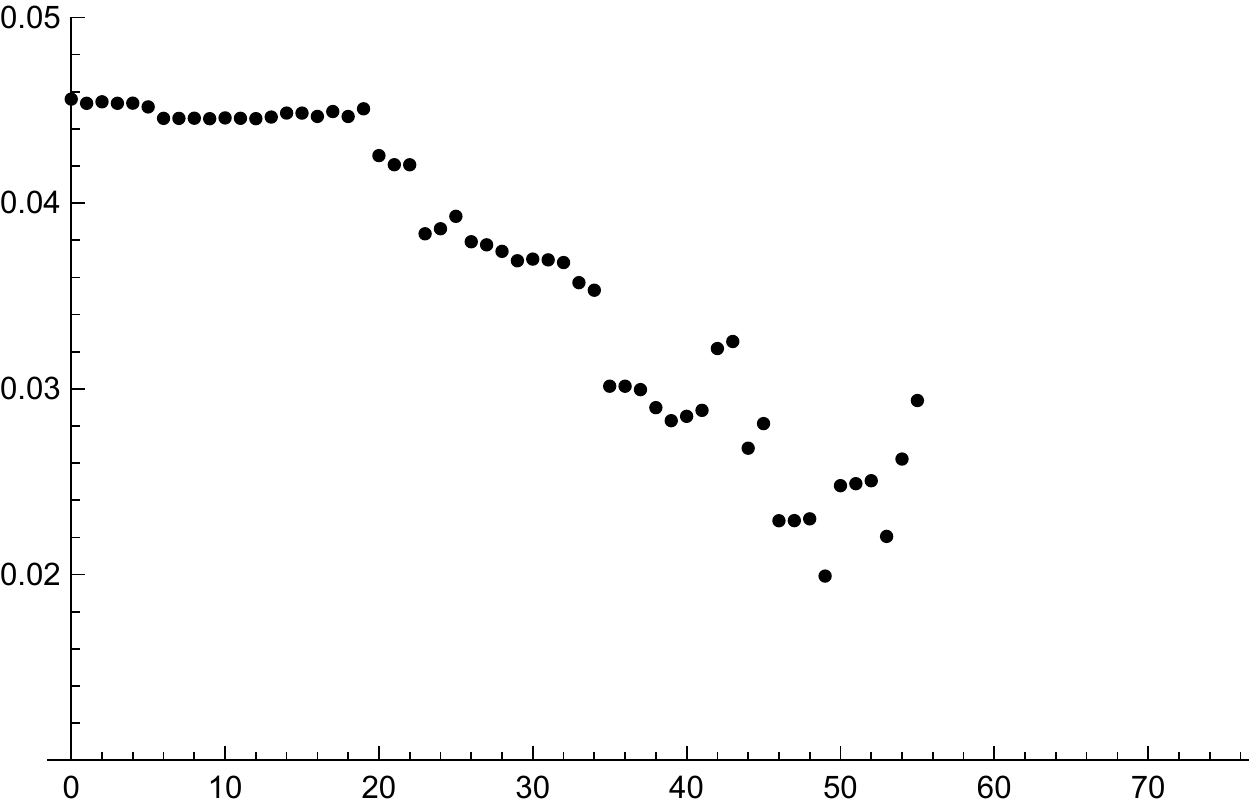}}

\vskip .5cm

\centerline{\includegraphics[width=2.9in]{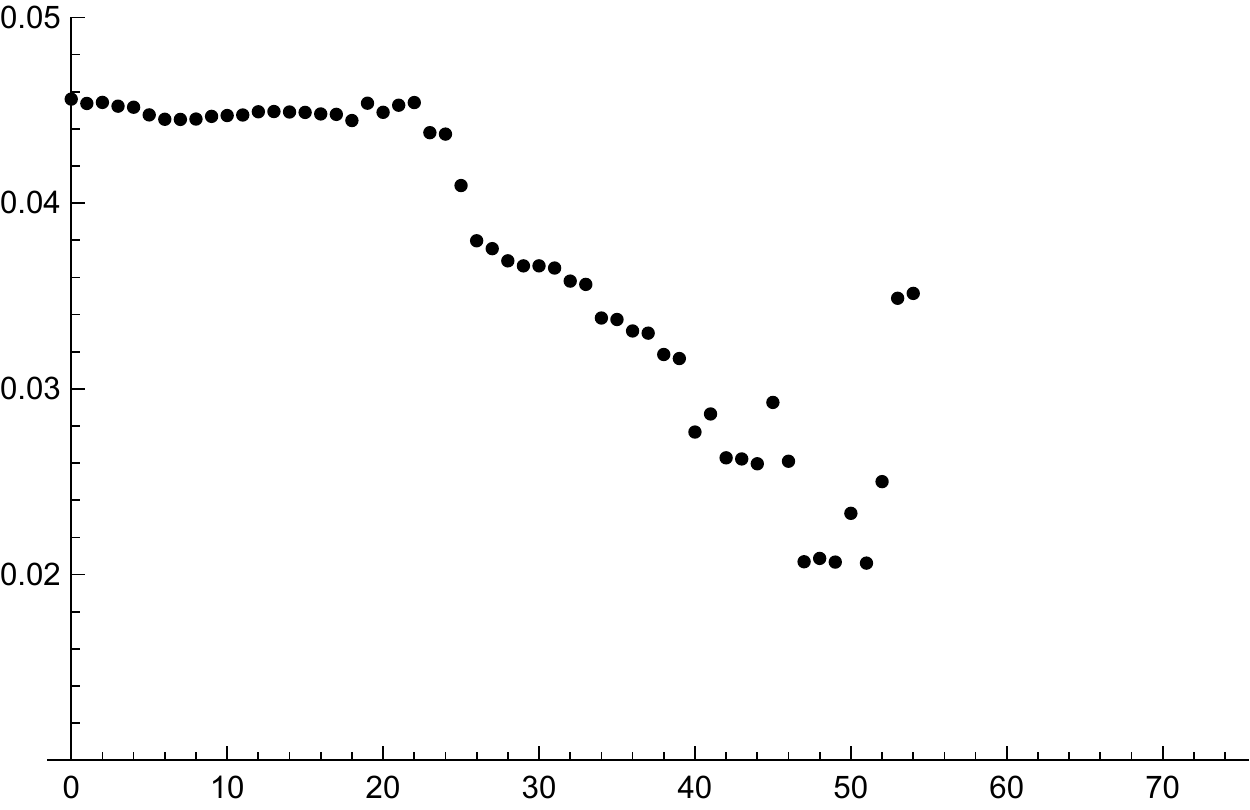}\hskip .2in\includegraphics[width=2.9in]{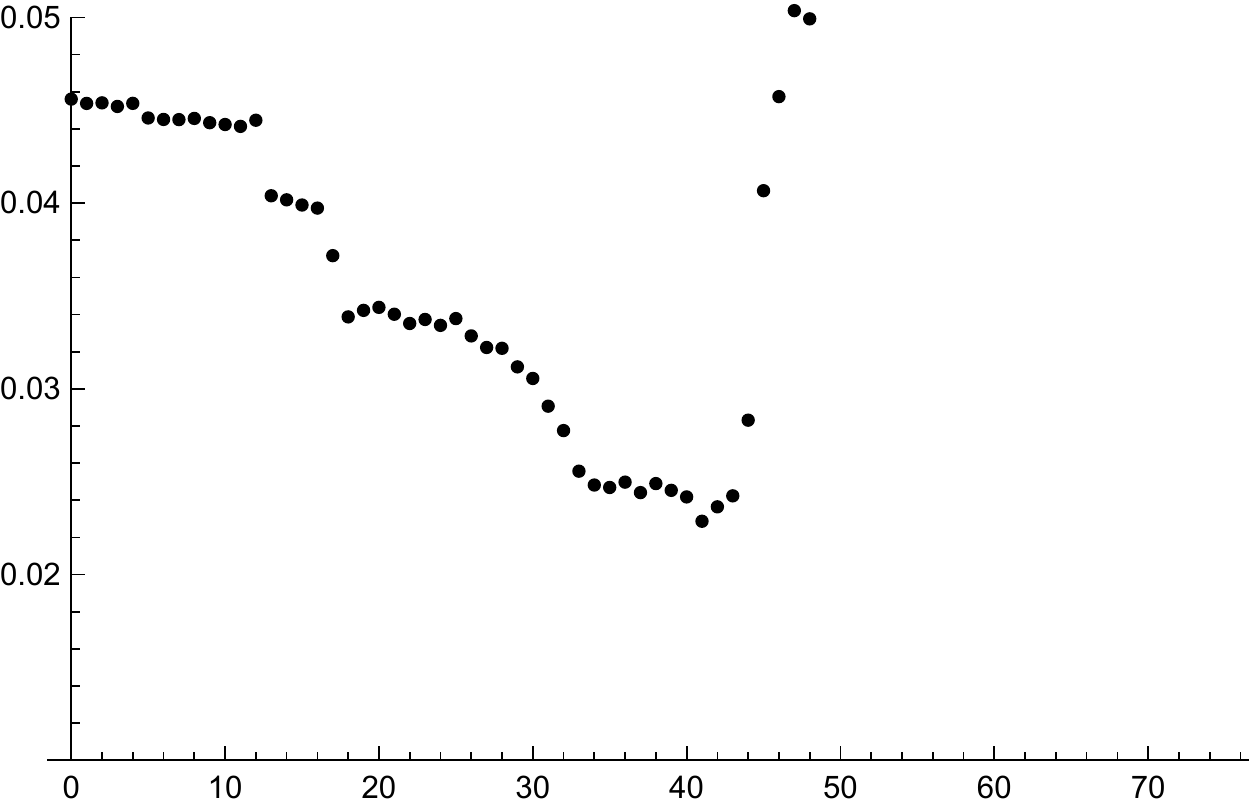}}

\caption{\label{fig19} The accuracy $\Delta(\bar n)$ as a function of $\bar n$, starting from the same approximate data set as in Fig.\ \ref{fig16} and \ref{fig17} and with $N=200$, $K=75$, $K_{\text{p}}=25$. Upper-left inset: we choose $l_{\delta}=\{1,2\}$, $l_{\alpha}={1/2}$ and $l_{u}=\{3,5,7\}$. This yields $\bar n_{0}=47$, $\Delta(\bar n_{0})\simeq 0.0207$ and $\bar w \simeq 2.20$. Upper-right inset: we choose $l_{\delta}=\{1,2\}$, $l_{\alpha}=\{1/2,1,3\}$ and $l_{u}=\{5\}$. This yields $\bar n_{0}=49$, $\Delta(\bar n_{0})\simeq 0.0199$ and $\bar w \simeq 2.29$. Lower-left inset: we choose $l_{\delta}=\{1,2\}$, $l_{\alpha}=\{1,2,3\}$ and $l_{u}=\{4,6,8\}$. This yields $\bar n_{0}=51$, $\Delta(\bar n_{0})\simeq 0.0206$ and $\bar w \simeq 2.21$. Lower-right inset: we choose $l_{\delta}=\{1,2\}$, $l_{\alpha}={1/4,1/2,3/4,1,5/4,3/2}$ and $l_{u}=\{3,7/2,4,9/2,5,11/2,6,13/2,7\}$. This yields $\bar n_{0}=41$, $\Delta(\bar n_{0})\simeq 0.0229$ and $\bar w \simeq 1.99$.}
\end{figure}

\begin{figure}
\centerline{\includegraphics[width=2.9in]{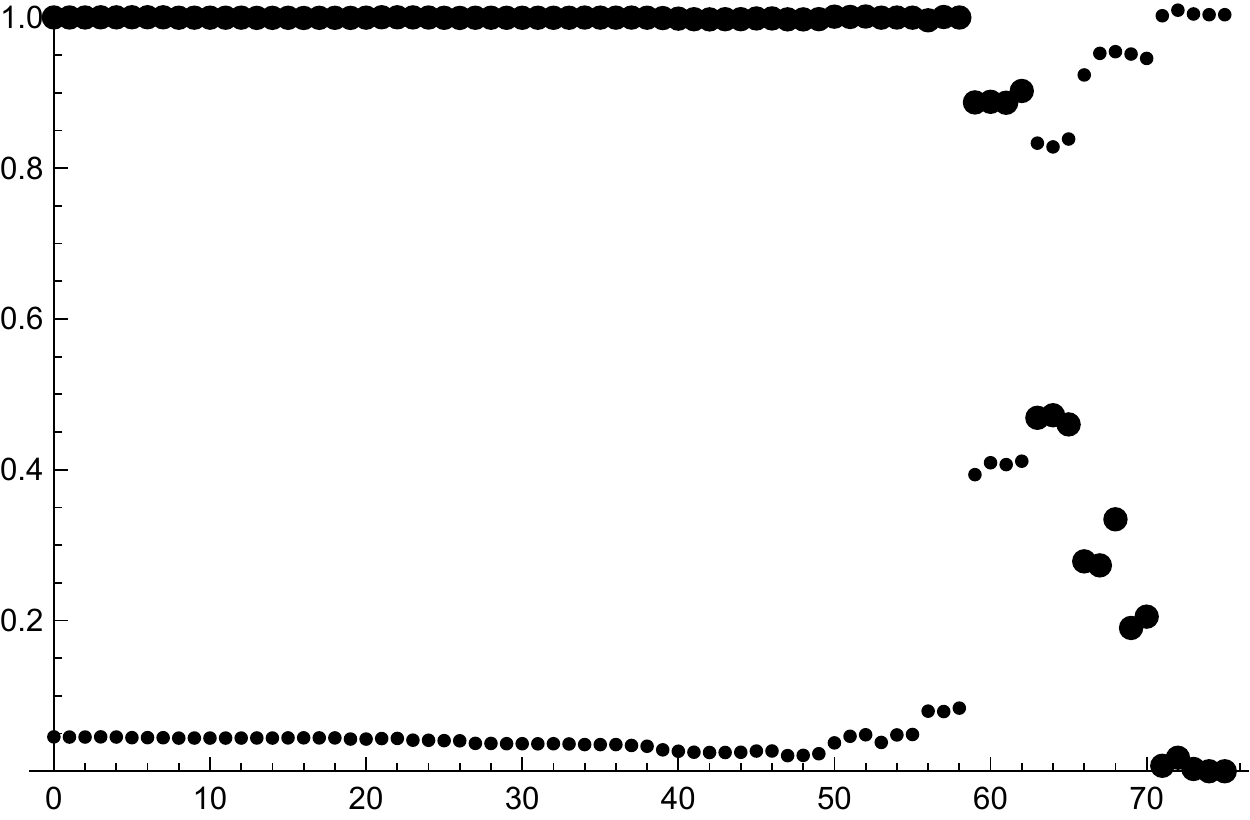}\hskip .2in\includegraphics[width=2.9in]{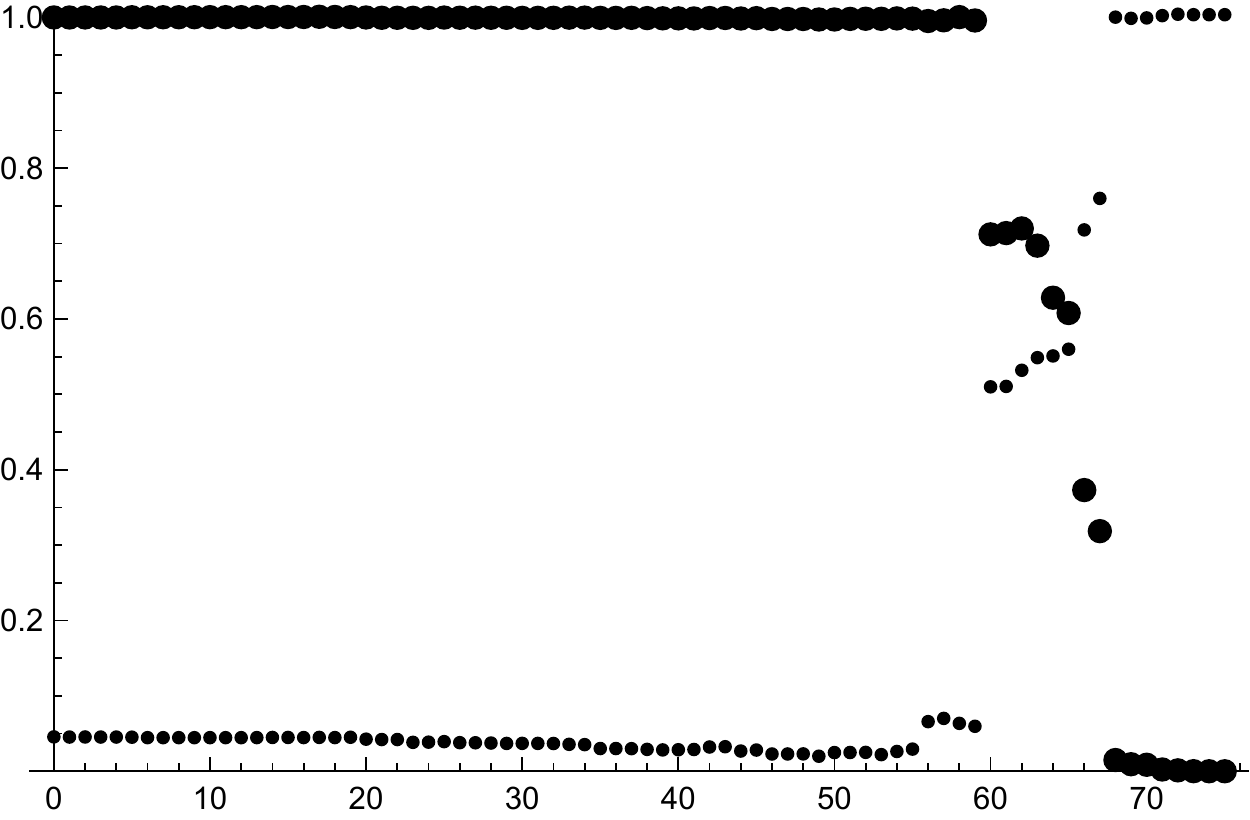}}

\vskip .5cm

\centerline{\includegraphics[width=2.9in]{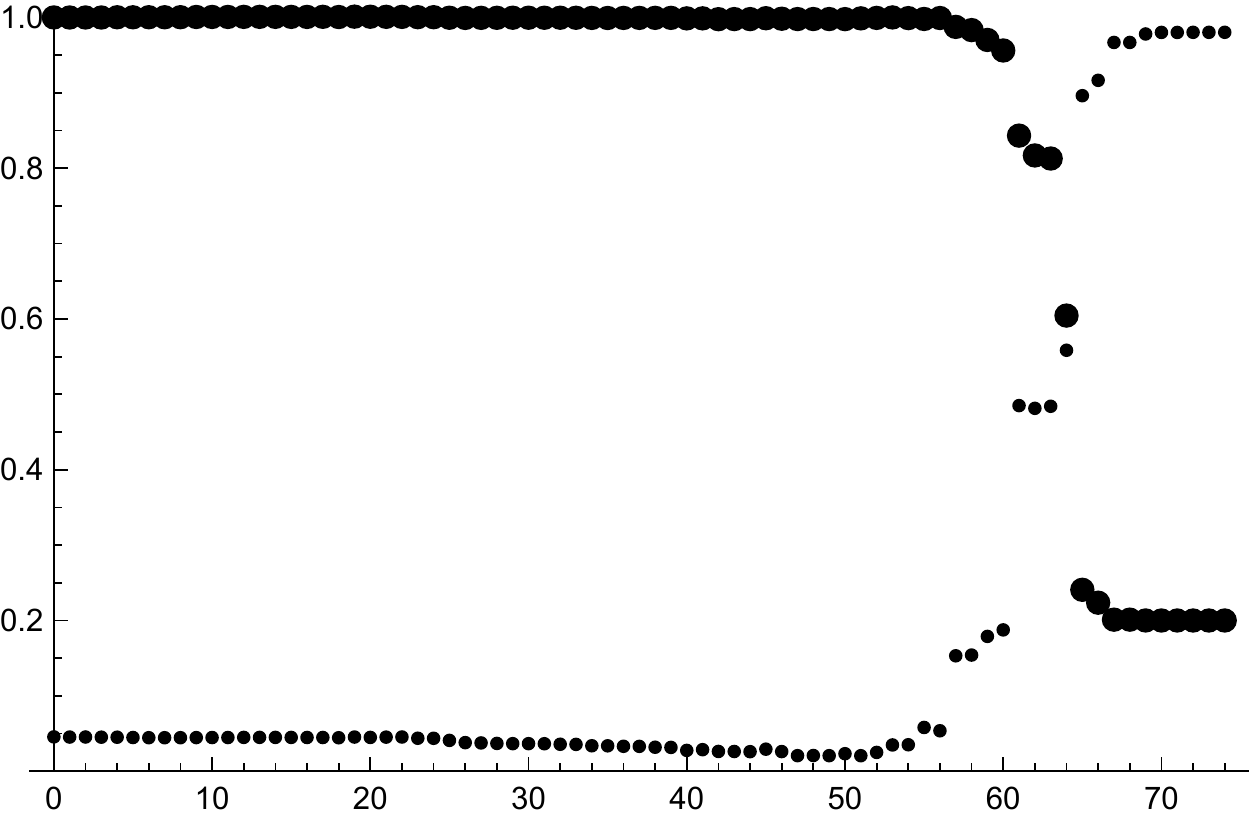}\hskip .2in\includegraphics[width=2.9in]{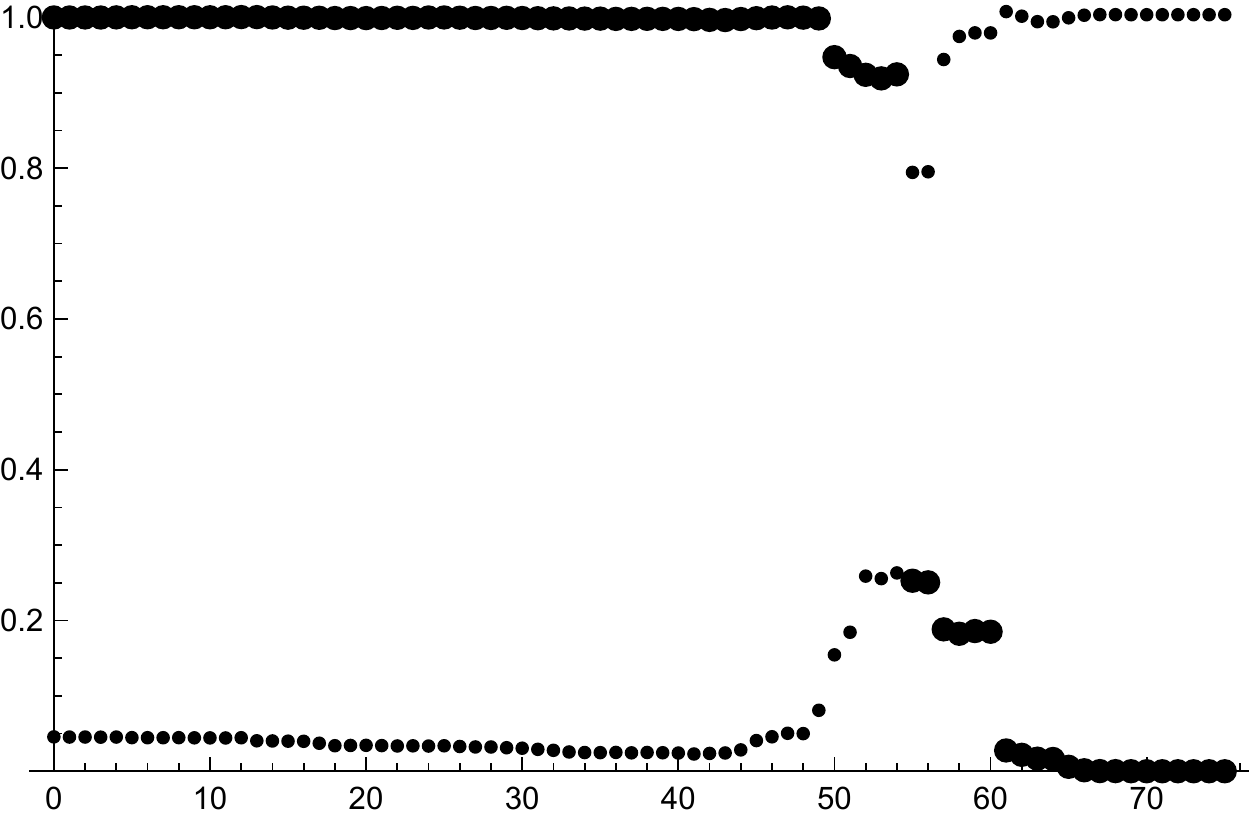}}

\caption{\label{fig20} The norm $\mu$ (thick dots) and the accuracy $\Delta$ (thin dots) as a function of $\bar n$, with the same data sets as in Fig.\ \ref{fig19}. In all cases we get a sudden sharp decrease of the norm, which yields rather good estimates $\tilde n_{0}$ and effective accuracy gain $\tilde w$. Upper-left inset: we get $\tilde n_{0}=47=\bar n_{0}$ and thus $\tilde w =\bar w\simeq 2.20$. Upper-right inset: we get $\tilde n_{0}=49=\bar n_{0}$ and thus $\tilde w =\bar w\simeq 2.29$. Lower-left inset: we get $\tilde n_{0}=48<51=\bar n_{0}$, $\Delta (\tilde n_{0}) \simeq 0.0209$ and thus $\tilde w = 2.19$ smaller but very near $\bar w$. Lower-right inset: we get $\tilde n_{0}=41=\bar n_{0}$ and thus $\tilde w=\bar w = 1.99$ in this case.}
\end{figure}

\begin{figure}
\centerline{\includegraphics[width=2.9in]{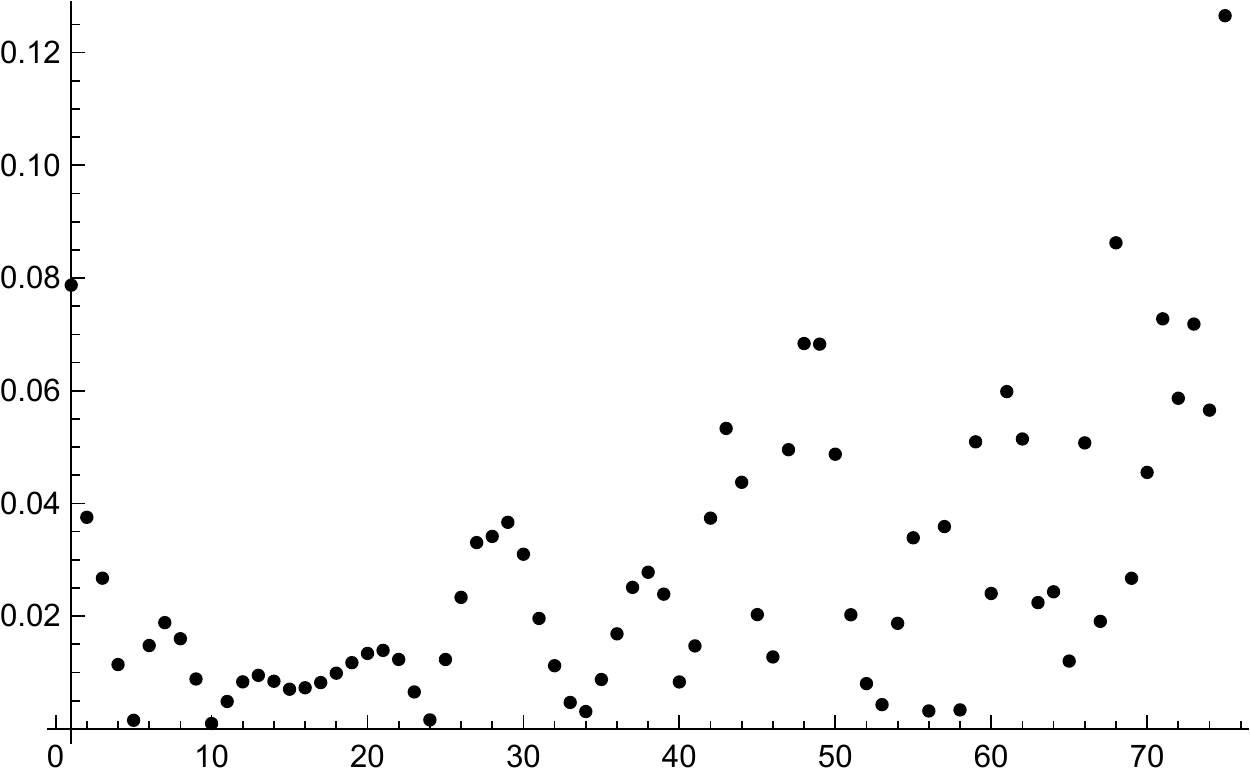}\hskip .2in\includegraphics[width=2.9in]{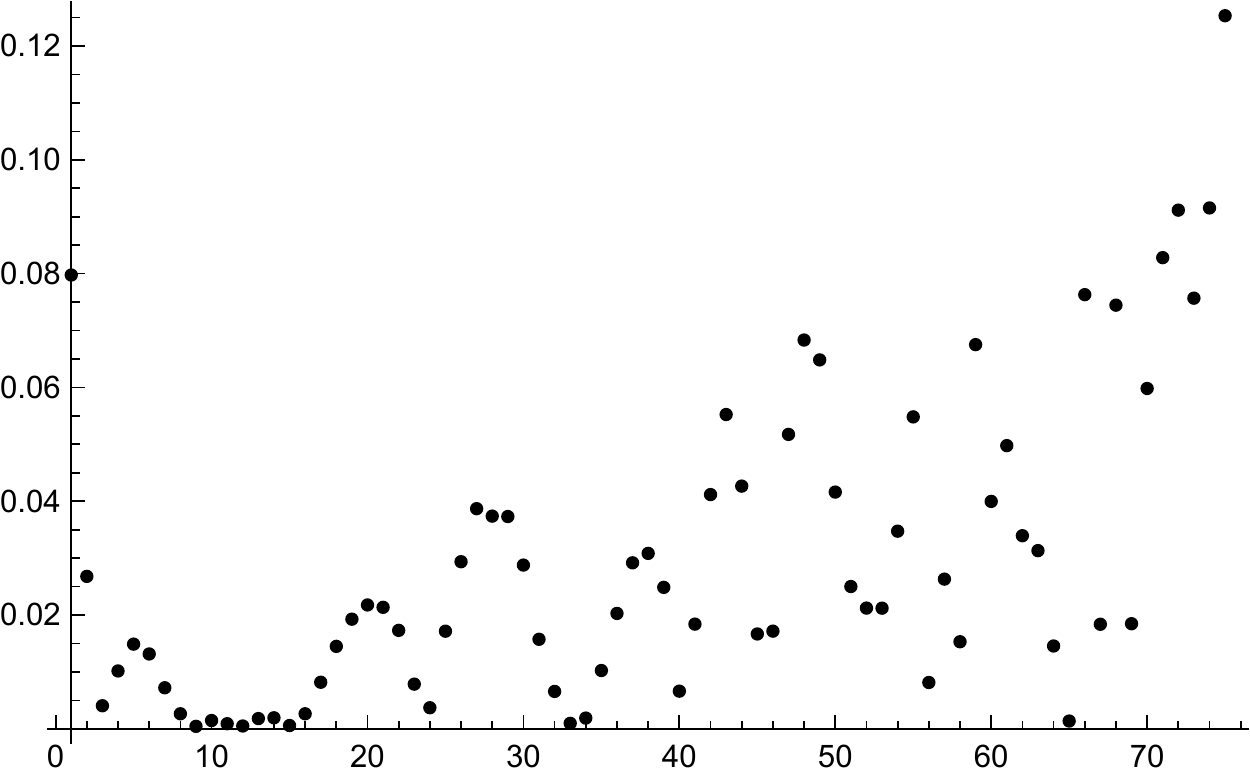}}

\vskip .5cm

\centerline{\includegraphics[width=2.9in]{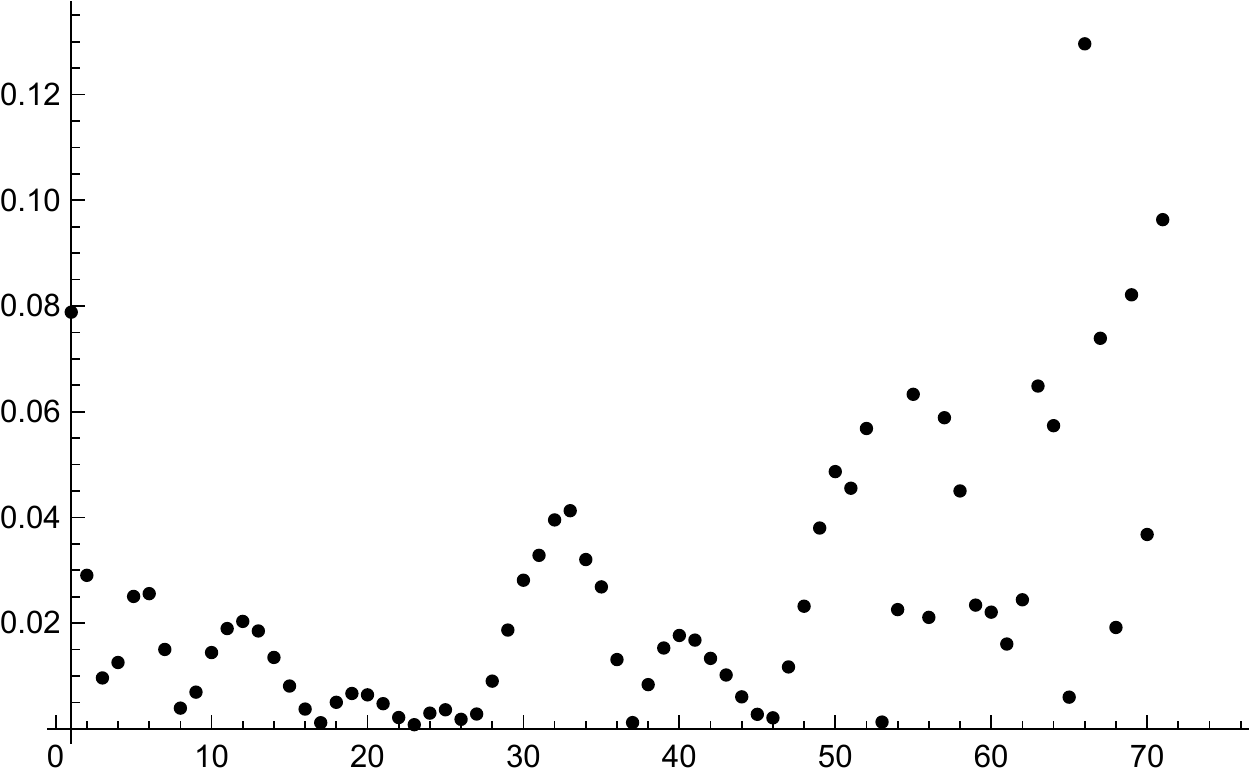}\hskip .2in\includegraphics[width=2.9in]{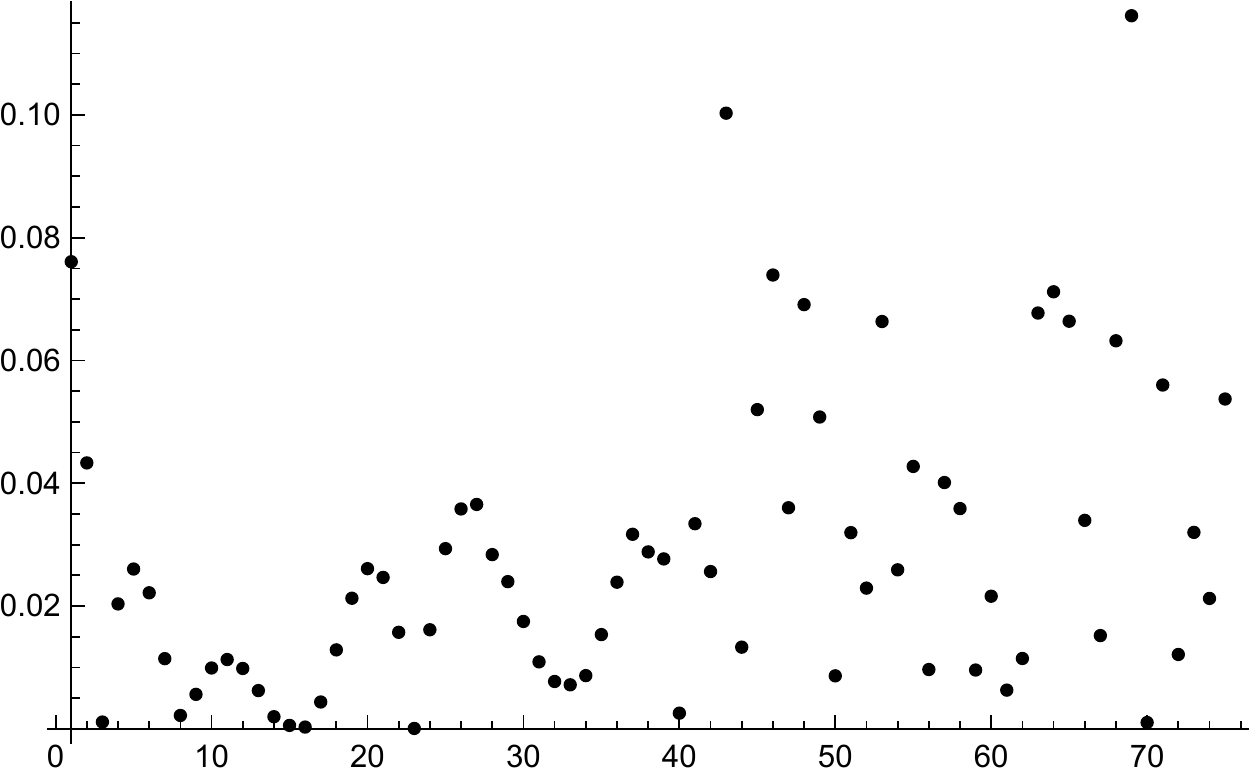}}

\caption{\label{fig21} Detailed accuracy, Fourier coefficients by Fourier coefficients, of the best improved data sets obtained from Fig.\ \ref{fig19}. The plot for the original (not improved) data set is depicted on the left inset in Fig.\ \ref{fig18}.}
\end{figure}

We now keep using the very same approximate data set $\Ga$ as in the previous subsection, but we run the algorithm with different parameters. Four typical results for $\Delta(\bar n)$, $\mu(\bar n)$ and the detailed accuracy of the best improved data sets are depicted on Fig.\ \ref{fig19}, \ref{fig20} and \ref{fig21}. As announced, all the plots look qualitatively the same and the accuracy gain produced by the algorithm is very similar in all cases. 


\subsubsection{Other data sets}

\begin{figure}
\centerline{\includegraphics[width=2.9in]{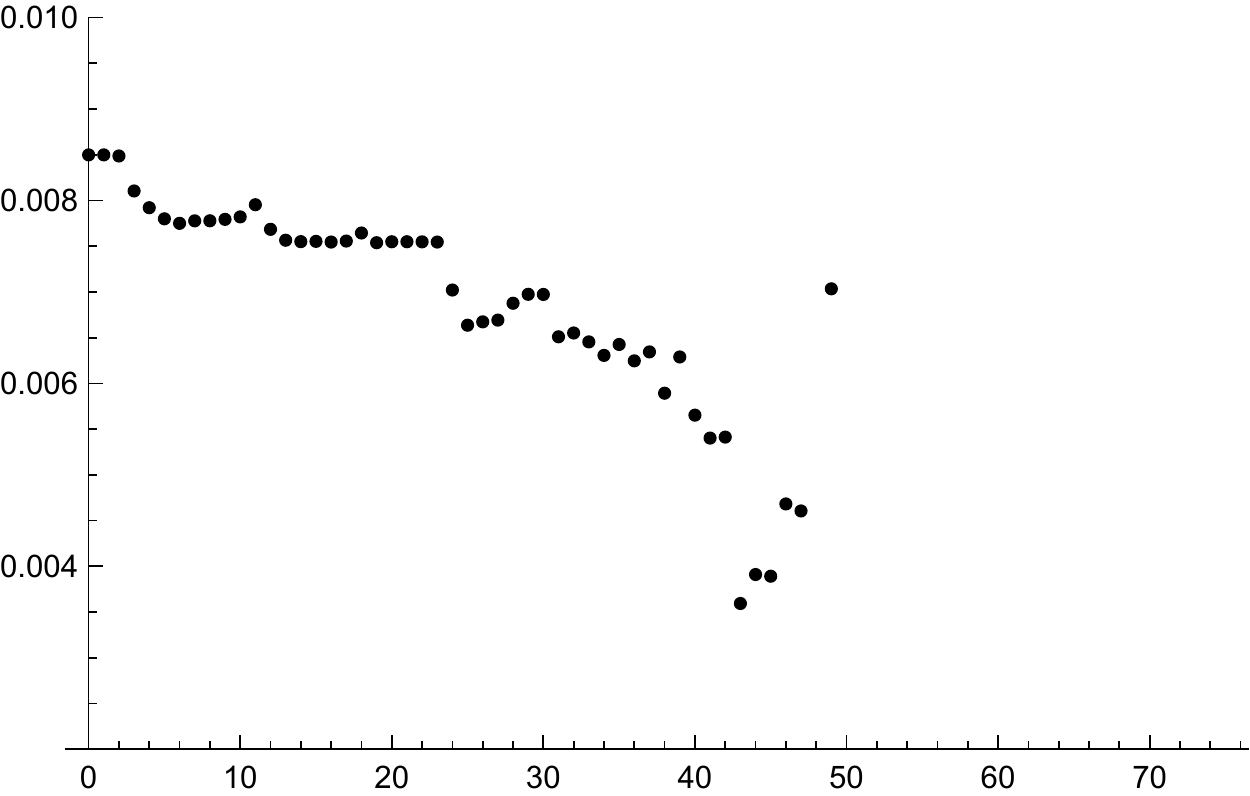}\hskip .2in\includegraphics[width=2.9in]{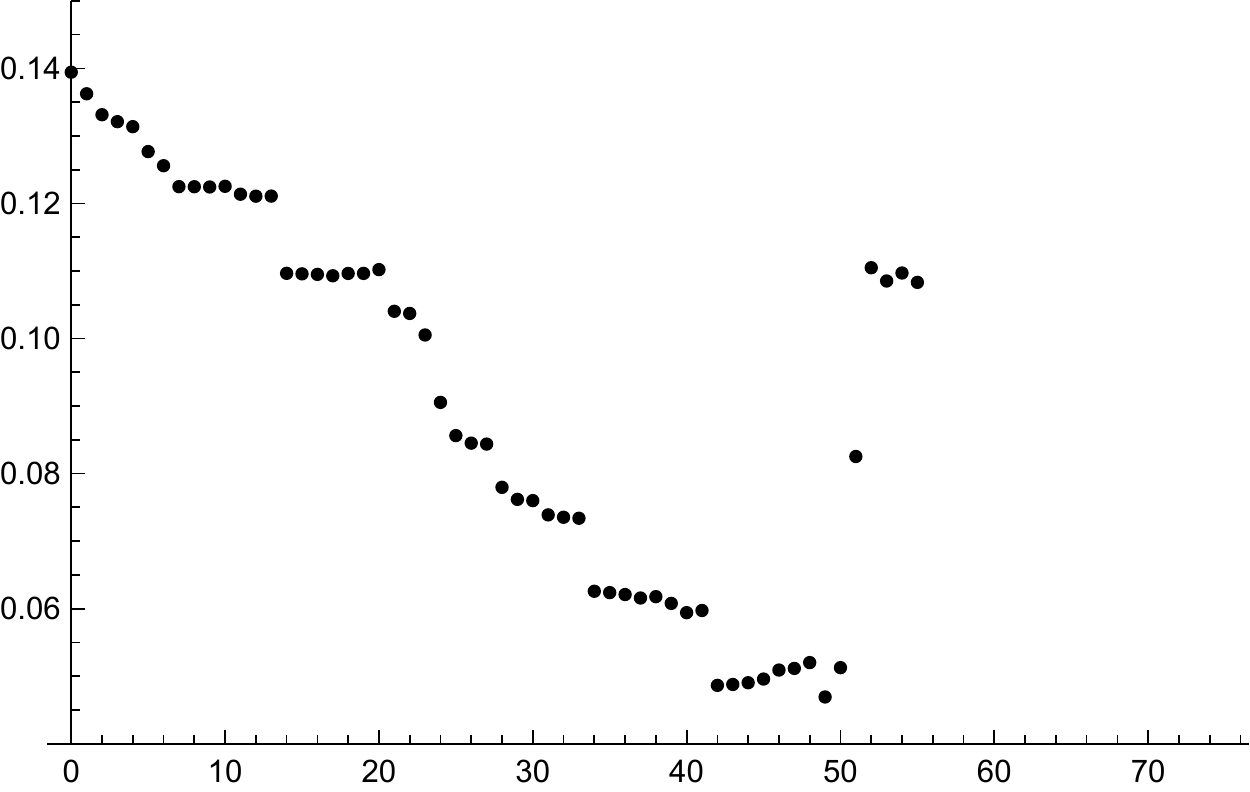}}

\vskip .5cm

\centerline{\includegraphics[width=2.9in]{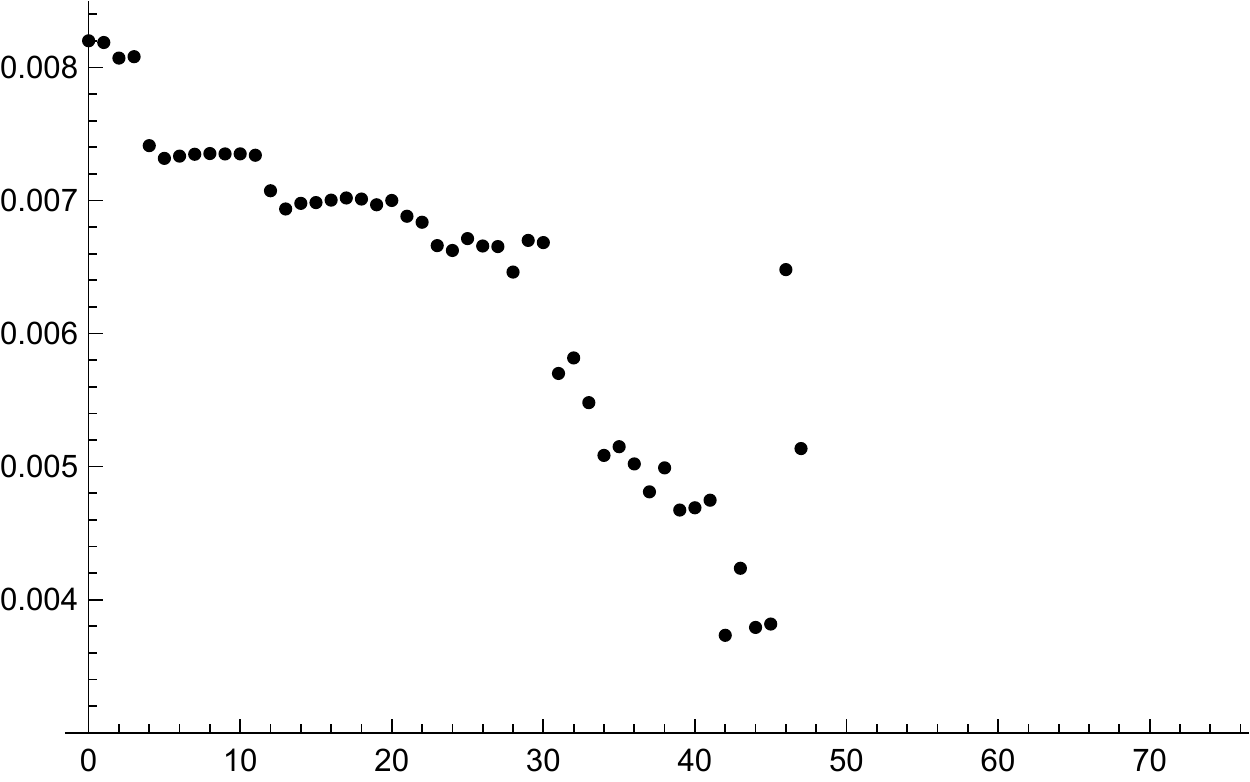}\hskip .2in\includegraphics[width=2.9in]{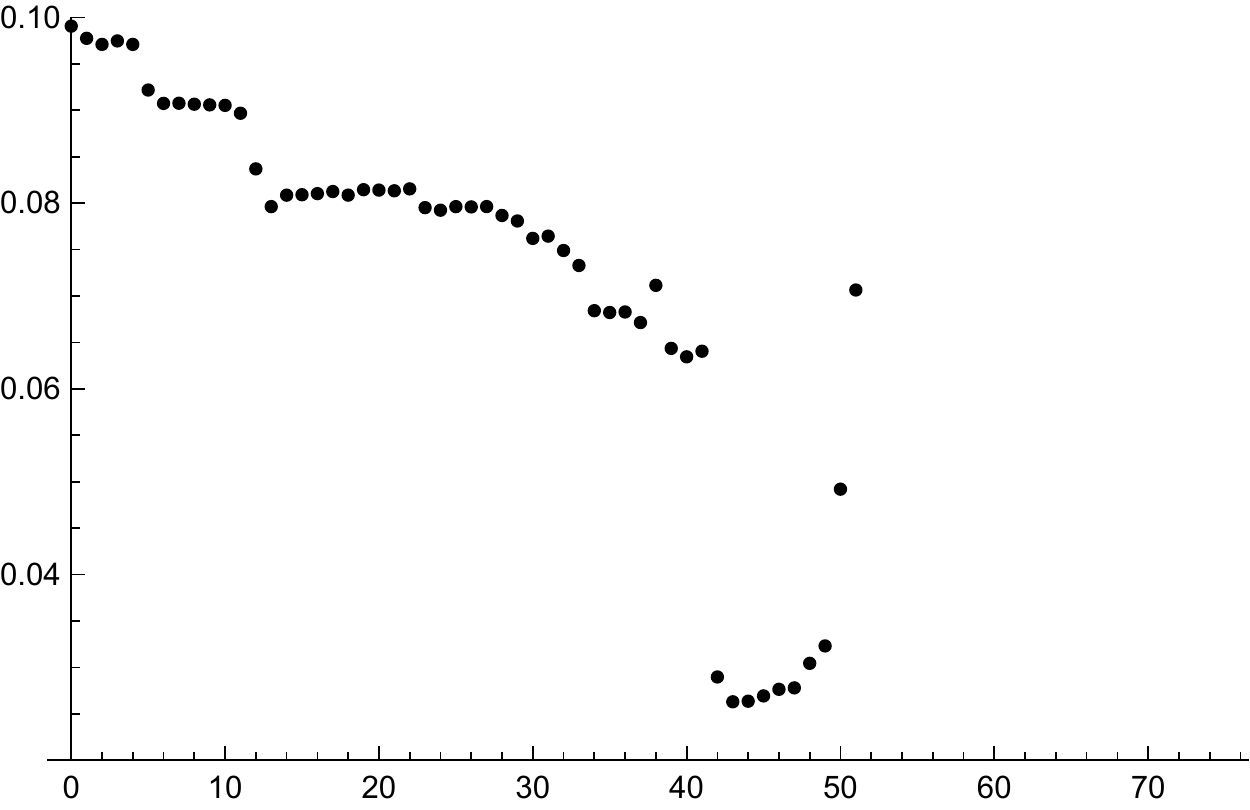}}

\caption{\label{fig22} The accuracy $\Delta(\bar n)$ as a function of $\bar n$, running the algorithm with the same parameters as in Fig.\ \ref{fig16}, but with different approximate data sets. Upper-left inset: the data set is for $(m,\Gamma)=(2,0.5)$ and $\sigma = 0.01$. The algorithm yields $\bar n_{0}=43$, $\bar w \simeq 2.36$, $\tilde n_{0}=44$ and $\tilde w \simeq 2.17$. Upper-right inset: the data set is for $(m,\Gamma)=(2,0.5)$ and $\sigma = 0.1$. The algorithm yields $\bar n_{0}=49$, $\bar w \simeq 2.97$, $\tilde n_{0}=44$ and $\tilde w \simeq 2.84$. Lower-left inset: the data set is for $(m,\Gamma)=(2,2)$ and $\sigma = 0.01$. The algorithm yields $\bar n_{0}=42$, $\bar w \simeq 2.20$, $\tilde n_{0}=44$ and $\tilde w \simeq 2.16$. Lower-left inset: the data set is for $(m,\Gamma)=(2,2)$ and $\sigma = 0.1$. The algorithm yields $\bar n_{0}=43$, $\bar w \simeq 3.77$, $\tilde n_{0}=44$ and $\tilde w \simeq 3.76$.}
\end{figure}

On Fig.\ \ref{fig22}, using the same values of $N$, $K$, $K_{\text{p}}$, $l_{\delta}$, $l_{\alpha}$ and $l_{u}$ as on Fig.\ \ref{fig16}, we plot the accuracy function starting from new data sets of various accuracies. In all cases, the algorithm yields an accuracy gain (optimal or effective) in the range 2--4.

\subsubsection{Working at precision $N=1000$}

\begin{figure}
\centerline{\includegraphics[width=2.9in]{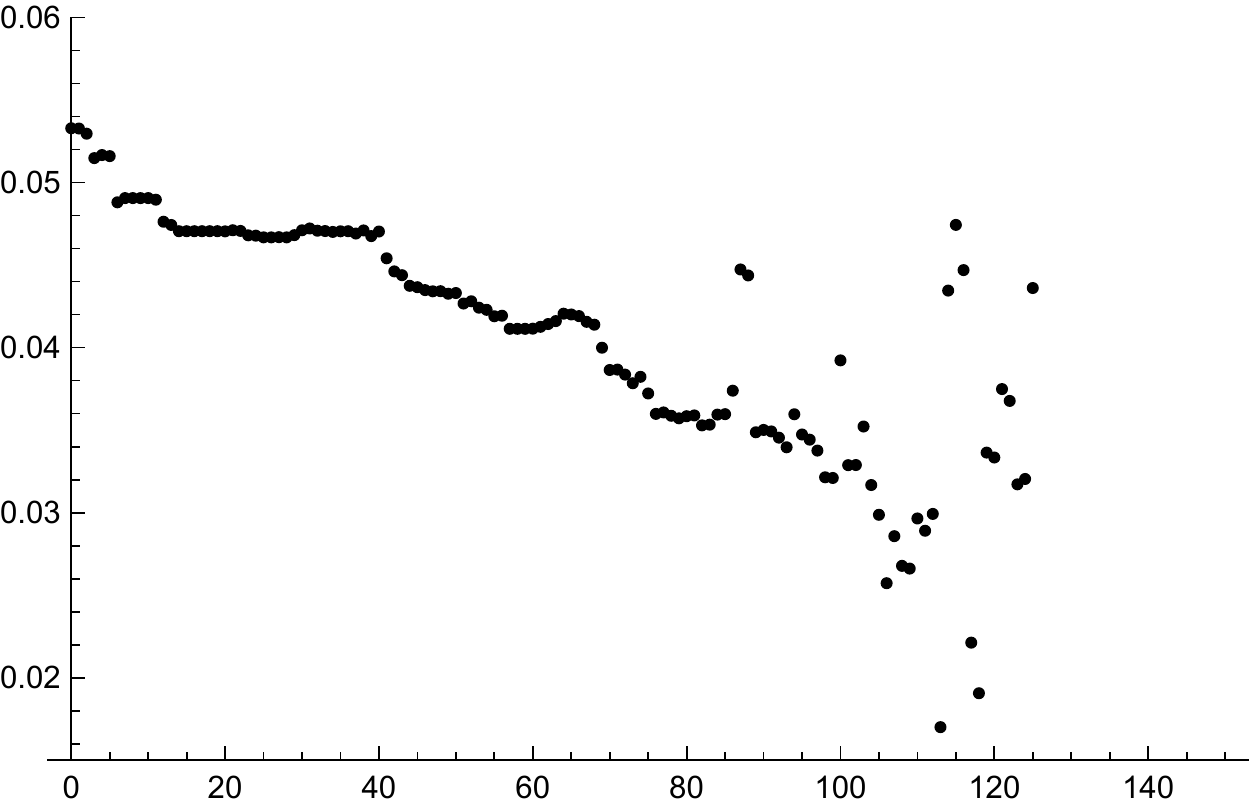}\hskip .2in\includegraphics[width=2.9in]{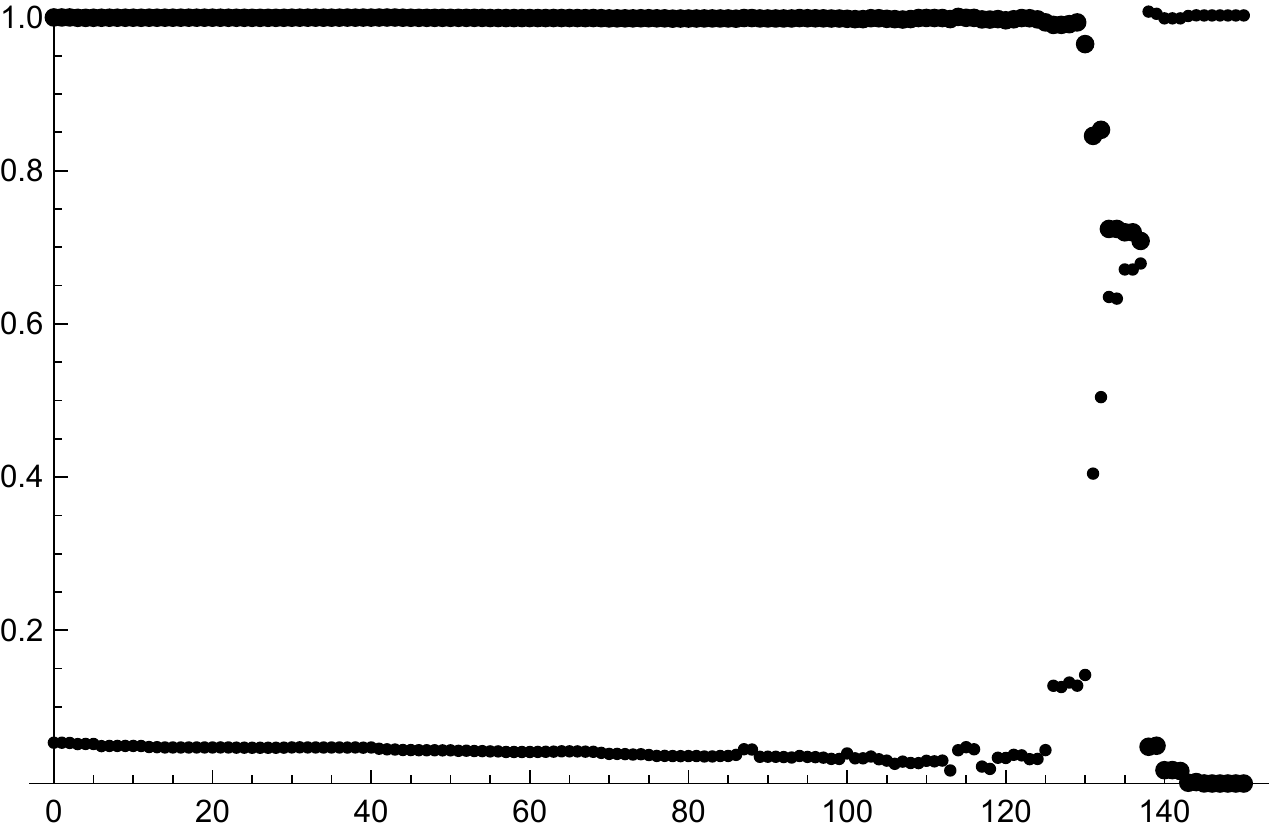}}

\vskip .5cm

\centerline{\includegraphics[width=2.9in]{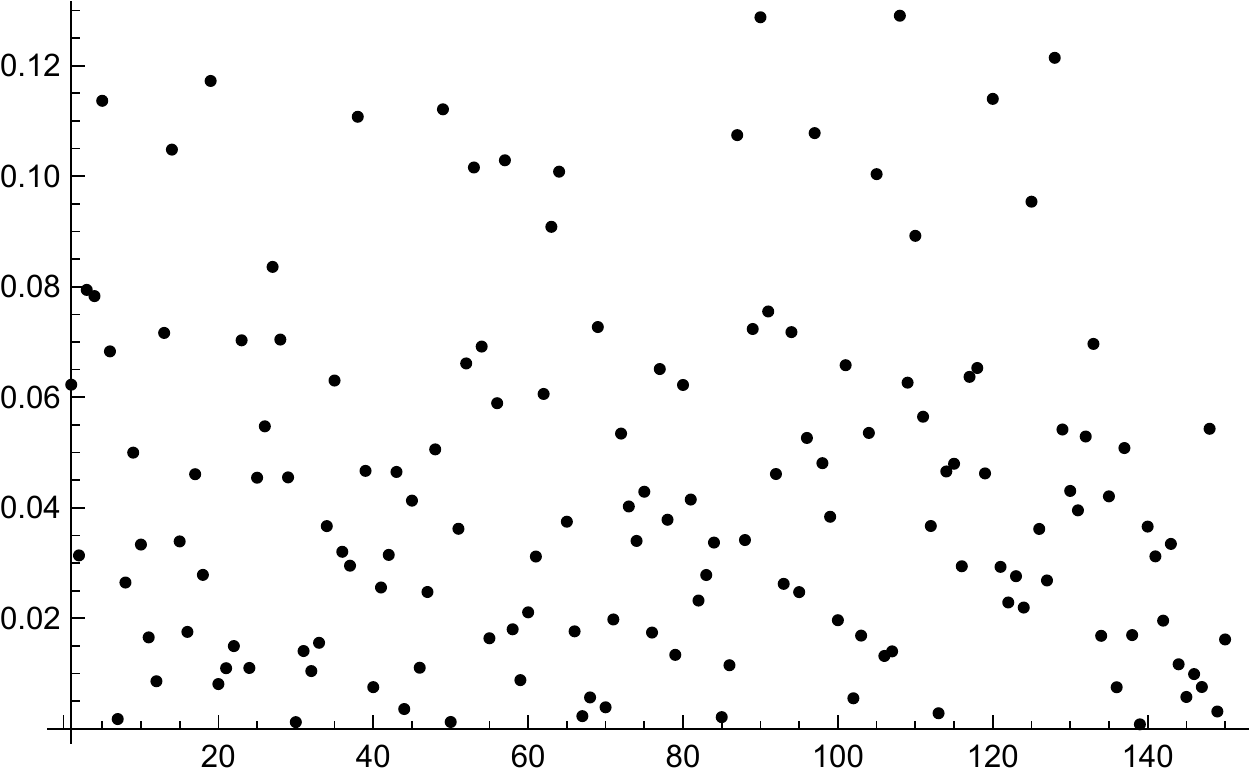}\hskip .2in\includegraphics[width=2.9in]{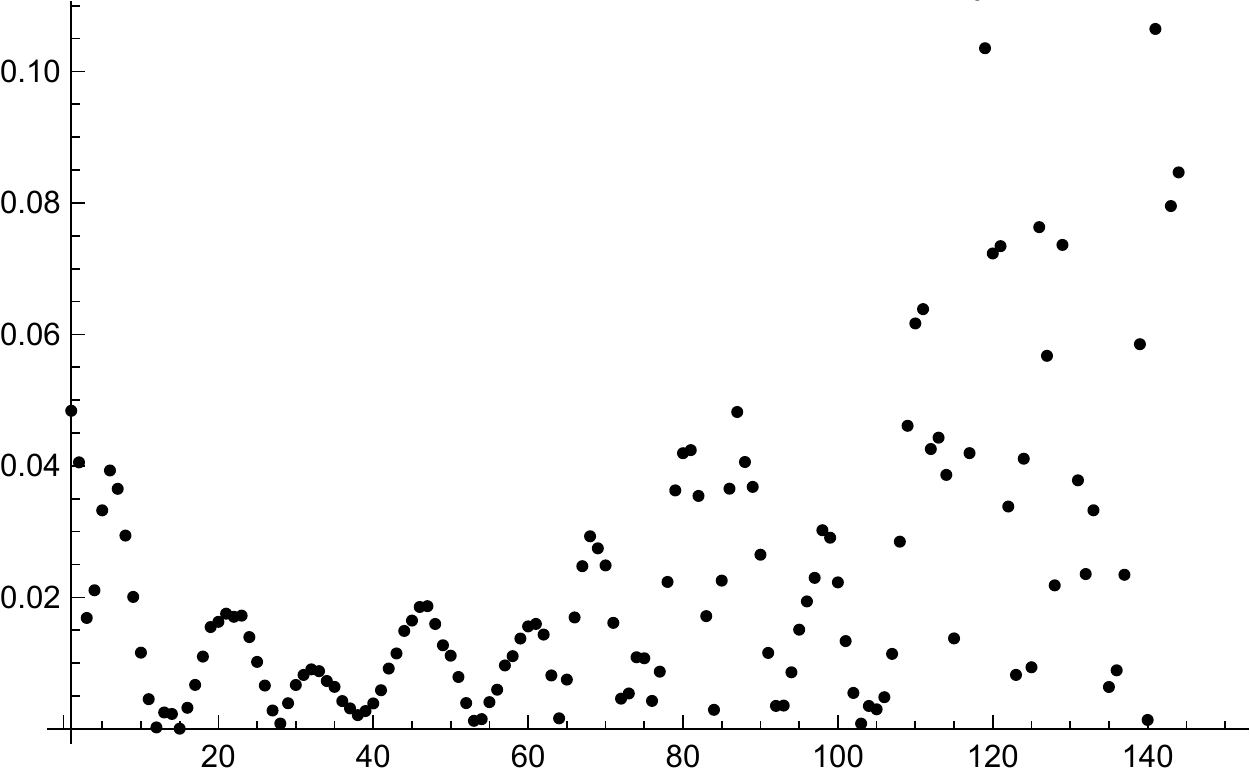}}

\caption{\label{fig23} The accuracy $\Delta(\bar n)$ (upper-left inset), the norm (thick dots) and the accuracy (thin dots) (upper-right inset), the detailed accuracy (coefficients by coefficients) of the original data set (lower-left inset) and the detailed accuracy of the best improved data set (lower-right inset) for $N=1000$, $K=150$, $K_{\text{p}}=50$, $l_{\delta}=\{1,2\}$, $l_{\alpha}=\{1/4,1/2,3/4,1\}$ and $l_{u}=\{3,5,7\}$. We get $\bar n_{0}=103$ which yields an optimal accuracy gain $\bar w\simeq 3.13$ and $\tilde n_{0}=103$ which yields an effective accuracy gain $\tilde w\simeq 1.51$.}
\end{figure}

\begin{figure}
\centerline{\includegraphics[width=2.9in]{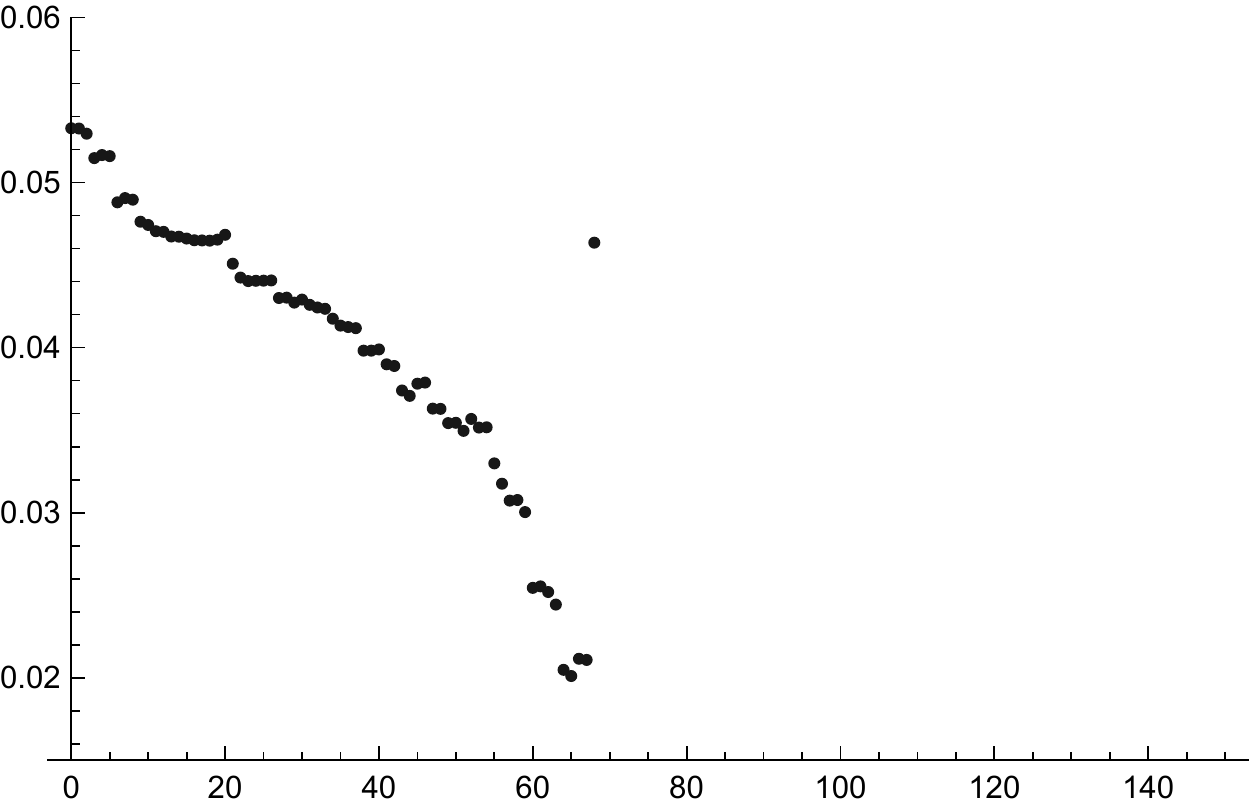}\hskip .2in\includegraphics[width=2.9in]{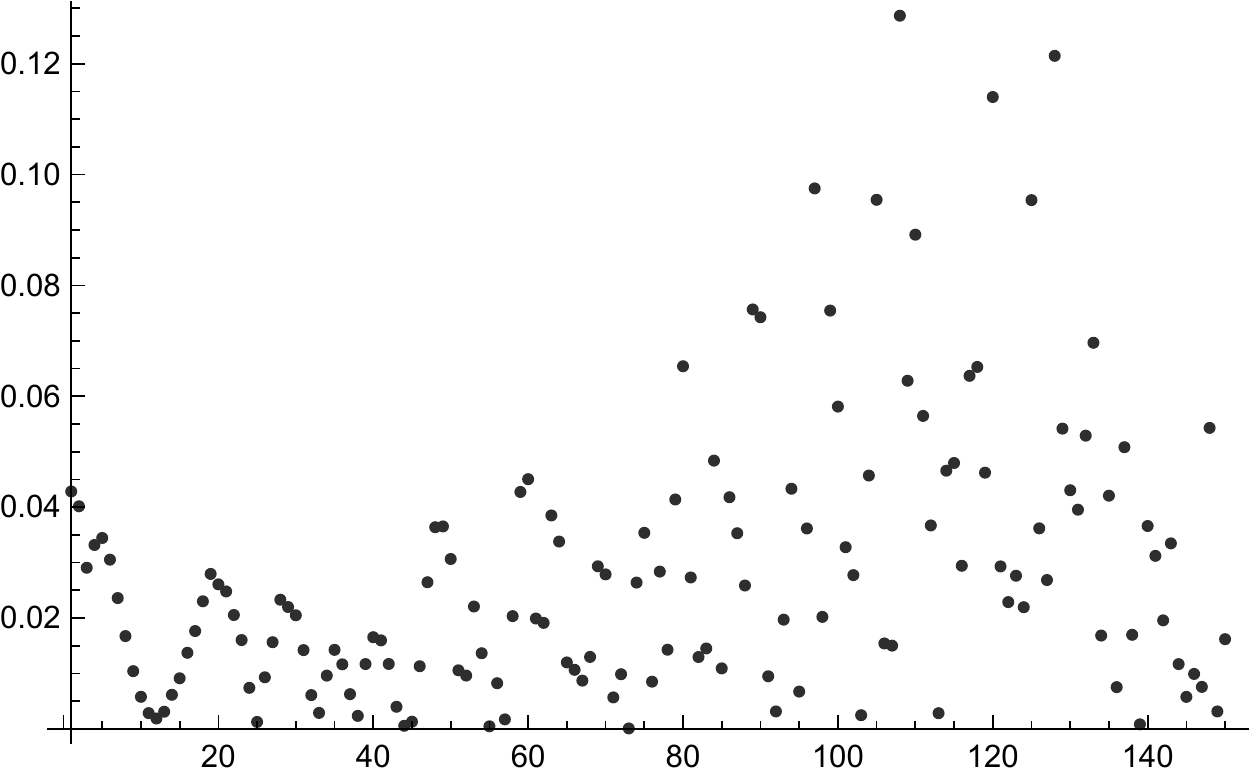}}

\caption{\label{fig24} Same as the upper-left and lower-right graphs of Fig.\ \ref{fig23}, except that we now use $l_{\delta}=\{1\}$ instead of $l_{\delta}=\{1,2\}$. The optimal accuracy gain is now $\bar w\simeq 2.65$, obtained for $\bar n_{0}=65$, which is a bit less than on Fig.\ \ref{fig23}, but the effective gain $\tilde w\simeq 2.09$, corresponding to $\tilde n_{0}=61$, is much better.}
\end{figure}

We now use the much improved precision $N=1000$. A UV cut-off $K=150$ is then adequate. We pick an approximate data set for our favorite values $(m,\Gamma)=(2,1/2)$ and $\sigma=0.05$. We run the algorithm with the same $l_{\delta}$, $l_{\alpha}$ and $l_{u}$ as in Fig.\ \ref{fig16}, but we now choose a larger $K_{\text{p}}=50$, consistently with the idea that the increased precision should allow to improve the Fourier coefficients up to some higher energy. The results are depicted on Fig.\ \ref{fig23}. The optimal accuracy gain $\bar w\simeq 3.13$ is excellent in this case, but the effective one $\tilde w\simeq 1.51$ is much lower. This is explained by the fact that the graph of the accuracy starts to become rather fuzzy for values of $\bar n$ as low as 85 and our simple algorithm to estimate $\bar n_{0}$ is not very good in such a case. The plot of the detailed accuracy of the improved data also shows that the Fourier coefficients are greatly improved up to $|k|$ in the range 60--70, which is much better than the value $\sim 40$ obtained when working with $N=200$, as expected. On Fig.\ \ref{fig24}, we have run the algorithm with $l_{\delta}=\{1\}$ instead of $l_{\delta}=\{1,2\}$. The graph of the accuracy is then sharper and the algorithm yields a better estimate $\tilde n_{0}$ of $\bar n_{0}$, with an effective accuracy gain of about $2.09$.

The conclusion is that using an improved precision does not seem to yield a much better accuracy gain for the algorithm. However, and as expected, the higher precision allows to work with a higher $K_{\text{p}}$.


\subsubsection{\label{BHlastSec} A last example}

We have mainly focused, in our applications, on the example of the damped harmonic oscillator, Eq.\ \eqref{resDO}--\eqref{chirDO2}. Even though this is a nice example capturing interesting physics, it is natural to ask whether the good results we have obtained might depend on the fact that the resolvent function is a very simple analytic function in this case. In more realistic examples, the resolvent is typically an extraordinarily complicated function for which no explicit closed-form formula is available. 

For this reason, we also include a more complicated and interesting example, corresponding to the large $\mathcal N$ solution of a quantum mechanical theory of $\mathcal N\times\mathcal N$ Hermitian matrices modeling some interesting properties of quantum black holes \cite{FerBH}. One can show that the full solution of the model is encoded in the Euclidean two-point function 
\be\label{GBHdef} G(\tau) =\frac{1}{\mathcal N}\sum_{i=1}^{\mathcal N} \bigl\langle \text{T} a_{\text E}^{i}(\tau) a^{\dagger}_{i}\bigr\rangle_{\beta}\, ,\ee
where the operators $a^{\dagger}_{i}$ and $a^{i}$ create and destroy strings interacting with the black hole.

Of course, our purpose here is not to discuss the physics of the model, which can be found in \cite{FerBH}, but instead to test our algorithm in a very non-trivial case. The Fourier coefficients $G_{k}$ for \eqref{GBHdef} cannot be found in closed form, but are determined in principle by a Schwinger-Dyson equation which is equivalent to the following infinite hierarchy of constraints on the coefficients,
\be\label{GkBHeqs} \frac{1}{G_{k}}+ i k - M = -\frac{\lambda}{2\pi}\sum_{k'\in\mathbb Z}\frac{G_{k'}}{(k-k')^{2}+m^{2}}\,\cdotp\ee
As usual, we chose the inverse temperature $\beta=2\pi$. The masses $m$, $M$ and the coupling $\lambda$ are parameters in the model. Note that the $G_{k}$ are not real but satisfy instead $G_{k}^{*}=G_{-k}$. Since the ARG equations are linear with real coefficients, we can use them to improve the real and imaginary parts $\re G_{k}$ and $\im G_{k}$ of the coefficients independently of each other.

\begin{figure}
\centerline{\includegraphics[width=2.9in]{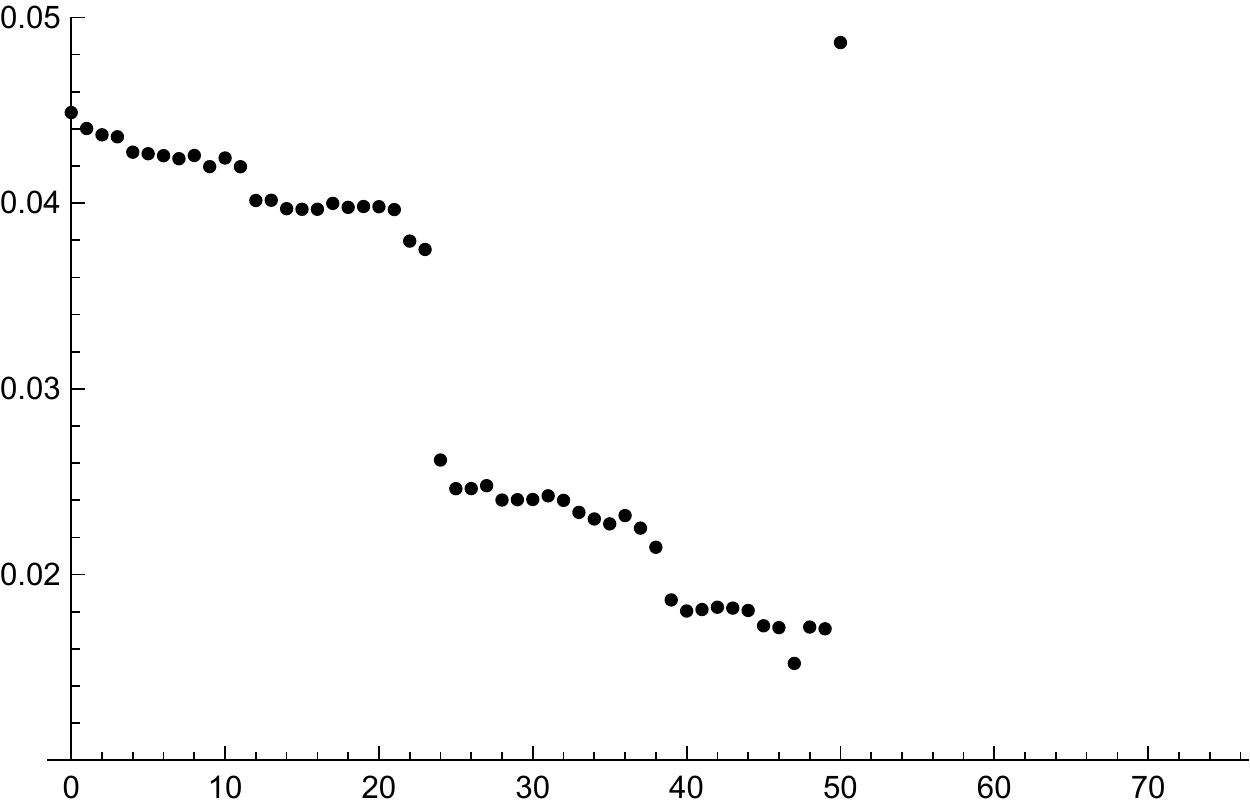}\hskip .2in\includegraphics[width=2.9in]{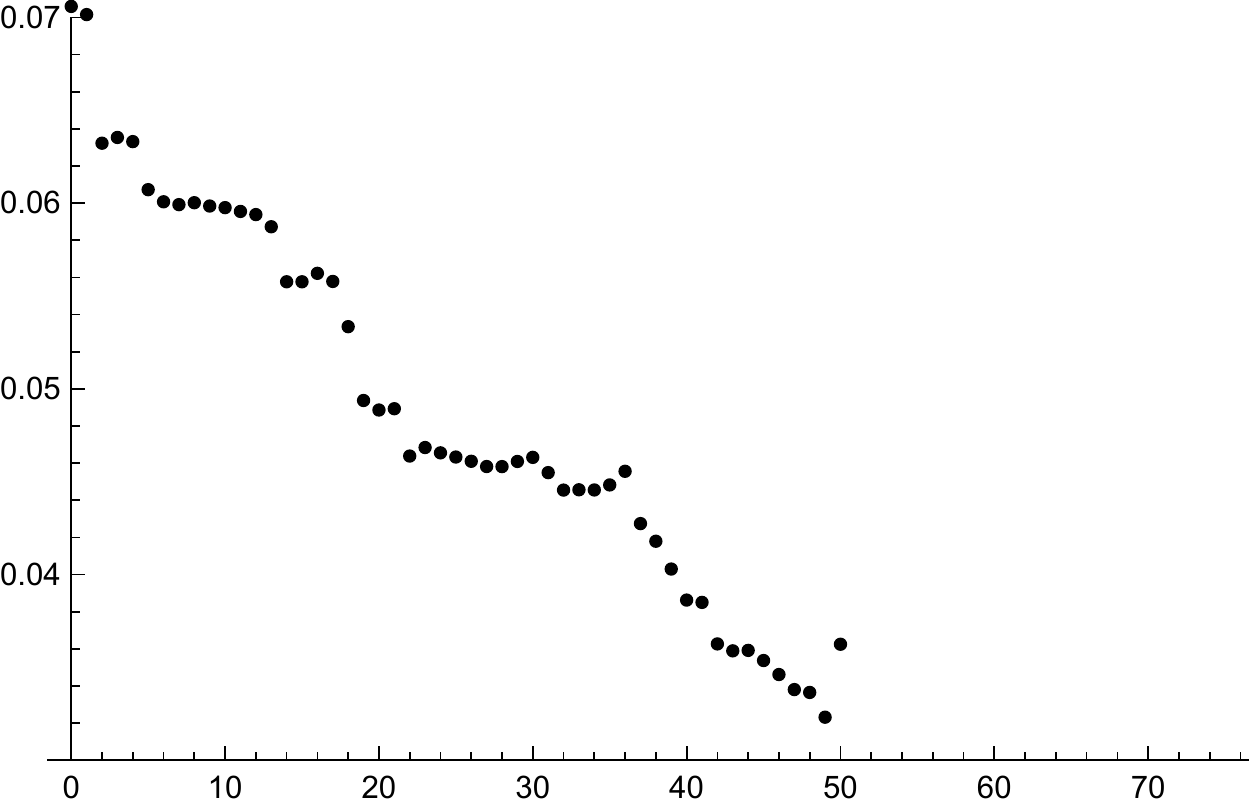}}

\caption{\label{fig25} The accuracies $\Delta(\bar n)$ for $\re G_{k}$ (left inset) and $\im G_{k}$ (right inset) for the model described in Section \ref{BHlastSec}, with the parameters $M=3$, $m=1$, $\lambda=1$ and $N=200$, $K=75$, 
$K_{\text{p}}=25$, $l_{\delta}=\{1,2\}$, $l_{\alpha}=\{1/4,1/2,3/4,1\}$ and $l_{u}=\{3,5,7\}$. The accuracy gains are $\bar w \simeq 2.95$ and $\tilde w\simeq 2.48$, corresponding to $\bar n_{0}=47$ and $\tilde n_{0}=44$ (left inset) and $\bar w \simeq 2.18$ and $\tilde w\simeq 1.97$, corresponding to $\bar n_{0}=49$ and $\tilde n_{0}=44$ (right inset).}
\end{figure}

It is possible to solve \eqref{GkBHeqs} numerically with great accuracy, see \cite{FerBH} for details. We have produced in this way an (almost) exact data set corresponding to the typical values $M=3$, $m=1$, $\lambda=1$ of the parameters and an approximate data set, using \eqref{GaGerel} for $\sigma=0.05$. On Fig.\ \ref{fig25} is depicted the result of the run of the algorithm, with our favorite values $N=200$, $K=75$, $K_{\text{p}}=25$, $l_{\delta}=\{1,2\}$, $l_{\alpha}=\{1/4,1/2,3/4,1\}$ and $l_{u}=\{3,5,7\}$. The outcome is excellent. We get accuracy gains (optimal or effective) in the range 2--3.

\section{\label{SecConc} Conclusion}

A central result of our work is to show that analyticity implies an infinite set of linear equations that any set of Fourier-Matsubara coefficients $G_{k}$ associated with a quantum mechanical finite temperature Euclidean two-point function must satisfy. Some of these equations admit an interesting Renormalization Group interpretation. Remarkably, these equations can be used to improve systematically any random approximate data set obtained, for example, from Monte-Carlo simulations.

Our main intention in this paper was to explain the main ideas and equations, with the physics applications in mind. It would be interesting to have a more complete, mathematically rigorous, presentation. In particular, a detailed discussion of the linear dependence between the ARG equations, that we have explicitly seen numerically (see e.g.\ Fig.\ \ref{fig15}), would be handy. Precise statements about how the finite dimensional spaces $\mathscr M_{K}\subset\mathscr F_{K}$ approximate $\mathscr M\subset\mathscr F$ when $K\rightarrow\infty$ would also be useful. Our results on the codimension $\bar n_{0}$ of $\mathscr M_{K}$ suggest that $\dim\mathscr M_{k}/\dim\mathscr F_{K} \sim 0.4$. Can we make this statement precise, in particular in the limit $K\rightarrow\infty$? More generally, a direct analysis of the geometry of $\mathscr M$ in infinite dimension, which we have avoided because the practical applications always deal with finite dimensional spaces, is desirable. Moreover, some fine aspects of our results, for example the curious but clearly visible oscillatory structure of the improved data sets seen in Fig.\ \ref{fig18}, \ref{fig21}, \ref{fig23} and \ref{fig24}, require a better understanding.

But the most compelling goal to pursue is probably to better assess how effective the use of the ARG equations can be in real-world problems. To do so, one has to apply our algorithm, or, better, some significantly improved version thereof, to the Monte-Carlo data found in interesting strongly coupled problems, including lattice QCD and condensed matter systems.

\subsection*{Acknowledgments}

This research is supported in part by the Belgian Fonds National de la Recherche Scientifique FNRS (convention IISN 4.4503.15, CDR grant J.0088.15 and MS grant) and the Advanced ARC project ``Holography, Gauge Theories and Quantum Gravity.''

%
%

%

%

%

\begin{thebibliography}{99}
%

%
\bibitem{FerBH}{F.~Ferrari, \emph{Black Hole Horizons and Bose-Einstein Condensation,} arXiv:1601.08120,\\
F.~Ferrari, \emph{Large $N$ Matrix Quantum Mechanics, Bose-Einstein Condensation and Black Hole Horizons,} arXiv:16mm.nnnn.}
%
\bibitem{Rubel}{L.A.~Rubel, {\it Trans.\ Amer.\ Math.\ Soc.\ }{\bf 83} (1956) 417.}
%
\bibitem{Cuniberti}{G.~Cuniberti, E.~De Micheli, G.A.~Viano, \cmp{216}{2001}{59}, cond-mat/0109175.}
%
\bibitem{Burnier}{Y.~Burnier, M.~Laine and L.~Mether, \epjc{71}{2011}{1619}, arXiv:1101.5534.}
%
\bibitem{Bateman}{H.~Bateman, \emph{Higher Transcendental Functions}, Volume II, McGraw-Hill, 1953, p.\ 191.}
%
\end{thebibliography}
\end{document}